\documentclass[epj]{svjour}
\pdfoutput=1
\usepackage{amsmath}

\newcommand{\beq}{\begin{equation}}
\newcommand{\eeq}{\end{equation}}

\newcommand{\di}{\displaystyle}

\newcommand{\si}{\sigma}

\newcommand{\D}{\displaystyle}

\newcommand{\gep}{G_{Ep}}
\newcommand{\gmp}{G_{Mp}}
\newcommand{\gmn}{G_{Mn}}
\newcommand{\gmpmu}{G_{Mp}/\mu_{p}}

\newcommand{\gen}{G_{En}}
\newcommand{\gmnmu}{G_{Mn}/\mu_{n}}

\usepackage{array, graphicx, subfigure}
\usepackage{graphics}
\begin{document}
\title{ Neutrino Quasielastic Scattering on Nuclear Targets}
\subtitle{Parametrizing Transverse Enhancement (Meson Exchange Currents)}
\author{A. Bodek\inst{1}, H.S. Budd\inst{1} and M. E. Christy\inst{2}}
\institute{Department of Physics and Astronomy, University of
Rochester, Rochester, NY  14627-0171 USA
\and Hampton University; Hampton, Virginia, 23668 USA}
%\author{Arie Bodek and Howard Budd}
%\institute{Department of Physics and Astronomy, University of
%Rochester, Rochester, NY  14627-0171}

%
\date{Received: date / arXiv:1106.0340. Revised  July 21, 2011, to be published in Eur. Phys. J. C}
%\date{Received: date / Revised version: date July 11, 2011}
\abstract{
We present a parametrization of  the observed  enhancement in the transverse electron  quasielastic (QE) response function for nucleons bound in carbon as a function of the square of the four momentum transfer ($Q^2$) in terms of a  correction to the magnetic form factors of bound nucleons. The parametrization should also be applicable to the transverse cross section in  neutrino scattering. If the transverse enhancement  originates from  meson exchange currents (MEC), then it is theoretically expected that any enhancement in the  longitudinal or axial contributions is small.    We present the predictions of the "Transverse Enhancement" model (which is based on electron scattering data only)  for the   $\nu_\mu, \bar{\nu}_\mu$ differential and total QE cross sections for nucleons bound in carbon. The $Q^2$ dependence of the transverse enhancement is observed to resolve much of the  long standing discrepancy in the QE total cross sections and differential distributions between
  low energy and high energy neutrino experiments on nuclear targets.
}
\maketitle

\section{Introduction}
\label{intro}
A reliable description of the neutrino ($\nu_\mu$) and antineutrino ($\bar{\nu}_\mu$) quasielastic (QE)  and inelastic scattering  
processes (particularly on nuclear targets)  is essential for precision studies of  $\nu_{\mu},\bar{\nu}_\mu$ oscillation~\cite{ATM,MINOS}
parameters such as mass splitting and  mixing angles. In addition to modeling the $\nu_{\mu},\bar{\nu}_\mu$ cross sections\cite{by}, a reliable  model of
the hadronic final states is needed because the hadronic energy
response of $\nu_\mu$ detectors is not the same
for  protons, neutrons,  pions, photons, 
and nuclear fragments.  Prescriptions
which can be readily incorporated into existing $\nu_\mu$ Monte Carlo
 generators\cite{GENIE} are preferable. 
% In this communication we
%study  the effects of several parametrizations of the contributions from  from meson exchange %currents (MEC) in nuclear targets.

%Experimental investigation of neutrino QE scattering at low energies
%indicates that the simple
 Models which assume that QE scattering on nuclear targets can be described in terms of scattering from independent nucleons bound in a nuclear potential (e.g. Fermi gas\cite{bodek-ritchie} 
or spectral functions) 
do not provide an adequate representation of measured differential and total 
QE cross sections for low energy ($\approx 1 GeV$) $\nu_\mu$ scattering on nucleons bound in  carbon\cite{MiniBooNE,new} (MiniBooNE) 
 and oxygen\cite{T2K,K2K} (K2K and T2K). The measured QE total cross sections are 20\% larger than the model and 
 the  differential distributions in $Q^2$ are also inconsistent. 
%and Iron\cite{MINOS}. 
The vector and axial form factors that are used in  independent nucleon models 
are the free nucleon form factors extracted from electron and $\nu_{\mu},\bar{\nu}_\mu$ scattering data on hydrogen and
deuterium\cite{quasi}.

Although there are more sophisticated calculations of quasielastic scattering (e.g.  relativistic distorted-wave impulse approximation\cite{other}),  it is the simple independent nucleon model that has been implemented in the currently
available neutrino cross section Monte Carlos\cite{GENIE}.

This disagreement between the measured low energy $\nu_\mu$  differential and total QE cross sections 
on nuclear targets and the predictions from the independent nucleon model has been attributed to an incomplete description of nuclear effects. 
These additional nuclear effects have been parametrized as an ad-hoc change in the 
the axial form factor mass parameter from the value measured for free nucleons\cite{quasi} ($M_A^{free}=1.014\pm0.014 ~GeV$) to $M_A^{eff}=1.20 \pm 0.12 ~GeV$ (K2K)  and
$M_A^{eff}=1.23  \pm 0.20  ~GeV$ (MiniBooNE). 
%The total and differential  QE cross sections of the independent nucleon model with the larger value of $M_A$ are in agreement with the low energy
 % data.

%FIGURE ! 
  \begin{figure*}
 \begin{center}
\includegraphics[width=6.6in,height=3.0in]{{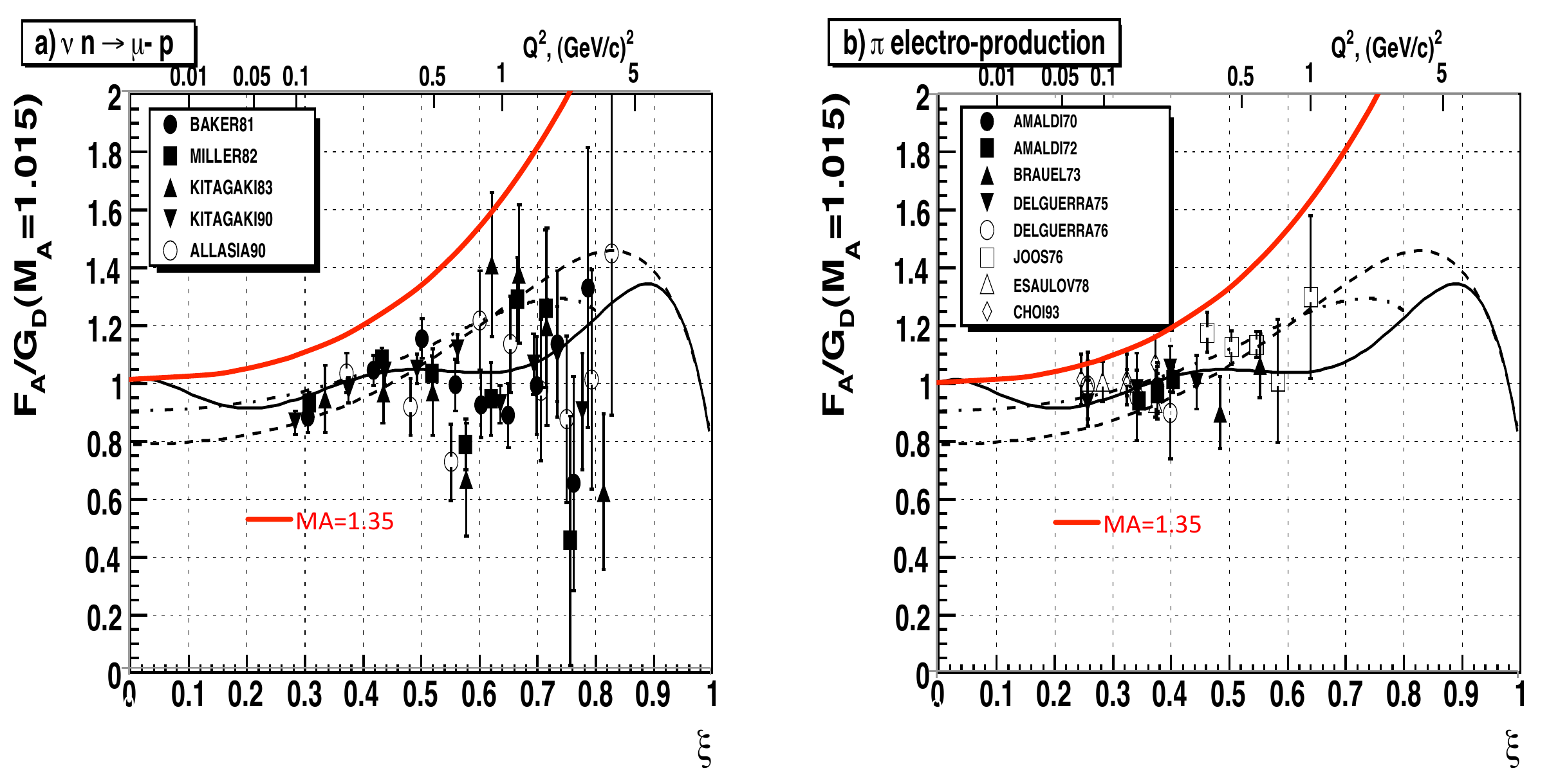}}
\caption[$F_{A}$:Axial Form factor ratios to $G_D^{A}$]{
 (a) $F_A(Q^2)$  extracted from $\nu_\mu$-deuterium
 data\cite{quasi} divided by $G_D^{A}(Q^2)$ with $M_A=1.015~GeV$. (b)  
$F_A(Q^2)$ from pion electroproduction (corrected for hadronic effects) 
  divided by $G_D^{A}(Q^2)$ with $M_A=1.015~GeV$.
  Thin solid line - duality based fit from reference \cite{quasi};
   Short-dashed line - $F_A(Q^2)_{A2=V2}$. 
 % The long-dashed line is $F_A(Q^2)_{A1=V1}$.
  Dashed-dot line - constituent quark model;
  Thick solid red line $F_A(Q^2)=G_D^{A}(Q^2)$=$ \frac {-1.267}{(1+Q^2/M_A^2)^2}$ with 
$M_A=1.35~GeV$. The  horizontal scale on top is $Q^2$. The horizontal scale
on the bottom is the target mass scaling variable   $\xi$ for elastic scattering ($x=1$). Here  
$\xi=\frac{2}{(1+\sqrt{1+1/\tau})}$, $\tau = Q^2/4M^2$, and M is the average nucleon mass.
 %  The values
%of $\xi$ and the corresponding
%values of $Q^2$ are shown on the bottom and top axis. 
}
 \end{center}
 %
 %---------------------------------------------
     \label{figaxial}  
\end{figure*}

A recent analysis\cite{new} of newly published differential QE cross sections from MiniBooNE (for nucleons bound in  carbon)  yields  larger values of $M_A^{eff}=1.350\pm0.066 ~GeV$
in the Fermi gas model and $M_A^{eff}=1.343\pm0.060 ~GeV$
in the spectral function model. In that analysis the free nucleon value
  $M_A^{free}=$ $1.014~GeV$ is excluded  at the confidence level greater than 5$\sigma$ (standard deviations).

Figure~\ref{figaxial} shows the  world's data \cite{quasi} for the nucleon axial
form factor  ($F_A (Q^2)$) extracted from
QE $\nu_{\mu},\bar{\nu}_\mu$ scattering
on  hydrogen and deuterium. Here, the data for  $F_A(Q^2)$ are shown as
a ratio to a nominal  dipole   $F_A(Q^2)=G_D^{A}(Q^2)$=$ \frac {-1.267}{(1+Q^2/M_A^2)^2}$ with 
$M_A=1.015~GeV$. On the left side we show the values extracted from 
$\nu_{\mu},\bar{\nu}_\mu$  experiments on hydrogen and  deuterium and on the right side we show the
values extracted from pion electro-production data on hydrogen.  
The average of the  measurements of $M_{A}$ from 
$\nu_{\mu},\bar{\nu}_\mu$  experiments on hydrogen and deuterium of  $M_{A}^{\nu_{\mu},\bar{\nu}_\mu}$ = $1.016 \pm 0.026$  $GeV$
is in  agreement with the average value
of $M_{A}^{pion}$=$1.014 \pm 0.016$  $GeV$ extracted 
from  pion electro-production experiments on hydrogen 
(after corrections for hadronic effects).
The average of the $\nu_{\mu},\bar{\nu}_\mu$  and electro-production 
values is \cite{quasi} 
$M_{A}^{world-av}$ = $1.014 \pm 0.014~GeV$.  The thin solid line is a duality based
parametrization\cite{quasi} of possible deviations from the dipole form. The dashed-dot line is
the prediction of a constituent quark model\cite{quark} and the short-dashed line
is the expectation for $F_A(Q^2)$ if the vector and axial-vector structure functions are equal (eg. ${\cal W}_{2}^{Qelastic-vector} = {\cal W}_{2}^{Qelastic-axial}$). 
 
It is clearly observed that a dipole axial form factor with $M_A=1.35~GeV$ (thick solid red line) is inconsistent with the measurements on hydrogen and deuterium.

It has been assumed
that an  $\it "effective"$ axial mass provides an adequate  description of the missing nuclear corrections.  However,  a large increase in the axial form factor  of bound nucleons is contrary
to theoretical expectations that  
$M_{A}$  in nuclear targets should  be smaller\cite{ma-nuclear} than 
 (or the same\cite{Tsushima_03})
 as in deuterium.

Additionally, the low energy neutrino data appear to be in disagreement with higher energy neutrino experiments on nuclear targets.   At high neutrino energies, the total and differential QE cross sections on nuclear targets are consistent with  models which assume that the scattering is on independent nucleons with free nucleon form factors.  For example,   $M_A$ of  $0.979 \pm 0.016$ GeV has been extracted  from a global analysis\cite{quasinuclear}of the differential distributions and total QE cross sections  measured in  $all$ high  energy $\nu_{\mu}$ experiments on nuclear targets.  

Recent measurements of  the differential and total QE cross section for nucleons bound in  carbon by the NOMAD\cite{NOMAD} collaboration for $\nu_{\mu},\bar{\nu}_\mu$ energies above  4 GeV are also consistent with models which assume that the scattering is from  independent nucleons with free nucleon form factors. 
The NOMAD analysis  yields a value of  $M_A$ ($1.05\pm0.02\pm0.06~GeV$)
% If  meson exchange currents contribute to the  electromagnetic
 % and $\nu_{\mu},\bar{\nu}_\mu$  structure functions, they contribute at low and high energies. 
 
 Therefore,  the results of the higher energy and low energy $\nu_{\mu}$ experiments on nuclear targets appear to be inconsistent with each other.

In this communication we investigate the transverse enhancement observed in QE electron scattering experiments on nuclear targets.  We obtain a parametrization of the enhancement and investigate its implication for   $\nu_{\mu},\bar{\nu}_\mu$ scattering.  We show that the $Q^2$ dependence of the transverse enhancement resolves much of the discrepancy between the low energy and high energy neutrino experiments, in addition to obviating the need for an ad-hoc nuclear modification to $M_A$.  
%We show that the low energy and
%high energy neutrino experiments are  consistent with each other. 
%nvestigated parametrization which is consistent with theoretical expectations,
%and
%
\section{Electron-nucleon scattering}
%
%In this section we define the kinematic variables for the case  of charged lepton
%scattering from neutrons and protons. 
The differential cross section for scattering of an
unpolarized charged lepton with an incident energy $E_0$, final energy
$E^{\prime}$ and scattering angle $\theta$ can be written in terms of
the structure functions ${\cal F}_1$ and ${\cal F}_2$ as:
\begin{tabbing}
$\frac{d^2\sigma}{d\Omega dE^\prime}(E_0,E^{\prime},\theta)  =
   \frac{4\alpha^2E^{\prime 2}}{Q^4} \cos^2(\theta/2)$  \\ \\
  $\times   \left[{\cal F}_2(x,Q^2)/\nu +  2 \tan^2(\theta/2) {\cal F}_1(x,Q^2)/M\right]$
\end{tabbing}
where $\alpha$ is the fine structure constant, $M$ is the nucleon
mass, $\nu=E_0-E^{\prime}$ is the energy of the virtual photon which
mediates the interaction, $Q^2=4E_0E^{\prime} \sin ^2 (\theta/2)$ is
the invariant four-momentum transfer squared, and $x=Q^2/2M\nu$ is the 
Bjorken scaling variable.  We define
${\cal F}_2=\nu {\cal W}_2$, ${\cal F}_1=M{\cal W} _1$ 
(and for $\nu_\mu,\bar{\nu}_\mu$ scattering  ${\cal F}_3=\nu {\cal W}_3$).

	Alternatively, one could view this scattering process 
as virtual photon
absorption.  Unlike the real photon, the virtual photon can have two
modes of polarization.  In terms of the cross section for the
absorption of transverse $(\sigma_T)$ and longitudinal $(\sigma_L)$
virtual photons, the differential cross section can be written as,
\begin{equation}
\frac{d^2\sigma}{d\Omega dE^\prime} =
   \Gamma \left[\sigma_T(x,Q^2) + \epsilon \sigma_L(x,Q^2) \right]
\end{equation}
where,
\begin{eqnarray}
 \Gamma &=& \frac{\alpha K E^\prime}{ 4 \pi^2 Q^2 E_0}  \left( \frac{2}{1-\epsilon } \right) \\
\epsilon &=& \left[ 1+2(1+\frac{Q^2}{4 M^2 x^2} ) \tan^2 \frac{\theta}{2} \right] ^{-1} \\
K &=& \frac{2M \nu - Q^2 }{2M}.
\end{eqnarray}

The quantities $\Gamma$ and $\epsilon$ represent the flux and the
degree of longitudinal polarization of the virtual photons
respectively, which the quantity $R$ is defined as the ratio
$\sigma_L/\sigma_T$, and is related to the structure functions by
\begin{equation}
 R(x,Q^2)
   = \frac {\sigma_L }{ \sigma_T}
   = \frac{{\cal F}_2 }{ 2x{\cal F}_1}(1+\frac{4M^2x^2 }{Q^2})-1
   = \frac{{\cal F}_L }{ 2x{\cal F}_1},
\end{equation}
where ${\cal F}_L$ is called the longitudinal structure function. The
structure functions are expressed in terms of $\sigma_L$ and
$\sigma_T$ as follows:
\begin{eqnarray}
 {\cal F}_1 &=& \frac{M K }{ 4 \pi^2 \alpha} \sigma_T, \\
 {\cal F}_2 &=& \frac{\nu K (\sigma_L + \sigma_T)}{4 \pi^2 \alpha (1 + 
 \frac{Q^2 }{4 M^2 x^2} )} \\
 {\cal F}_L(x,Q^2) &=& {\cal F}_2 \left(1 + \frac{4 M^2 x^2 }{ Q^2}\right) - 2x{\cal F}_1
\end{eqnarray}
or,
\begin{equation}
2x{\cal F}_1 = {\cal F}_2 \left(1 + \frac{4 M^2 x^2 }{ Q^2}\right) -  {\cal F}_L(x,Q^2).
\label{eq:fl-rel}
\end{equation}
In addition, $2x{\cal F}_1$ is given by
\begin{eqnarray}
2x{\cal F}_1 (x,Q^{2}) &=& {\cal F}_2 (x,Q^{2}) 
%\times  \\ && 
\frac{1+4M^2x^2/Q^2}{1+R(x,Q^{2})}  \nonumber 
\end{eqnarray}
or equivalently
\begin{eqnarray}
{\cal W}_1 (x,Q^{2}) &=& {\cal W}_2(x,Q^{2})
% \times  \\&& 
\frac{1+\nu^2/Q^2}{1+R(x,Q^{2})}  \nonumber 
\end{eqnarray}
In the case of elastic scattering from free nucleons ($x=Q^2/2M\nu$=1) the structure functions
are related to the nucleon form factors by the following expressions\cite{steffens}:
$${\cal W}_{1p}^{elastic} =\delta(\nu-\frac{Q^2}{2M})\tau |\gmp (Q^2)|^2$$
$${\cal W}_{1n}^{elastic} =\delta(\nu-\frac{Q^2}{2M})\tau |\gmn (Q^2)|^2$$
and 
$${\cal W}_{2p}^{elastic} =  \delta(\nu-\frac{Q^2}{2M})
 \frac{[\gep(Q^2)]^2+ \tau [\gmp(Q^2)]^2}{1+\tau}$$
 $${\cal W}_{2n}^{elastic} =
 \delta(\nu-\frac{Q^2}{2M})
 \frac{[\gen(Q^2)]^2+ \tau [\gmn(Q^2)]^2}{1+\tau}$$
 $$ R_{p,n}^{elastic}(x=1, Q^2)=  \frac {\sigma_L^{elastic} }{ \sigma_T^{elastic}}= \frac{4M^2}{Q^2}\left(\frac{G_E^2}{G_M^2}\right)$$
 Here, $\tau= Q^2/4M_{p,n}^2$, where  $M_{p,n}$ are the masses of  proton and neutron.
 Therefore,  $\gmp$ and  $\gmn$ contribute to the transverse virtual
 photo-absorption cross section, and  $\gep$ and  $\gen$ contribute
 to the longitudinal cross section. 
 % 
 %fFigure 2
 \begin{figure*}
 \begin{center}
\includegraphics[width=6.6in,height=4.8in]{{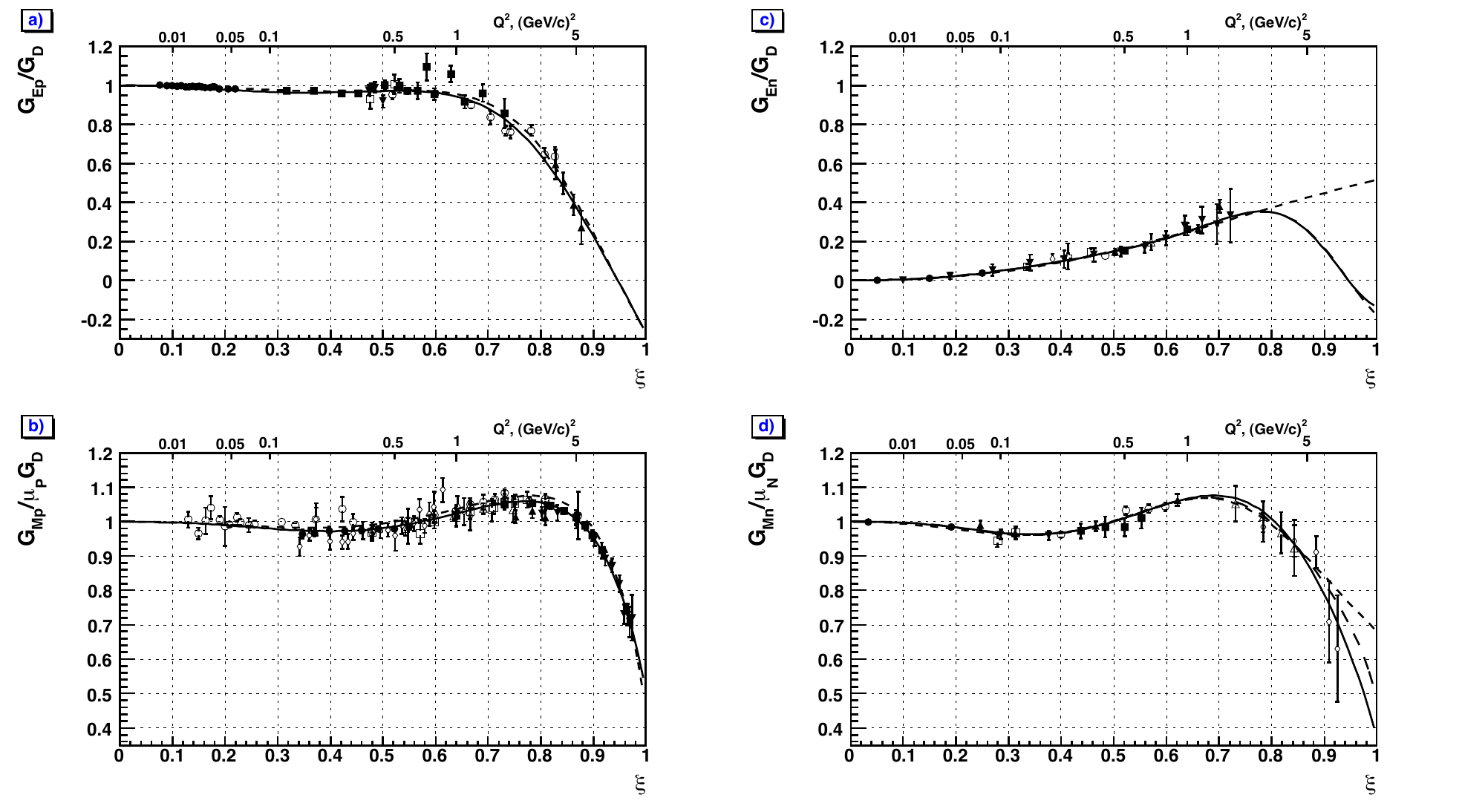}}
   \caption[Ratios to $G_D^V$]{Ratios of $\gep$ (a), $\gmpmu$ (b),
   $\gen$ (c) and
$\gmnmu$ (d) to $G_D^V = \frac {1}{(1+Q^2/M_V^2)^2}$ with $M_V=0.8426~GeV$.   
The short-dashed line in each plot is
the old Kelly\cite{kelly}  parameterizations (old Galster\cite{galster}  for $\gen$). 
The solid line is the $BBBA07_{25}$ and the long-dashed line is
 $BBBA07_{43}$ parametrizations\cite{quasi}, respectively.
The values
of $\xi$ and the corresponding
values of $Q^2$ are shown on the bottom and top axis. 
}
 \end{center}
   \label{subfig1}
\end{figure*}
 \section{ Nucleon form factors}
The nucleon  electromagnetic form factors are best
described by the  $BBBA2007_{25}$  duality based parametrization\cite{quasi}.
The $deviations$ from the dipole form factors are parametrized
by multiplicatives  functions   $A_N(\xi)$ for each of the proton and neutron
form factors ($A_{Ep}(\xi^{p})$, $A_{Mp}(\xi^{p})$, $A_{En}(\xi^{n})$, and 
$A_{Mn}(\xi^{n})$).  Here, $A_N(\xi)=1$ for pure dipole form factors. 
% and $A_{FA} (\xi^{N})$).
The variable $\xi$ is the target mass scaling variable for elastic scattering ($x=1$), where 
$$\xi^{p,n}=\frac{2}{(1+\sqrt{1+1/\tau_{p,n}})},$$
and  $\tau_{p,n} = Q^2/4M_{p,n}^2$. Here
$M_{p,n}$ are the proton (0.9383 $GeV/c^2$) and neutron (0.9396 $GeV/c^2$) masses,
respectively.
%and average nucleon mass (for proton,
%neutron, and axial form factors, respectively).
\begin{eqnarray}
 %\begin{equation}
%\lefteqn  
G_D^V(Q^2) &\equiv &  \frac{1 }{ (1+Q^2/M_V^2)^2} \nonumber \\
     {G_{Ep}(Q^2)} &=& A_{Ep-dipole}(\xi^{p})\times {G_D^V(Q^2)} \nonumber
\\
  {G_{En}(Q^2)} &=& A^{25}_{En}(\xi^{n})\times {G_{Ep}(Q^2)} \times
   \left( {\frac{a\tau_{n}}{1+b\tau_{n}}} \right)
  \nonumber \\
  {G_{Mp}(Q^2)}/{\mu_{p}} &=&  A_{Mp-dipole}(\xi^{p})\times  {G_D^V(Q^2)} \nonumber \\
%  \end{equation}
    {G_{Mn}(Q^2)}/{\mu_{n}} &=& A^{25}_{Mn}(\xi^{n})
    \times  {G_{Mp}(Q^2)}/ {\mu_{p}} \nonumber 
      \end{eqnarray}
    %    F_A (Q^2)&=&A^{25}_{FA} (\xi^{N}) \times G_D^{A}(Q^2). 
 % {G_{En}(Q^2)} &=& A^{a,b}_{En}(\xi) \times   {\left( \frac
% {G_{Mn}(\xi)}{\mu_{n}} 
% \right) }
%     \times { \left( {\frac{a\tau}{1+b\tau}} \right) }
%   \nonumber \\, Here $\mu_p = 2.7928$, $\mu_n = -1.913$,
 Here $\mu_p = 2.7928$, $\mu_n = -1.913$, and 
 $M_{V}^2$ = 0.71 $GeV^2$ 
  ($M_V=0.8426~GeV$).
  The parameters for the multiplicative functions $A_N(\xi)$ which
  describes the ratio to dipole are given
  in reference\cite{quasi}. The parametrizations
  are compared to  experimental
  data in Figure~\ref{subfig1}. 
  
  For the axial form factor we use
$$  F_A (Q^2) =G_D^{A}(Q^2)=  \frac {g_a }{ (1+Q^2/M_A^2)^2} $$ 
where 
 $g_{A}$ = -1.267,  and $M_{A}=1.014 \pm0.014~GeV $ is the axial mass for free nucleons. 

The ratio of longitudinal and transverse cross sections for free nucleons
is given by:
 $$ R_{p}^{elastic}= \frac{4M^2/\mu_p^2}{Q^2}\frac{A_{Ep-dipole}^2}{A_{Mp-dipole}^2}= \frac{0.481}{Q^2}\frac{A_{Ep-dipole}^2}{A_{Mp-dipole}^2}$$
 $$ R_{n}^{elastic}= \frac{\mu_p^2}{\mu_n^2}R_{p}^{elastic} \frac{(A_{En}^{25})^2}{(A_{Mn}^{25})^2}\left( {\frac{a\tau_{n}}{1+b\tau_{n}}} \right)^2$$
 In the dipole approximation with  $G_{En}=0$
 \begin{equation}
 R_{deuteron}^{elastic}\approx  \frac{4M^2/(\mu_p^2+\mu_n^2)}{Q^2} = \frac{0.328}{Q^2}
 \label{relastic}
 \end{equation}

  \section {Quasielastic electron scattering from nuclear targets}
For electron-nucleon and muon-nucleon scattering, scattering from free nucleons (with no pions in the final state) is called elastic scattering, and scattering from nucleons
bound in a nuclear target (with no pions in the final state) is called QE  scattering because the
scattering is from quasi-free nucleons.

For charged-current $\nu_{\mu}$ -nucleon and $\bar{\nu}_\mu$-nucleon scattering (with no pions in the final state),  the term QE  scattering is used  to describe scattering from either  free  or  bound nucleons because the neutrino is transformed to a final state muon.  For neutrino processes,
the term elastic scattering is only used when there is a neutrino in the final state.

Studies of QE  electron  scattering on nuclear targets indicate
that only the longitudinal part of the QE cross section  can be described in terms of a universal response function of 
independent nucleons bound in a nuclear potential\cite{MEC1} (and free nucleon form factors).  In contrast, a significant additional enhancement with respect to the model is observed in the transverse part of the QE cross section.

The enhancement  in the transverse QE cross section has been attributed to meson exchange currents (MEC) in a nucleus\cite{MEC1,MEC2,MEC3,MEC4,MEC5}. Meson exchange currents originate from nucleon-nucleon correlations (predominantly neutron- proton). The final state
for the  MEC process can include 
one or two nucleons.  If  no final state pions are produced, the process is considered as an enhancement of the QE cross section. If one or more final state pions are produced,  the process enhances the  inelastic cross section.

Within models of meson exchange currents the enhancement is primarily in the transverse part of the QE cross section, while the  enhancement in the longitudinal QE cross section is small (in agreement with the electron scattering  experimental data).
 The conserved vector current hypothesis (CVC) implies that  the  corresponding  vector structure function for  the QE cross section in $\nu_{\mu},\bar{\nu}_\mu$ scattering can be expressed in terms of the  structure functions measured in electron scattering on nuclear targets. Therefore,
 there should also be a transverse enhancement in neutrino scattering.

In addition, for some models of  meson exchange currents\cite{MEC4} the enhancement in the 
axial part of $\nu_{\mu},\bar{\nu}_\mu$ QE cross section on nuclear targets is also small. Therefore, the  axial form factor for bound nucleons is expected to be the same as the axial form factor for free nucleons. 

%We  QEetrize the transverse
%enhancement for electron  scattering on nuclear targets in terms of an $\it "effective"$ vector %mass (only in the transverse part of the vector QE cross section). 
% The same parameters  should also provide a good description of  vector part of 
% $\nu_{\mu},\bar{\nu}_\mu$  QE scattering on nuclear targets. 
\subsection{Measuring the transverse enhancement at low $Q^2$}

 The longitudinal response scaling functions extracted by Donnely et. al.\cite{MEC1} for different momentum scales and different nuclei (A=12 ,40 and 56) are essentially described by one universal curve\cite{MEC1}  which is a function of the nuclear scaling variable $\psi'$ only.  The function  peaks
at $\psi'$=0 and ranges from $\psi'=-1.2$ to  $\psi'=2$. In contrast, the transverse response scaling function  is larger and increases with momentum transfer. 
The response function of the  transverse enhancement $excess$ is shifted
to higher $\psi'$  and peaks at   $\psi'\approx 0.2$.
%which is  below pion threshold on a nucleus.  Therefore, %the hadronic final state for the excess includes only nucleons.

Carlson et. al.\cite{MEC4}  uses the measured longitudinal and
transverse response functions to extract the  ratio (${\cal R}_{T}$)
 of the integrated response functions for
the  transverse and transverse components of the QE response functions 
for values of  $\psi'<0.5$ and  $\psi'<1.2$.   

For nucleons bound in carbon, the ratios for $\psi'<0.5$ 
are 1.2, 1.5, 1.65 for values of the 3-momentum transfer $q_3$ of 0.3, 0.5, and 0.6  $GeV/c$, respectively ($q_3^2= Q^2+\nu^2$ where $\nu=Q^2/2M$ at the QE peak).

 The ratios for  $\psi'<1.2$  are 1.25, 1.6, 1.8 for $q_3$ values of 0.3, 0.5, and 0.6  $GeV$, respectively. (These correspond to $Q^2$ values of 0.09, 0.15, and 0.33). At  higher
 values of $\psi'$  the  transverse response functions include both
 QE scattering and pion production processes (e.g. $\Delta$ production with Fermi motion). 
 
 Therefore, we use the measured values of ${\cal R}_{T}$ for $\psi'<0.5$, where the contribution from pion production process  is small, and apply  correction to extract the ratio for the
 entire range of $\psi'$, as described below.
 
The excess transverse response function peaks at  $\psi'\approx0.2$, while the  longitudinal
response function peaks at $\psi'=0$. A fit of an asymmetric gaussian to the
longitudinal response function indicates that 
the  ${\cal R}_{T}$ values for the total response functions integrated over all $\psi'$ are
related to the ratio for  $\psi'<0.5$ by the following expression:
$${\cal R}_{T} (all-\psi') =  1+ 1.18~ [{\cal R}_{T}(\psi'<0.5)-1]$$ 
 We obtain ${\cal R}_{T} (all-\psi')$ values of  $1.24\pm0.1$, $1.59\pm0.1$, and $1.77\pm0.1$ for $Q^2$ values of 0.09, 0.15, and 0.33  $(GeV/c)^2$, respectively.  We use the difference in the measured values of  ${\cal R}_{T}$  for $\psi'<0.5$ and  $\psi'<1.2$ as an estimate of the systematic error. Since the longitudinal  response function is equal to the response
 function for independent nucleons, the ratio ${\cal R}_{T} (all-\psi')$ is equivalent to the ratio 
 of the integrated transverse response function in a nucleus to the response function for  independent nucleons (as a function of $Q^2$).
 
\begin{figure}[ht]
\includegraphics[width=3.3in,height=2.5in]{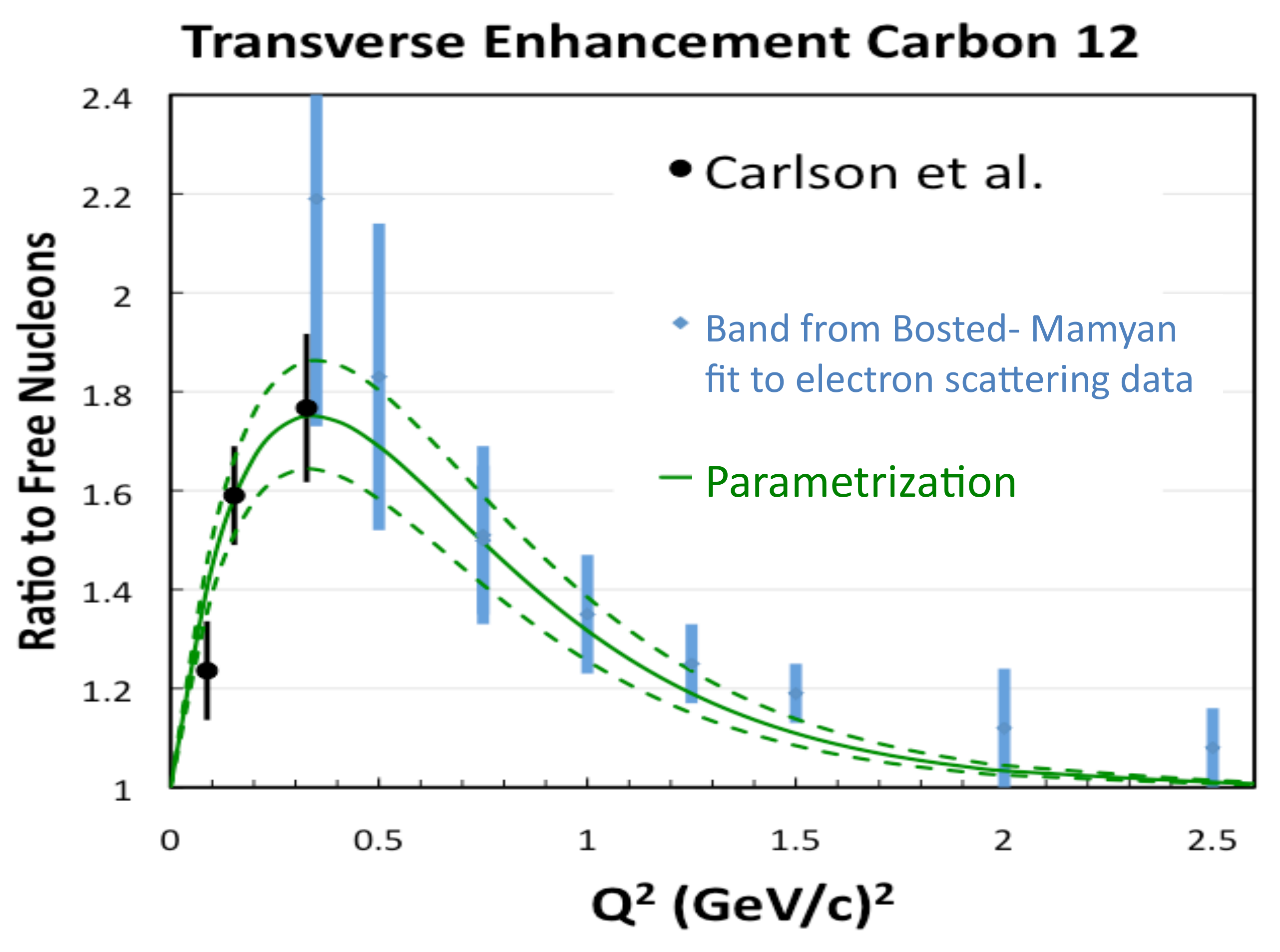}
\caption{ The transverse enhancement ratio (${\cal R}_{T}$) as a function of $Q^2$. Here, ${\cal R}_{T}$ is  ratio of the  integrated transverse response function
for QE  electron scattering on nucleons bound in  carbon divided by 
the integrated response function for independent nucleons. 
 The black points are extracted from Carlson $et~al$\cite{MEC4}, and
the blue bands are extracted from a fit\cite{vahe-thesis} to QE data from the JUPITER\cite{JUPITER} experiment (Jlab experiment E04-001). The curve is a fit to the data of the form  ${\cal R}_{T}=1+AQ^2e^{-Q^2/B}$. The dashed lines are the upper and lower error bands.
}
\label{GMPN}
\end{figure}

The values of  ${\cal R}_{T}$ extracted from the data of from Carlson $et~al$ are shown  as a function of $Q^2$ (black points) in    Figure~\ref{GMPN}.
 %figure 3

 \begin{figure}
\includegraphics[width=3.3 in,height=2.4in]{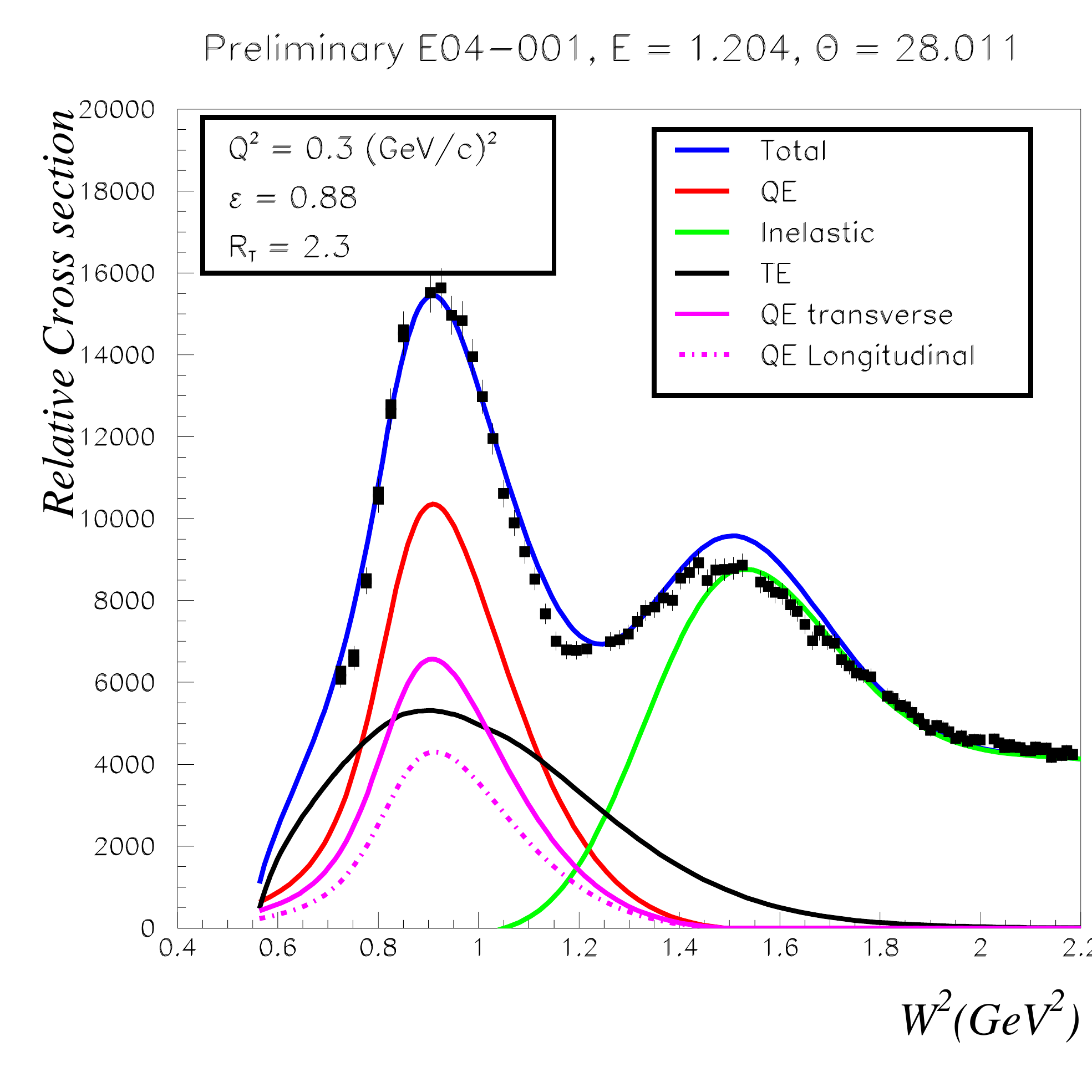}
\includegraphics[width=3.3 in,height=2.4in]{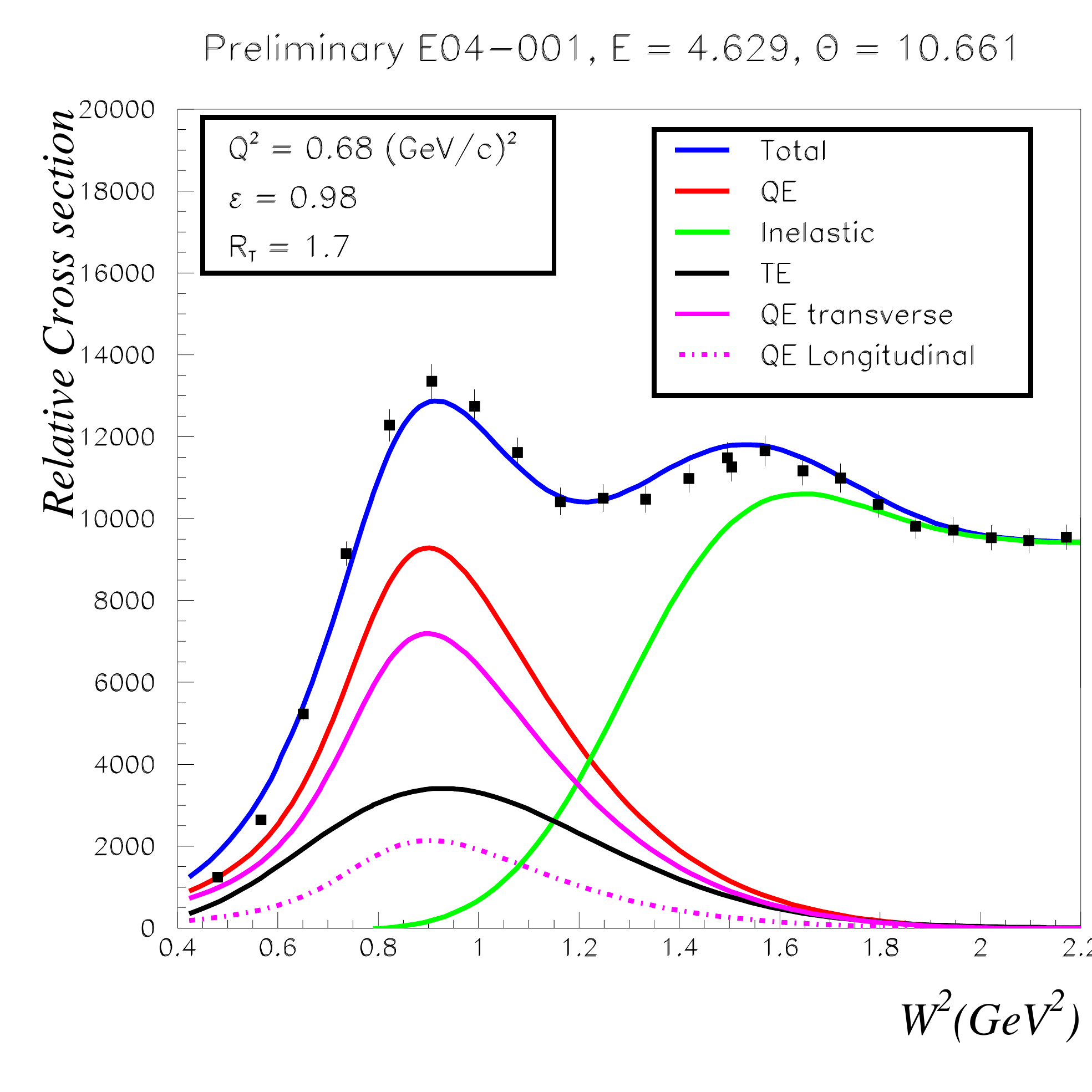}
\caption{Samples of fits\cite{vahe-thesis} to preliminary electron scattering data from the JUPITER collaboration\cite{JUPITER} (Jefferson Lab experiment E04-001) on a carbon target. Shown are the contributions from the transverse QE (solid pink), longitudinal QE (dashed pink), total QE (solid red), inelastic (pion production) processes (solid green), and a  transverse excess (TE) contribution (solid black line).
Top:  $Q^2=$~0.3 $GeV/c^2$ at the QE peak.  Bottom: $Q^2=$~0.68 $GeV/c^2$ at the QE peak.  }
\label{jlabfit1}
\end{figure}
%--Peter Bosted and Vahe Mamyan,
\begin{figure}
\includegraphics[width=3.3 in,height=2.4in]{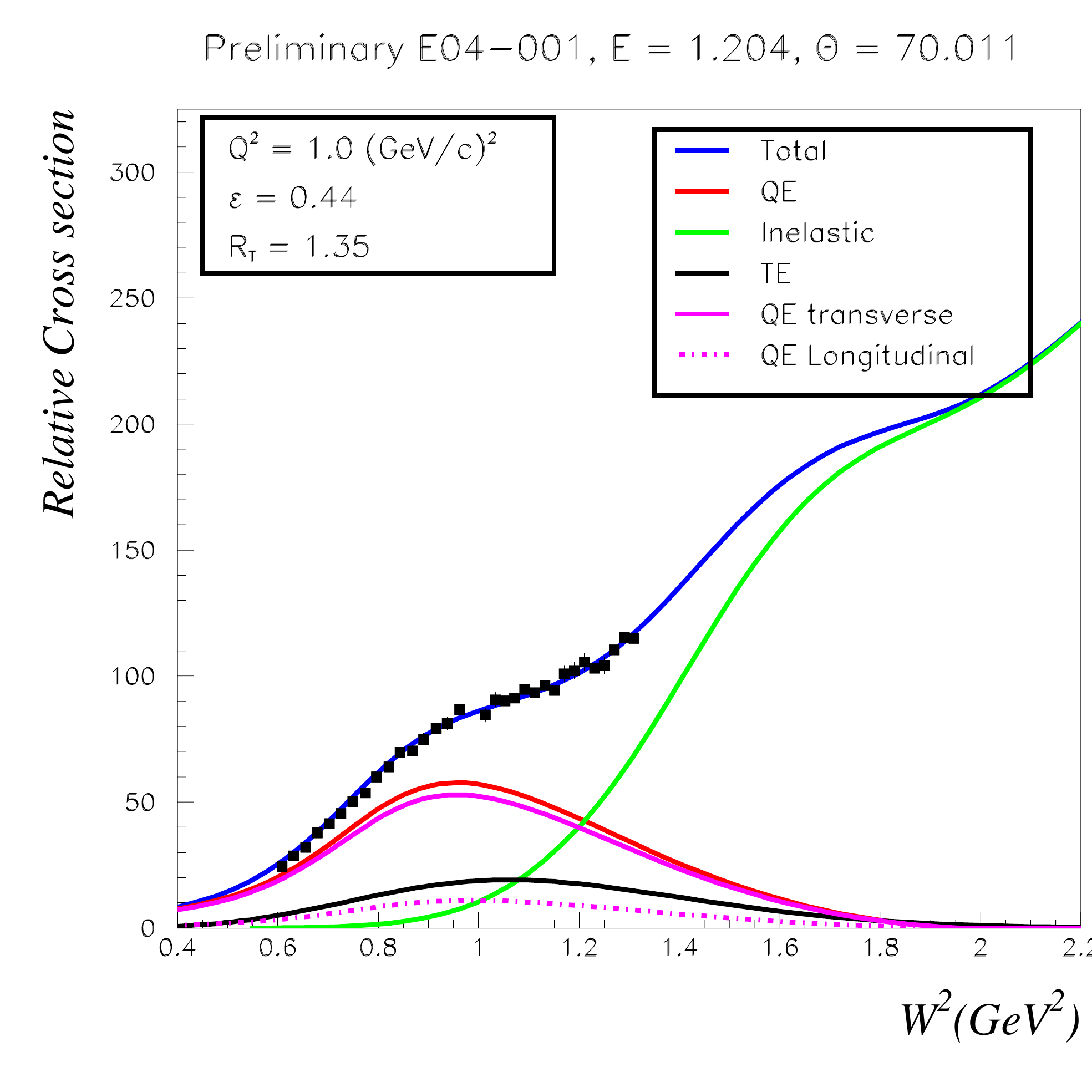}
\includegraphics[width=3.3 in,height=2.4in]{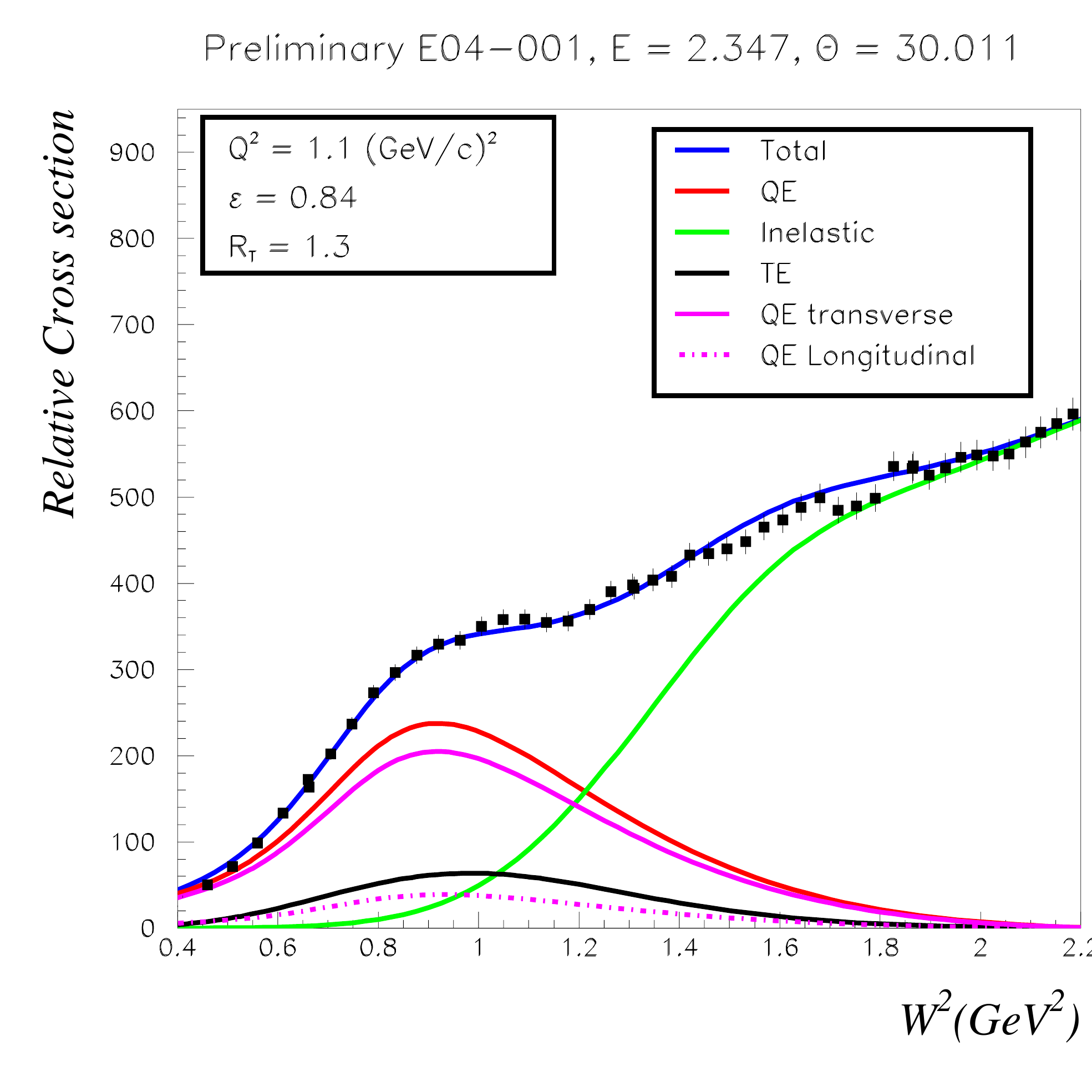}
\caption{Same as Fig.~\ref{jlabfit1}. Top: $Q^2=$~1.0 $GeV/c^2$ at the QE peak. Bottom: $Q^2=$~1.1 $GeV/c^2$ at the QE peak).
  }
\label{jlabfit2}
\end{figure}

 \subsection{Measuring the transverse enhancement at high $Q^2$}
The technique of using the ratio of  longitudinal and transverse QE structure functions
to determine the transverse enhancement in the response functions for QE scattering is less reliable for $Q^2>0.5~(GeV/c)^2$, because at  high values of $Q^2$ the  longitudinal contribution to the QE cross section is small (as illustrated in equation~\ref{relastic}).

Since the transverse cross section dominates  at large $Q^2$  one can extract the transverse enhancement by comparing the measured QE cross sections to the predictions of the independent nucleon model directly. However, because there is overlap between pion production processes and QE scattering, the contribution from pion production processes should be accounted for in the extraction process.

We extract the transverse enhancement at higher values of $Q^2$ from a fit to both 
existing electron scattering data on nuclei and  preliminary data from the JUPITER 
collaboration\cite{JUPITER}  (Jefferson lab experiment E04-001).  The fit (developed 
by P. Bosted and V. Mamyan) provides a description of inclusive electron scattering cross sections on a range of nuclei with $A > 2$.  It is an extension of fits to the 
free proton~\cite{resp} and deuteron~\cite{resd} and was utilized for calculations 
of the radiative corrections for the JUPITER analysis~\cite{vahe-thesis}.  Experiment   
E04-001 was designed to provide separations of the longitudinal and transverse 
structure functions from a range of nuclei.  These data, therefore, provides a 
significant constraint on this separation in both the quasi-elastic and resonance 
regions, which is of critical importance for the current study.  A brief description 
of the fit is given in \cite{vahe-thesis}, which also provides plots of the fit 
residuals to the data sets utilized. 

The inclusive fit is a sum of four components: 
\begin{itemize}
\item The longitudinal QE contribution calculated for independent nucleons (smeared by Fermi motion in carbon)
\item  The transverse  QE contribution calculated for independent nucleons (smeared by Fermi motion in carbon)
\item  A transverse excess (TE) contribution 
\item  The contribution of inelastic pion production processes (smeared by Fermi motion in carbon).
%\item The contribution of higher resonances and inelastic scattering (smeared by % Fermi motion in carbon)
\end{itemize}

The QE model used in the Bosted-Mamyan fit is the super-scaling model\cite{super} of  Sick, Donnelly, and  Maieron. 

 %This is possible because the contribution of the pion production process is included in the fit.
Figures~\ref{jlabfit1} and \ref{jlabfit2} show samples of Bosted-Mamyan fits to preliminary electron scattering data from JUPITER on a carbon target. Shown are the contributions from the transverse QE (solid pink), longitudinal QE (dashed pink), total QE (solid red), inelastic pion production processes (solid green), and a transverse excess (TE) contribution  (solid black line).  

We extract the transverse enhancement ratio as a function of $Q^2$ by integrating the various contributions to the fit up to $W^2=1.5~ GeV^2$. Here
 $${\cal R}_{T} =\frac {QE_{transverse}+TE}{QE_{transverse}}$$
 We assign a conservative systematic error to ${\cal R}_{T}$ to account for
 the possibility that a fraction of the transverse excess  events may  be produced with  a pion in the final state. 
  
 Figure~\ref{GMPN} shows the values of ${\cal R}_{T}$  as a function of $Q^2$.  The black points are extracted from Carlson $et~al$\cite{MEC4}, and
the higher $Q^2$ blue bands are from the fit to QE data from the JUPITER collaboration\cite{JUPITER}.  The data are parametrized by the expression:
  $${\cal R}_{T}=1+AQ^2e^{-Q^2/B}$$
   with $A=6.0$ and $B=0.34~(GeV/c)^2$.  The electron scattering data indicates that the transverse enhancement is maximal near $Q^2=0.3~(GeV/c)^2$ and is small for $Q^2$ greater
than $1.5~(GeV/c)^2$. 
The upper error band is given by  $A=6.7$ and $B=0.35~(GeV/c)^2$, and
the lower error band is given by $A=5.3$ and $B=0.33~(GeV/c)^2$. This parametrization     is valid for carbon (A=12)  and higher A  nuclei.

% Figure~\ref{GMPN} shows  the integrated  ${\cal R}_{T}$  
%as a function of $Q^2$. The curve is 
%${\cal R}_{T}=2-e^{-Q^2/A}$ with $A=0.3~(GeV/c)^2$
%

%We now investigate what this
%parametrization implies for $\nu_{\mu},\bar{\nu}_\mu$ QE scattering on nuclear targets.
%
%
%Figure 4
	 \begin{figure}[ht]
\includegraphics[width=3.3in,height=2.3in]{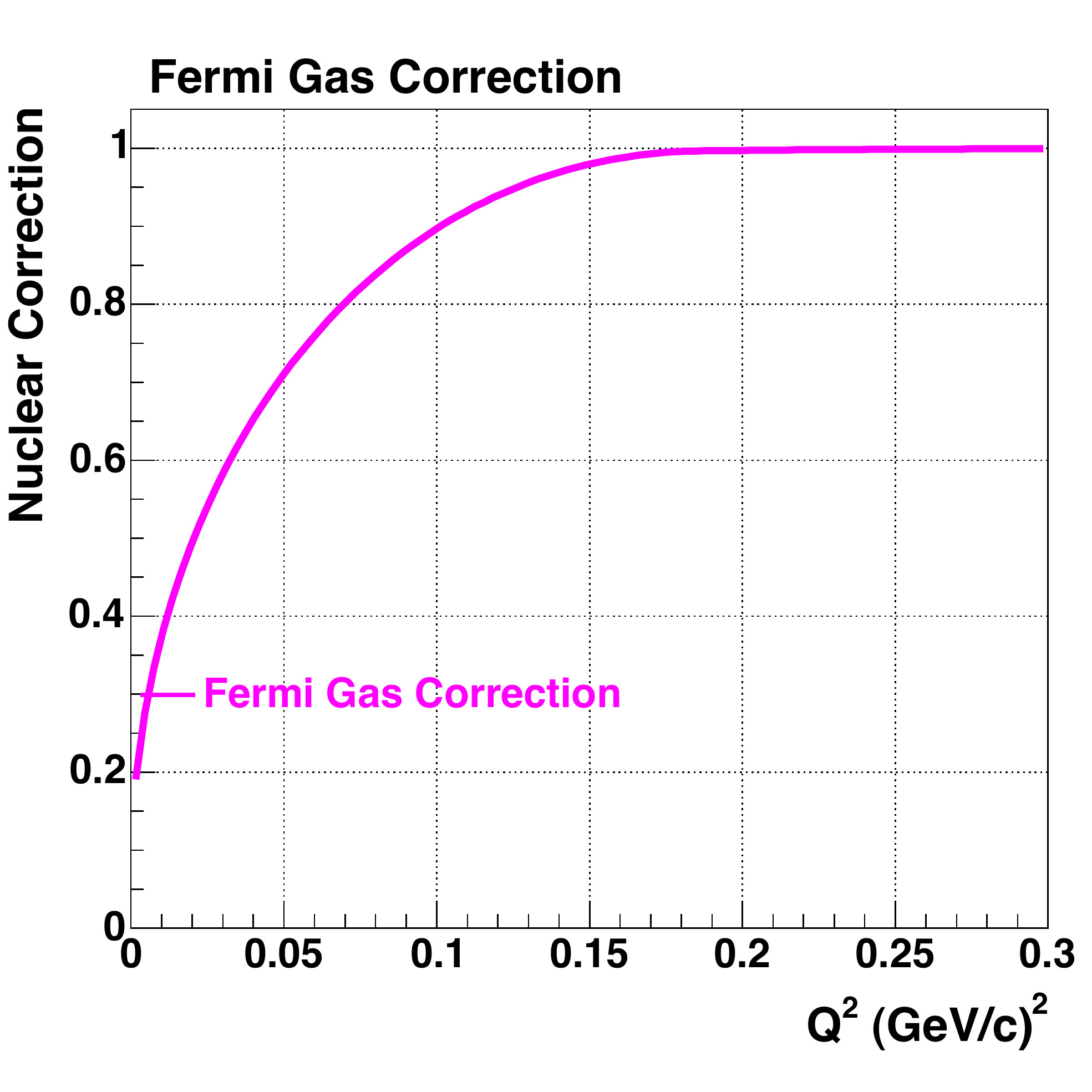}
\caption{ The Fermi suppression factor (Pauli blocking) used in our studies as a function of $Q^2$. We use the Pauli blocking factor which is implemented in the NUEGEN Monte Carlo\cite{GENIE}. 
}
\label{pauli}
\end{figure}
 \section{Consequences for $\nu_{\mu},\bar{\nu}_\mu$  charged-current  QE scattering
 on carbon}
%
%\subsection{Transverse enhancement  in ${\cal W}_1^{Qelastic}$ and  ${\cal W}_2^{Qelastic}$ }
%
We assume that there is  a corresponding  transverse enhancement  
in the   $\nu_{\mu},\bar{\nu}_\mu$ QE cross sections on nuclear targets. Although motivated by
 MEC,  the analysis
is model independent since the parameters are taken from electron scattering data. 

In the rest of this paper, the terms  cross sections and differential distributions refer to scattering
from  nucleons bound in carbon. 

\subsection{The "Independent Nucleon ($M_A$=1.014)" baseline model 
%(shown as orange dotted lines on plots) 
}
In modeling  $\nu_{\mu},\bar{\nu}_\mu$  QE scattering on nuclear targets we use   $BBBA2007_{25}$  free nucleon electromagnetic form factors  (with $M_V^2=0.71$), and  a dipole axial form factor with  $M_A=1.014~GeV$.    We apply  Pauli blocking corrections to the differential QE cross section,  as implemented in the $NEUGEN$ Monte Carlo\cite{GENIE}. The Pauli blocking factor as a function of $Q^2$ is shown in figure \ref{pauli}.  We do not apply Fermi motion corrections since we only study the total  integrated QE cross section. We refer to this baseline model, which is shown as  orange dotted lines on plots, as the "Independent Nucleon ($M_A$=1.014)" model.
\subsection{The "Transverse Enhancement" model 
%(shown as a solid red lines on plots)
}
We use our parametrization of ${\cal R}_{T}$ to modify $G_{Mp}$ and $G_{Mn}$ for bound 
nucleons as follows.  First, we assume that the enhancement in the transverse QE cross 
section modifies ${\cal G}_M^V = G_{Mp}-G_{Mn}$ for nucleons bound in carbon with a 
form given by
 \begin{eqnarray}
  {G_{Mp}^{nuclear}(Q^2)}&=&  G_{Mp}(Q^2) \times \sqrt {1+AQ^2e^{-Q^2/B}} 
     \nonumber \\
%  \end{equation}
    {G_{Mn}^{nuclear}(Q^2)} &=&  G_{Mn}(Q^2) \times  \sqrt {1+AQ^2e^{-Q^2/B}}. 
 \nonumber 
      \end{eqnarray}   
In all of the studies  we keep $G_{Ep}(Q^2)$, $G_{En}(Q^2)$ and $F_A(Q^2)$ for bound nucleons the same as for free nucleons.  The transverse enhancement leads to an enhancement in the structure functions  ${\cal W}_1^{Qelastic} $,  ${\cal W}_2^{Qelastic}$ and  ${\cal W}_3^{Qelastic}$.  
 The expressions for the  $\nu_{\mu},\bar{\nu}_\mu$ differential QE cross sections are given in the Appendix.  We also apply Pauli blocking as a function of $Q^2$ as shown in figure \ref{pauli}.  We refer to this model as the "Transverse Enhancement" model. The predictions on the plots for the "Transverse Enhancement Model"
are shown with  solid red lines. The error bands are shown
as dotted dashed red lines.  The ratio of calculated quantities for the "Transverse Enhancement model"
divided by "Independent Nucleon (ma=1.014) are also shown as 
solid red lines.

    %
      %Figure 5
         \begin{figure}
\includegraphics[width=3.503in,height=2.8in]{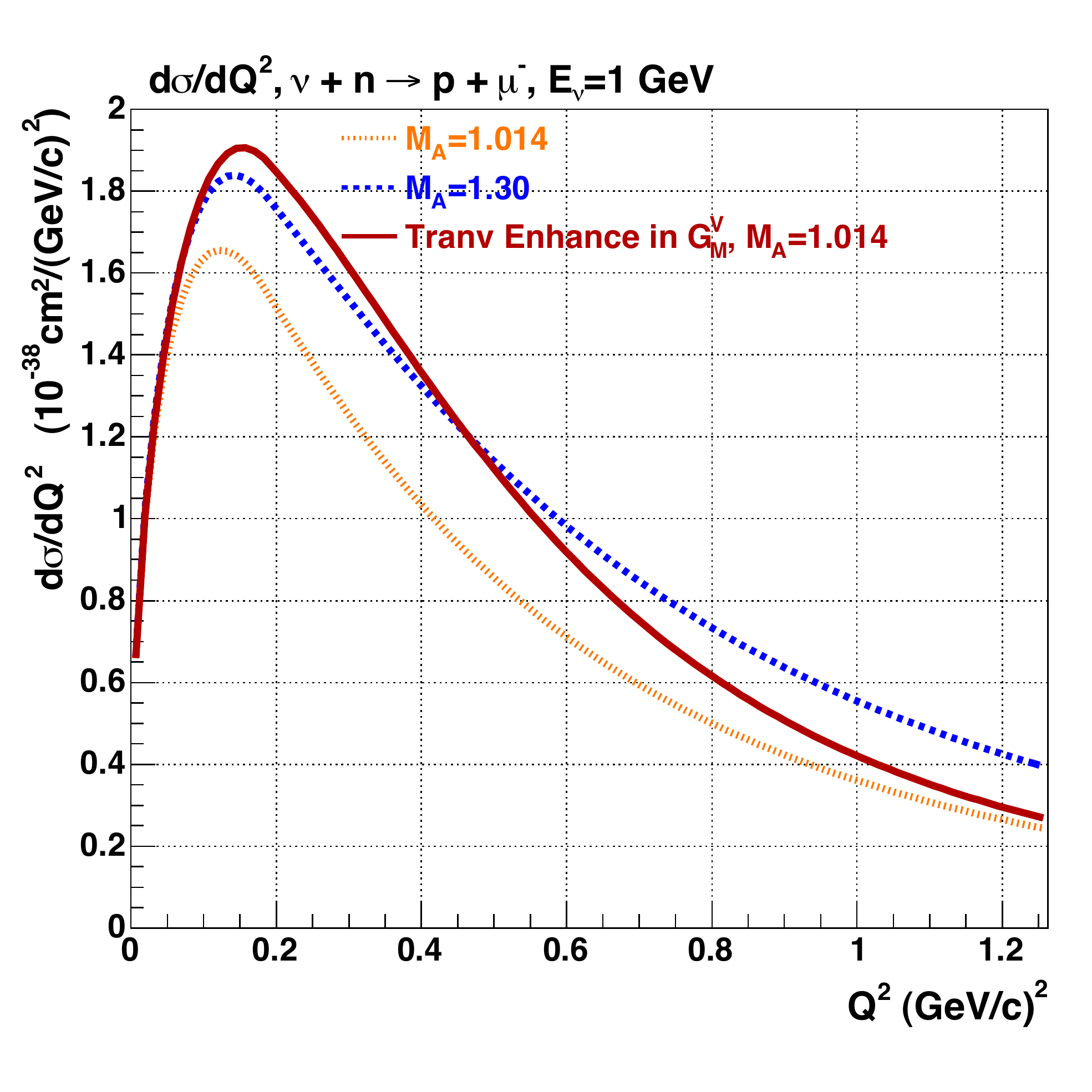}
\vspace{-0.15in}
\includegraphics[width=3.503 in,height=2.8in]{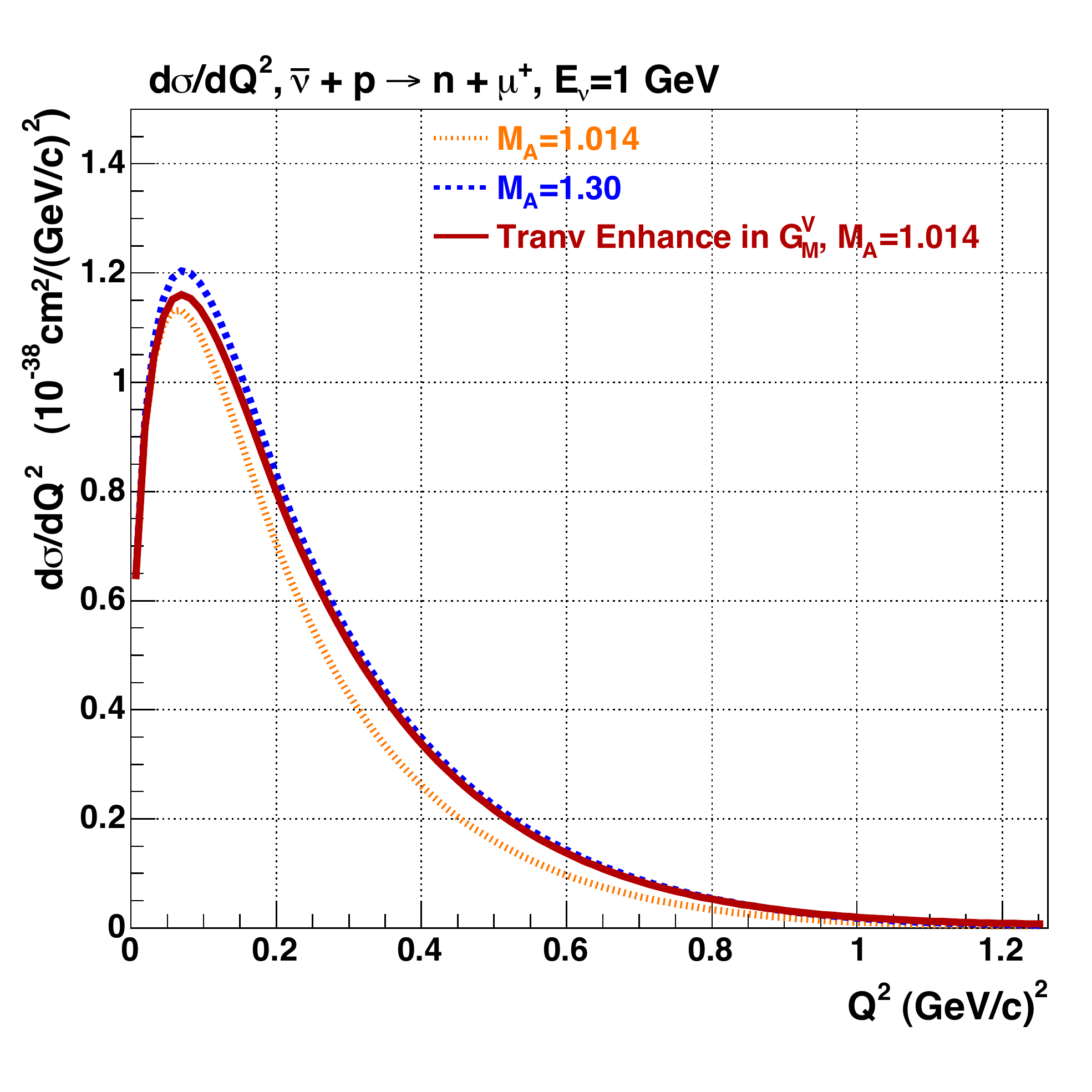}
\vspace{-0.1in}
\caption{The QE differential cross section (d$\sigma$/d$Q^2$) 
%for nucleons bound in carbon 
 as a function of $Q^2$ for $\nu_{\mu},\bar{\nu}_\mu$  energies of 1.0 GeV (maximum accessible $Q^2_{max} = 1.3~ (GeV/c)^2$).
  Here, the orange dotted line is the prediction of the  "Independent Nucleon  ($M_A$=1.014)"  model.   
The blue dashed  line is the prediction of the   the "Larger $M_A$  ($M_A$=1.3)" model.
The red line is prediction of the  "Transverse Enhancement" model. This color and line
style convention is used in all subequent plots. 
Top (a): $\nu_{\mu}$  differential QE cross sections.  Bottom (b): $\bar{\nu}_\mu$  differential QE cross sections. }
\label{diff1}
\end{figure}
%
%2.6133 to become 3.503
%Figure 6
\begin{figure}
\includegraphics[width=3.503in,height=2.8in]{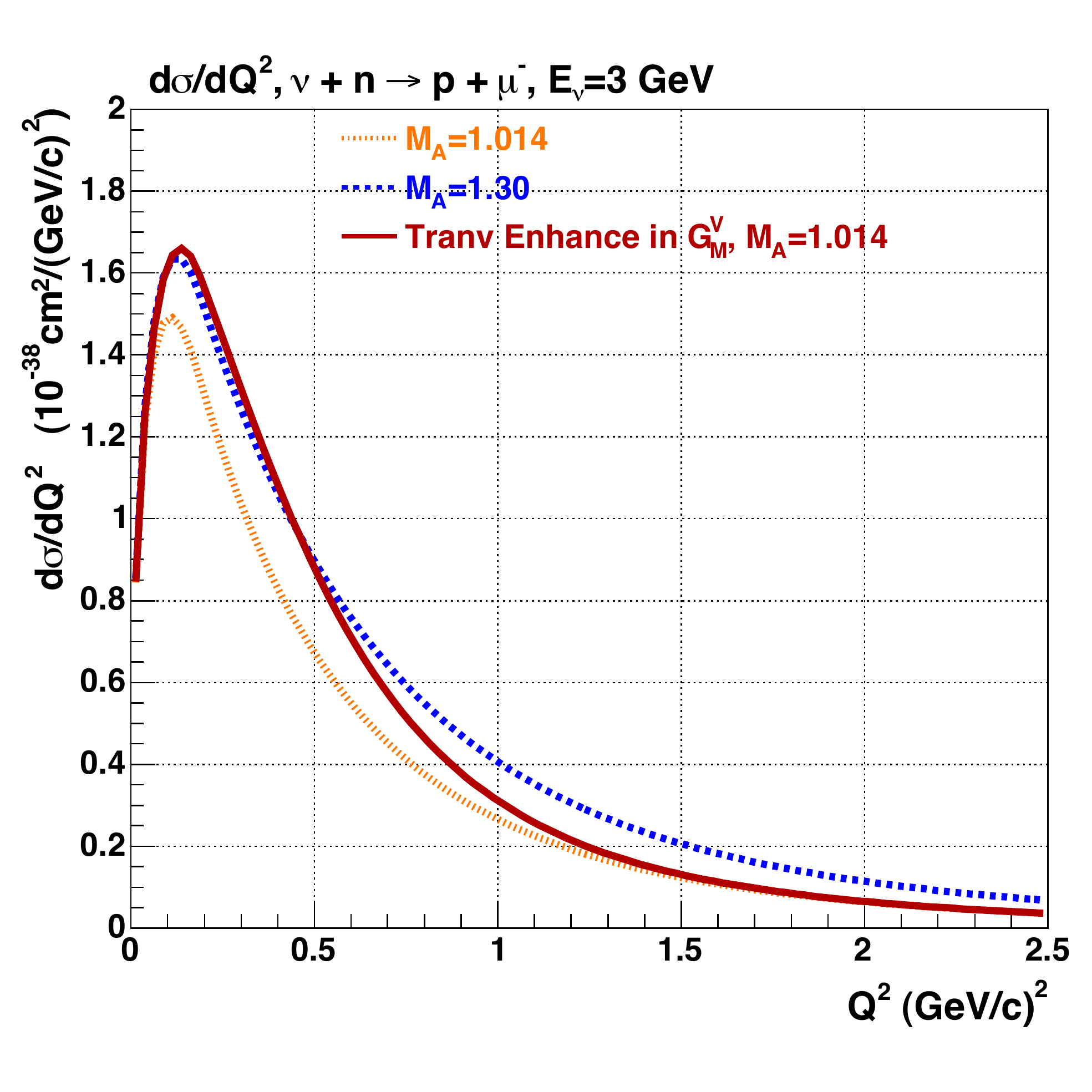}
\vspace{-0.15in}
\includegraphics[width=3.503in,height=2.8in]{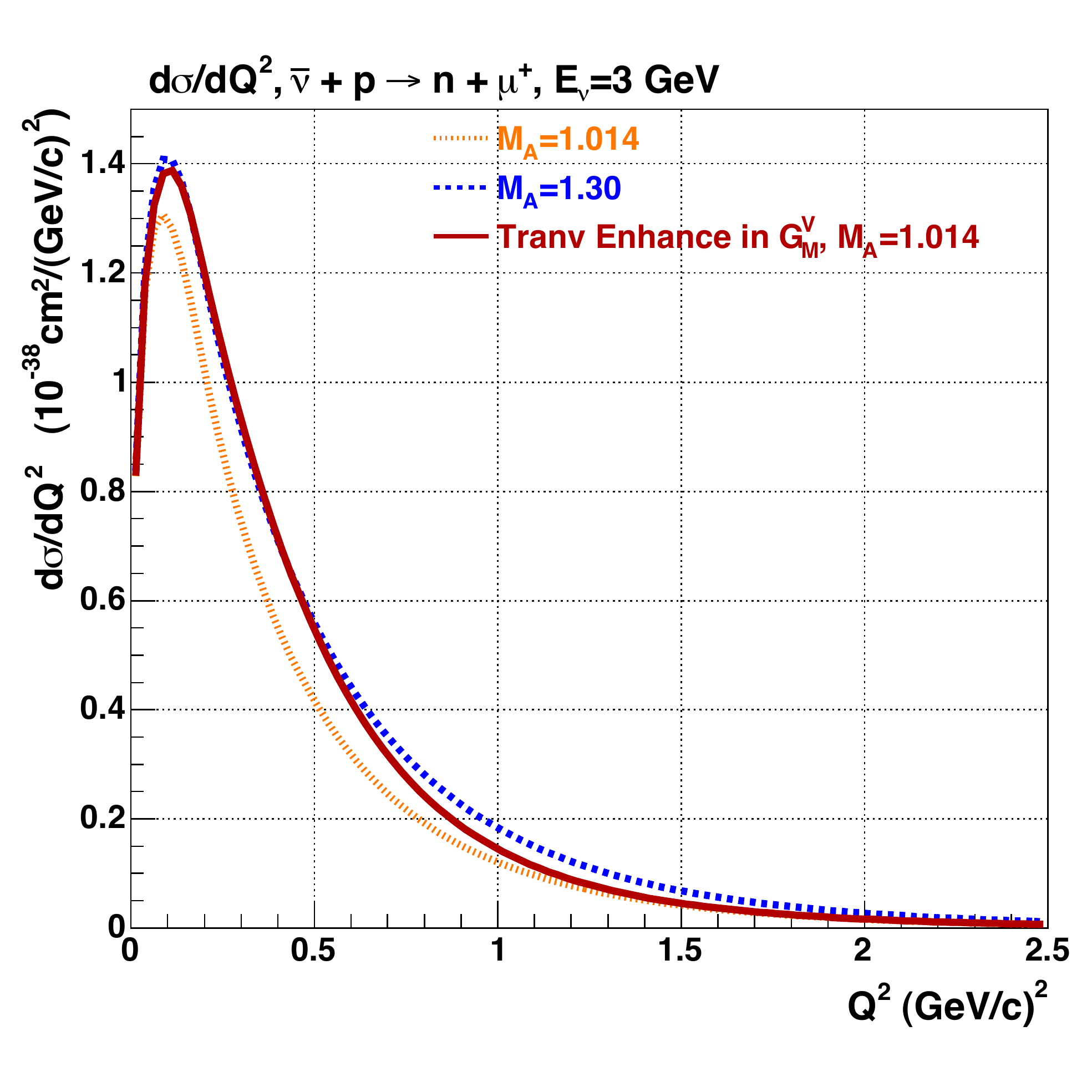}
\vspace{-0.1in}
\caption{ Same as figure~\ref{diff1} for $\nu_{\mu},\bar{\nu}_\mu$  energies of 3.0 GeV (maximum accessible $Q^2_{max} = 4.9~ (GeV/c)^2$). }
\label{diff2}
\end{figure}

 \subsection{The "Larger $M_A$  ($M_A$=1.3)" model 
 %(shown as dashed blue lines on plots)
  }
 
 Since low energy neutrino experiments have  used an ad-hoc  $M_A^{eff}\approx 1.3~GeV$ to
 account for additional nuclear effects, we also
 compare  our results to the differential and total QE cross sections  calculated for independent nucleons  with  $M_A^{eff}=1.3~GeV$  in the following expression:
       \begin{eqnarray}
              F_A ^{nuclear}(Q^2)&=& \frac{1}{ (1+Q^2/M_{A}^2)^2} 
 %      F_A ^{nuclear}(Q^2)&=&A^{25}_{FA} (\xi^{N}) \times \frac{(1+Q^2/1.014^2)^2 }{ (1+Q^2/M_%  {A}^2)^2} 
  \end{eqnarray}
  For this model, we use the electromagnetic form factors for free nucleons, and apply Pauli blocking as described above.
 We refer to this model, which is shown as dashed blue lines on plots,  as the "Larger $M_A$  ($M_A$=1.3)" model.
   The ratio of calculated quantities for the  "Larger $M_A$  ($M_A$=1.3)" model  
divided by  the predictions of the  "Independent Nucleon  ($M_A$=1.014) model are also shown as
dashed blue lines. 

\subsection{Results}  
  Figures \ref{diff1} and \ref{diff2}  show the QE differential cross section (d$\sigma$/d$Q^2$) 
  % for nucleons bound in carbon 
   as a function of $Q^2$ for $\nu_{\mu},\bar{\nu}_\mu$  energies of 1.0   and 3.0  GeV, respectively.  The orange dotted line is the prediction of the "Independent Nucleon  ($M_A$=1.014)" model, the blue dashed line is the prediction of  the "Larger $M_A$ ($M_A$=1.3)" model, and the solid red line is the prediction of the  "Transverse Enhancement" model.
  The top panels (a) show $\nu_{\mu}$  differential QE cross sections, and the bottom panels (b) show the $\bar{\nu}_\mu$  differential QE cross sections.

 %2.6133  to become 3.503
 % Figure 7
  \begin{figure}
\includegraphics[width=3.5 in,height=2.6in]{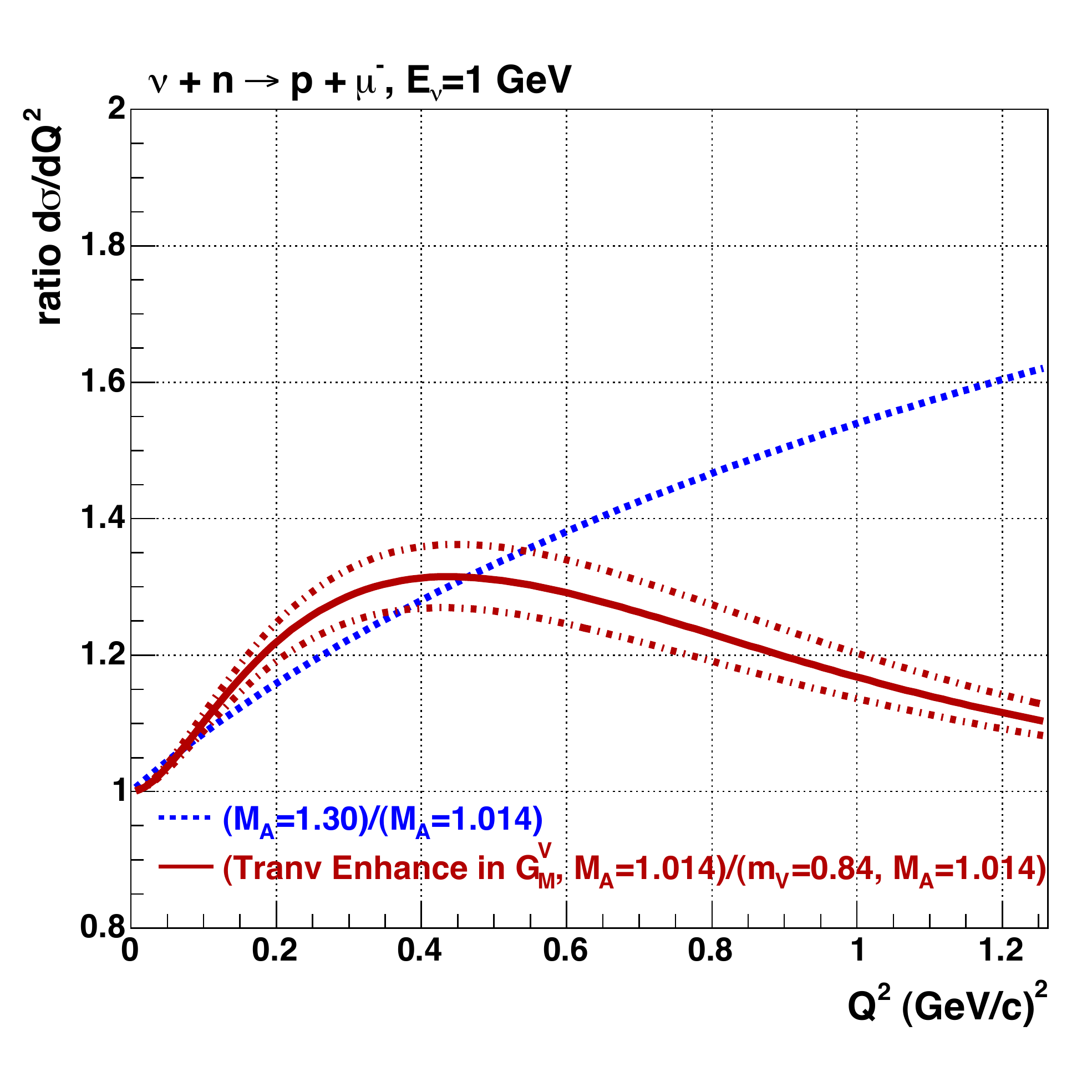}
\vspace{-0.15in}
\includegraphics[width=3.5in,height=2.6in]{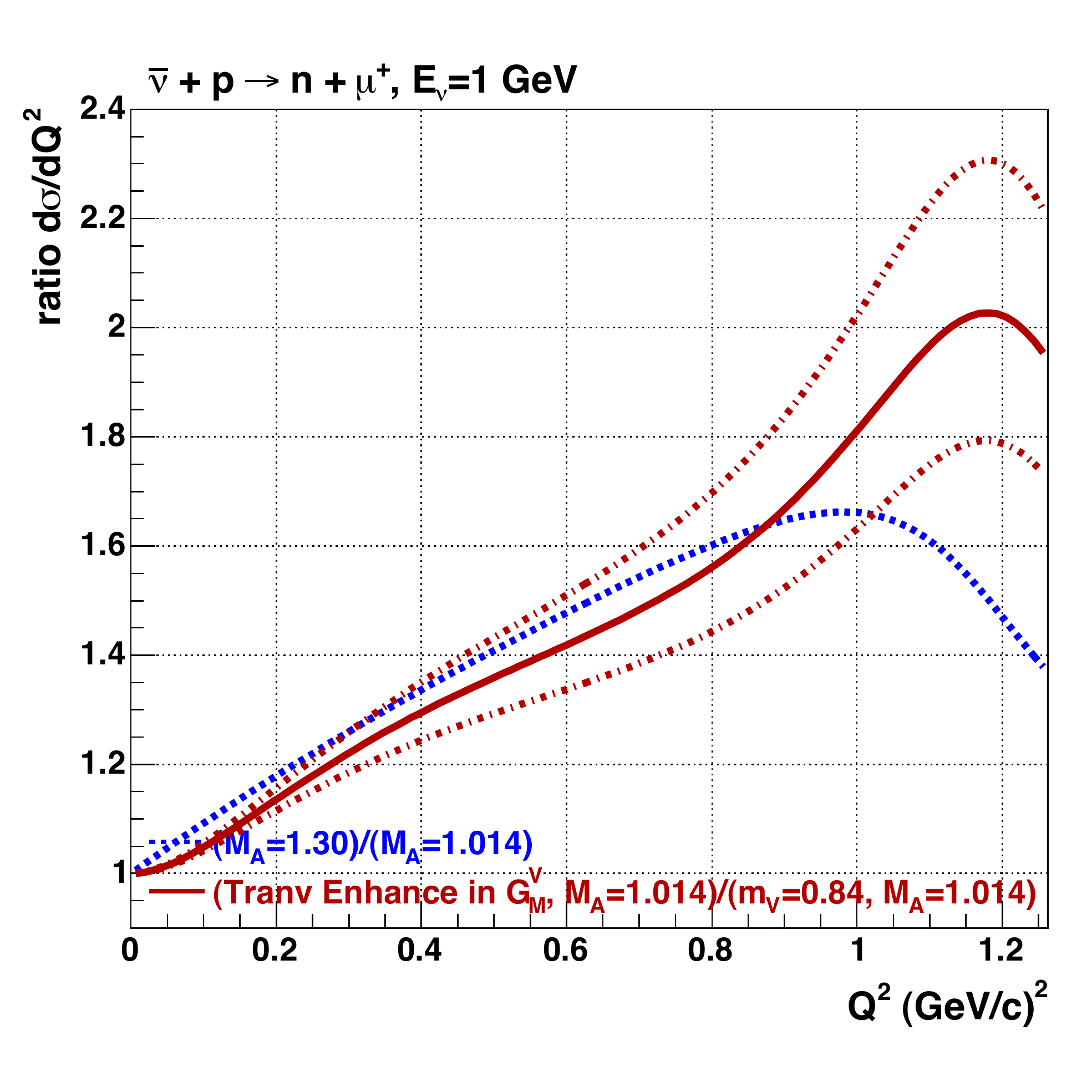}
\vspace{-0.15in}
\includegraphics[width=3.5in,height=2.6in]{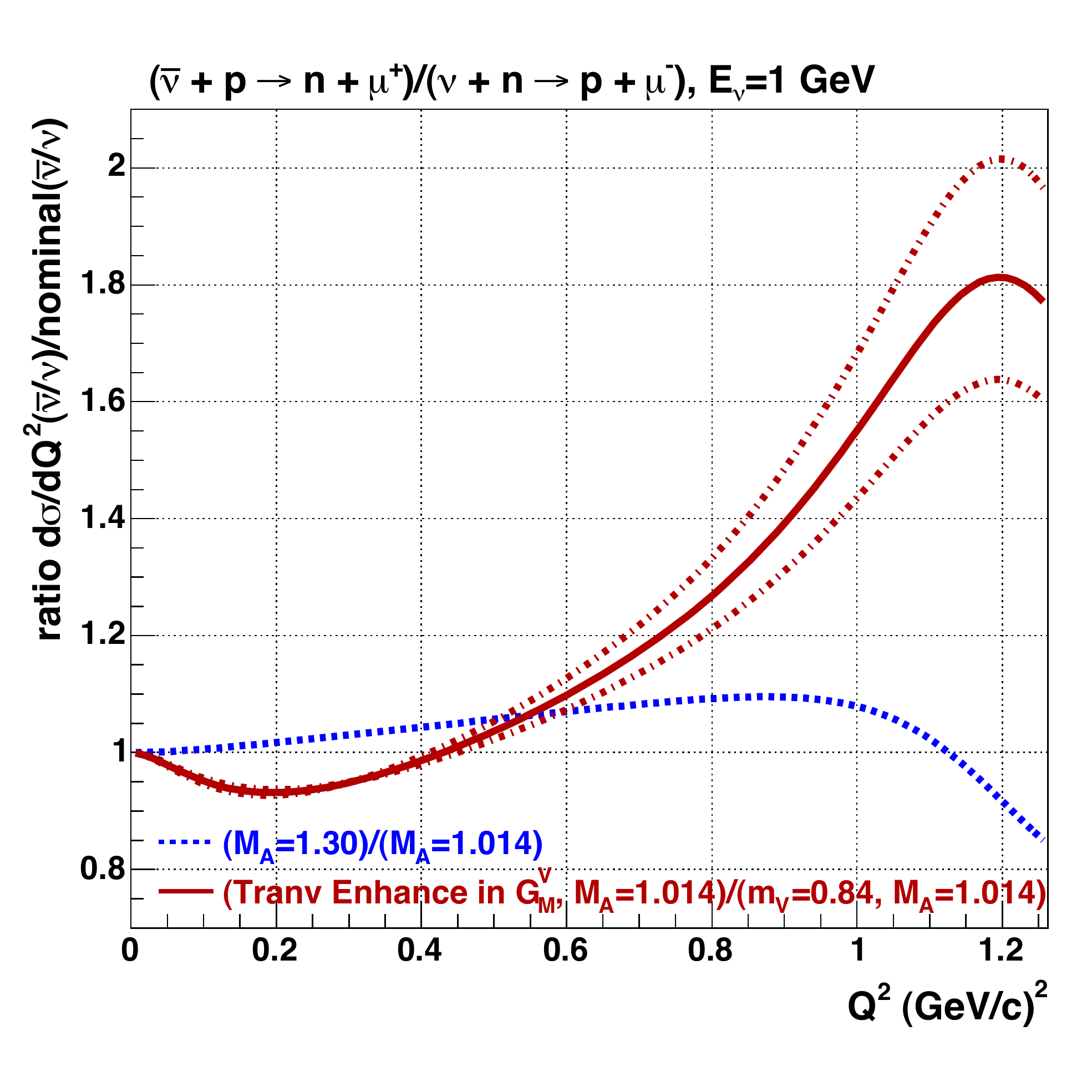}
\vspace{-0.1in}
\caption{The ratio of the prediction of the two models for the QE differential cross section d$\sigma$/d$Q^2$ 
% for nucleons bound in carbon 
 to the prediction of the  "Independent Nucleon  ($M_A$=1.014)" model as a function
of $Q^2$ for $\nu_{\mu},\bar{\nu}_\mu$  energies of 1.0 GeV (maximum accessible $Q^2_{max} = 1.3~ (GeV/c)^2$). 
The blue dashed  line is  the ratio for the  "Larger $M_A$  ($M_A$=1.3)" model.
The red  line is the ratio for the  "Transverse Enhancement" model (with error bands
shown as dotted red lines). Top (a): ratio for $\nu_{\mu}$  differential QE cross sections. Middle (b): ratio for $\bar{\nu}_\mu$ differential QE cross sections. Bottom (c):  The  $\bar{\nu}_\mu/\nu_{\mu}$ ratio for the differential QE cross sections divided by the corresponding  $\bar{\nu}_\mu/\nu_{\mu}$ ratio for the "Independent Nucleon  ($M_A$=1.014)" model).}
\label{ratio1}
\end{figure}
%
%
% Figure 8
\begin{figure}
\includegraphics[width=3.5in,height=2.6in]{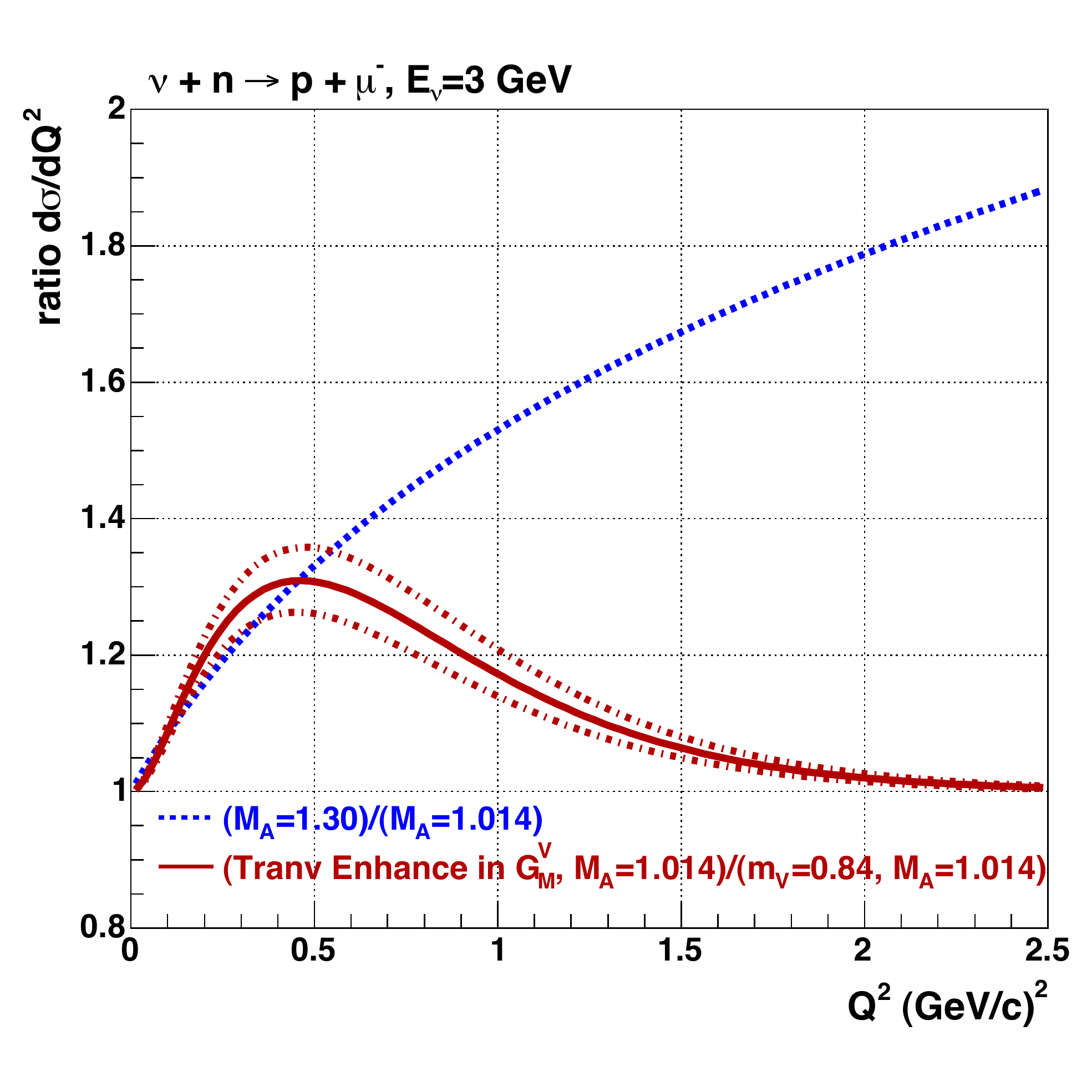}
\vspace{-0.15in}
\includegraphics[width=3.5in,height=2.6in]{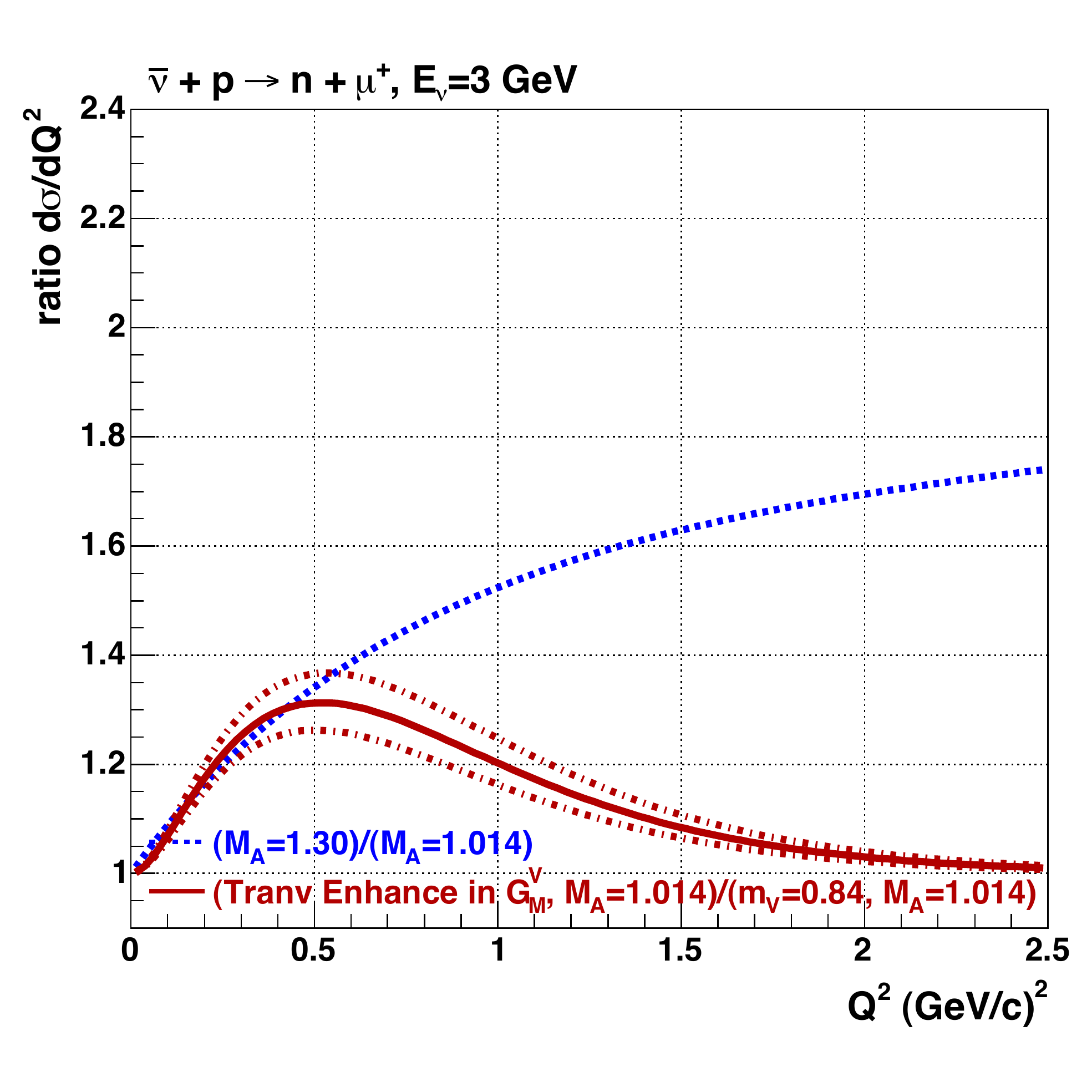}
\vspace{-0.15in}
\includegraphics[width=3.5in,height=2.6in]{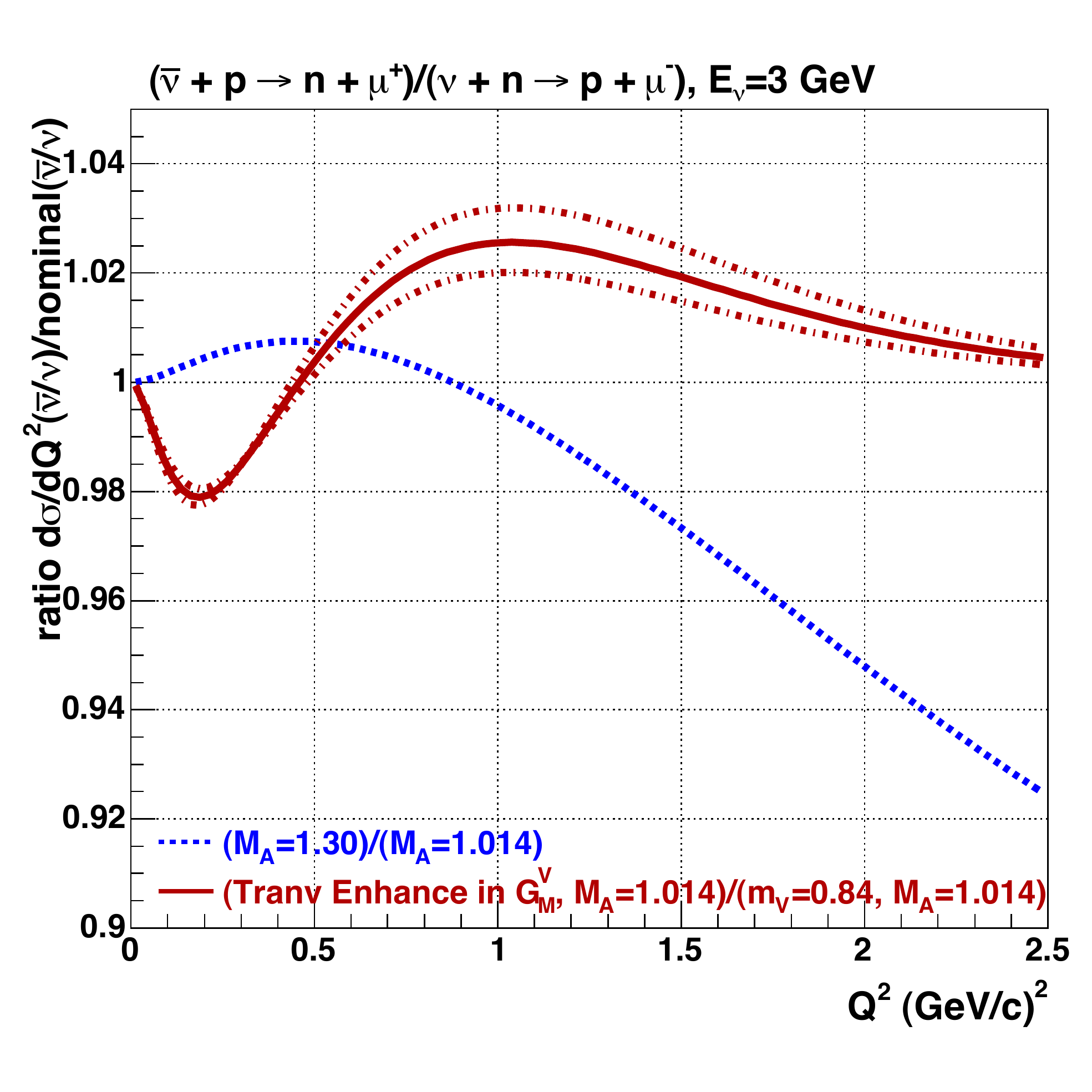}
\vspace{-0.1in}
\caption{ Same as figure~\ref{ratio1} for $\nu_{\mu},\bar{\nu}_\mu$  energies of 3.0 GeV (maximum accessible $Q^2_{max} = 4.9~ (GeV/c)^2$).}
\label{ratio3}
\end{figure}

 Figures~ \ref{ratio1} and~ \ref{ratio3} show the ratio of the predictions of the two models 
  % the predicted QE differential cross sections
 % d$\sigma$/d$Q^2$ to the 
  % d$\sigma$/d$Q^2$ 
to the  predictions of the "Independent Nucleon ($M_A$=1.014)" model as a function of $Q^2$ for $\nu_{\mu},\bar{\nu}_\mu$  energies of 1.0 GeV, and 3.0 GeV,  respectively.   
The blue dashed  line is  the ratio for the  "Larger $M_A$ ($M_A$=1.3)" model. The red  line is the ratio for the  "Transverse Enhancement" mode (with error bands
shown as dotted red lines).
 The top (a) panels shows the ratio for d$\sigma$/d$Q^2$  for  ${\nu}_\mu$. The 
 middle (b) panels  shows the ratio for d$\sigma$/d$Q^2$  for $ \bar{\nu}_\mu$.  The bottom (c) panels shows  the ratio of  predicted ratio of  $\bar{\nu}_\mu/\nu_{\mu}$ d$\sigma$/d$Q^2$ cross sections for the two models  (divided by the $\bar{\nu}_\mu/\nu_{\mu}$ ratio predicted by the "Independent Nucleon  ($M_A$=1.014)" model).

  For $Q^2<0.6~(GeV/c)^2$ the differential QE cross section for the "Transverse Enhancement" model is close to the  "Larger $M_A$  ($M_A$=1.3)" model. 
 The  maximum accessible  $Q^2$ for 1 GeV neutrinos is $1.3~GeV/c)^2$  (as shown in figure~\ref{q2max}). Therefore,  fits to  the neutrino differential QE cross sections for an incident energy of 1 GeV (e.g. MiniBooNE)  would yield $M_A \approx 1.2~GeV$.  The extracted value of $M_A$  depends on the specific
  model parameters that are used for Pauli blocking and the variation of the
  statistical errors in the data with $Q^2$.  For a neutrino energy of 1 GeV, the total integrated QE cross section  predicted by the   the "Transverse Enhancement"  model is  is also  larger than the total QE cross section prediction of the  "Independent Nucleon  ($M_A$=1.014)" model.

 In the high $Q^2$ region  ($Q^2>1.2~(GeV/c)^2$),
  the predicted differential QE cross section for the "Transverse Enhancement" model  is similar to the prediction of the  "Independent Nucleon  ($M_A$=1.014)"  model.  The maximum accessible  $Q^2$ for 3 GeV neutrinos is $4.9~GeV/c)^2$.  In order to reduce the sensitivity to modeling of 
  Pauli blocking,  experiments at higher energy\cite{ma-nuclear}  typically remove the
  lower $Q^2$ points in fits for $M_A$.   Consequently,  fits
 for  the neutrino differential QE cross sections measured in high energy experiments would yield a value of 
 $M_A$ which is smaller than $1.014~GeV$ because for $Q^2>0.5~(GeV/c)^2$ the slope of the differential QE cross section in the transition region between low and high $Q^2$ 
 is steeper than for $M_A=1.014~GeV$. This is consistent  with the fact that the average 
 $M_A$ extracted from high energy data on nuclear targets\cite{ma-nuclear}  is  $0.979 \pm 0.016$.

\begin{figure}
\includegraphics[width=3.503 in,height=2.6in]{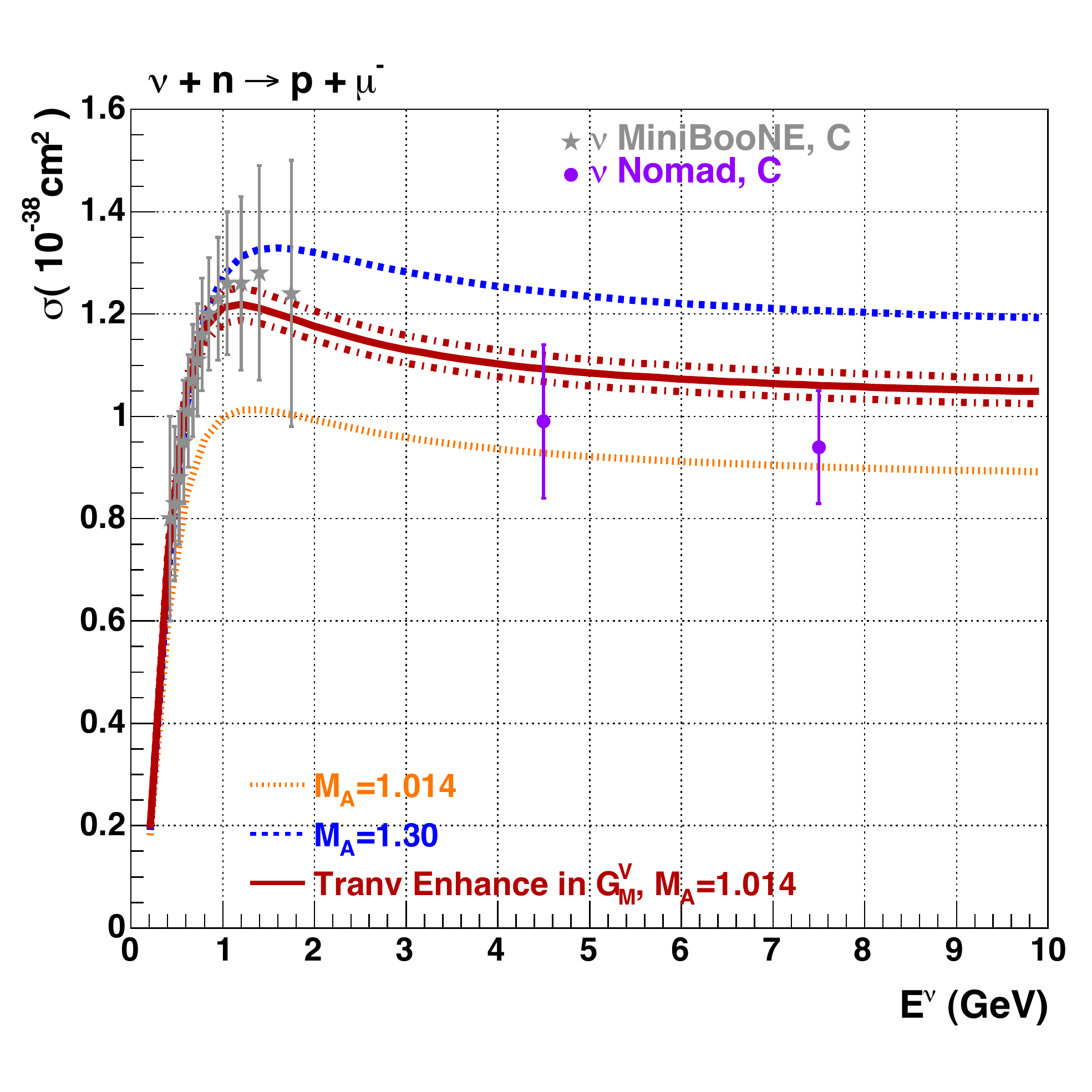}
\vspace{-0.15in}
\includegraphics[width=3.503 in,height=2.6in]{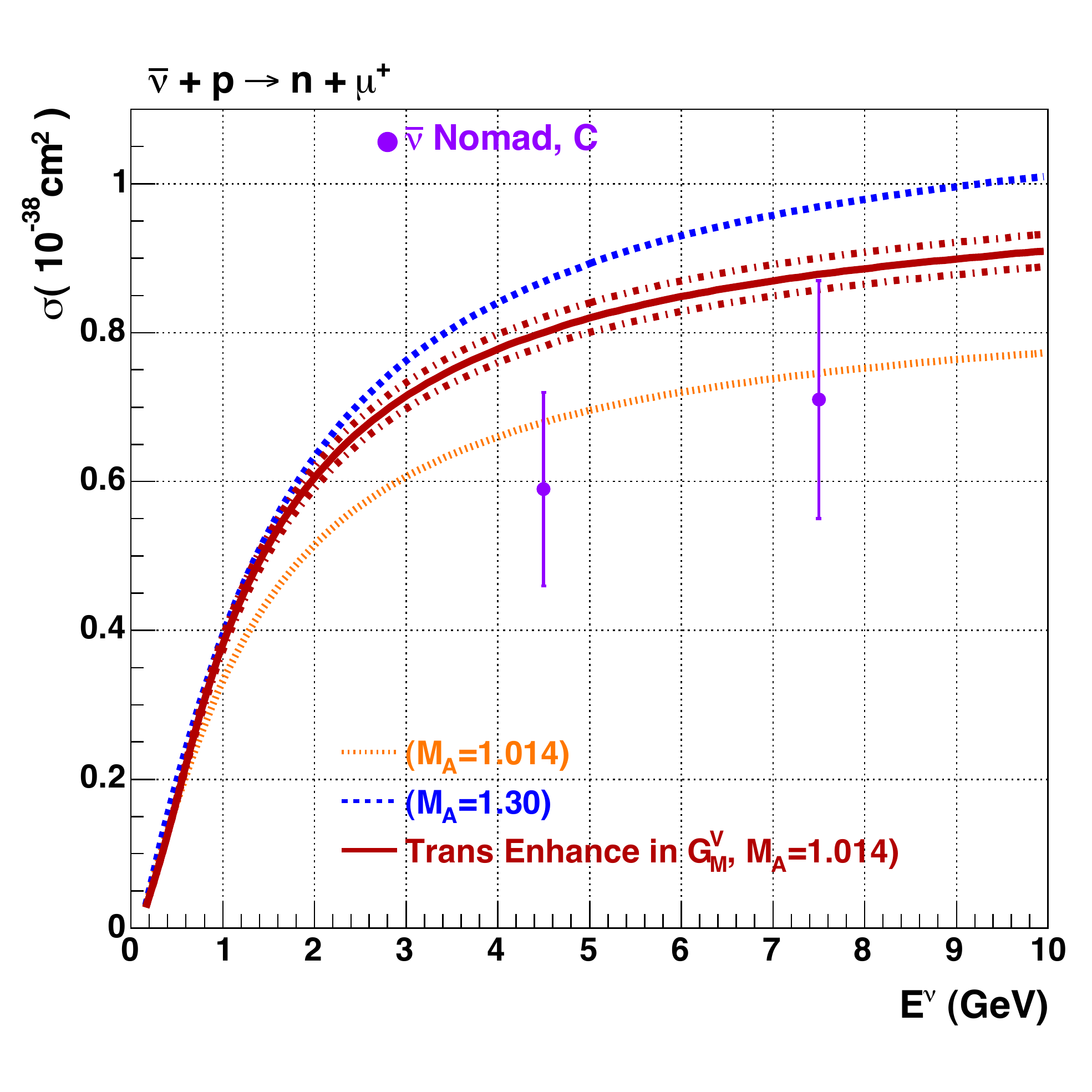}
\vspace{-0.15in}
\includegraphics[width=3.503 in,height=2.6in]{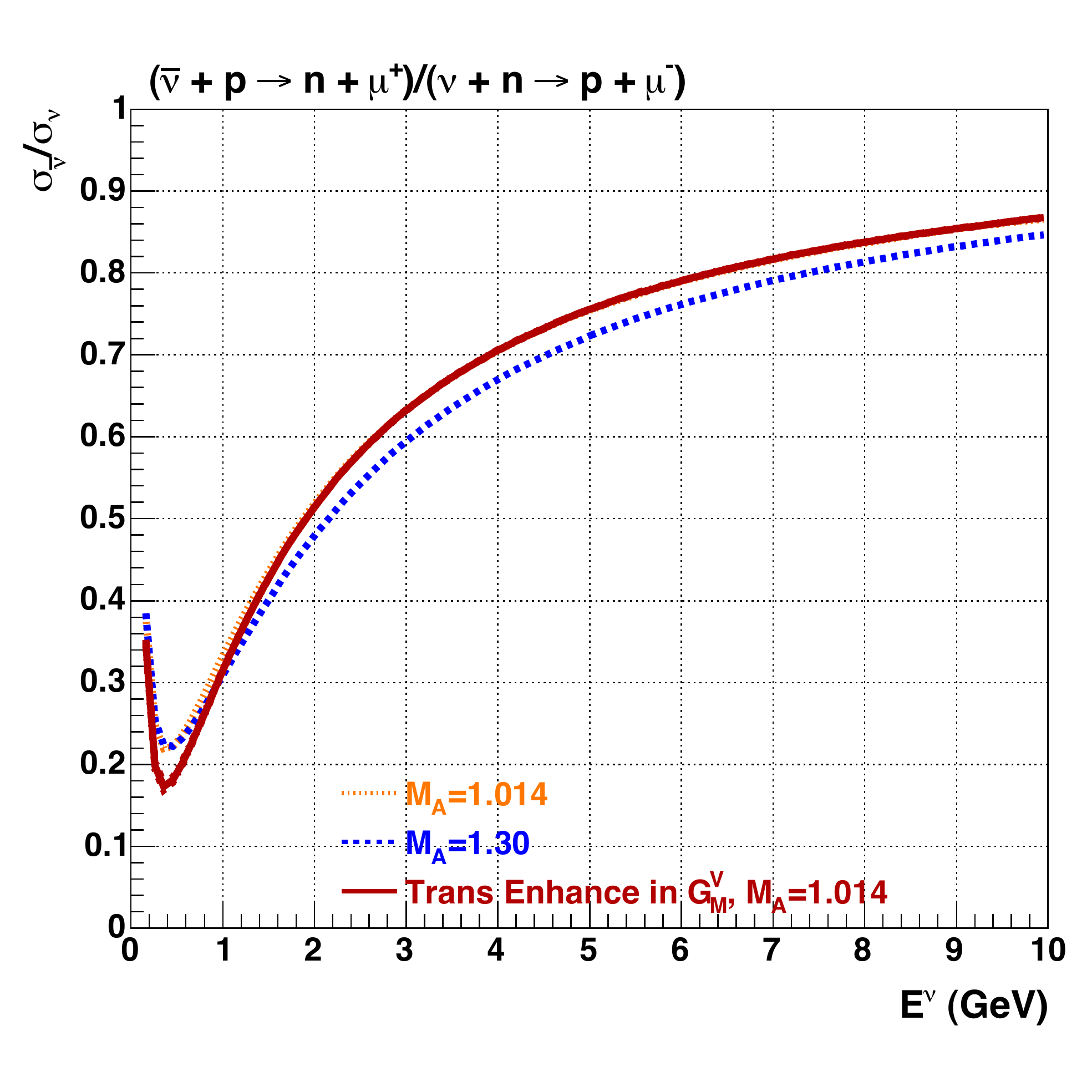}
\vspace{-0.1in}
\caption{The  total QE cross section 
%for nucleons bound in carbon
 as function of energy.
The data points are measurements of MiniBooNE\cite{MiniBooNE} (gray stars)  and NOMAD\cite{NOMAD} (purple circles).
The predictions for the "Independent Nucleon (MA=1.024)"  model,
"Larger $M_A$  ($M_A$=1.3) model", and "Transverse Enhancement model"  are shown. 
%The  orange dotted line is the prediction of the  "Independent Nucleon  ($M_A$=1.014)" model, the  blue dashed  line is  the prediction of the  "Larger % $M_A$  ($M_A$=1.3)" model, and the red  line is the prediction of the  "Transverse Enhancement" model (with error bands
%shown as dotted red lines).
Top (a): $\nu_{\mu}$  total QE cross section.  Middle (b): $\bar{\nu}_\mu$ total  QE cross section.
 Bottom (c):  QE $\bar{\nu}_\mu/\nu_{\mu}$  total cross section ratio..
  }
\label{total}
\end{figure}

Figure~\ref{total} shows the total QE cross section 
%for nucleons bound in  carbon 
%integrated (over $\nu$) including Pauli suppression 
%(but no Fermi motion) 
as function of energy. 
The data points are the measurements from MiniBooNE\cite{MiniBooNE} and  NOMAD\cite{NOMAD}.
The orange dotted  line is the prediction of the  "Independent Nucleon  ($M_A$=1.014)" model. The blue dashed  line is  prediction of the "Larger $M_A$  ($M_A$=1.3)".  The red  line is the prediction of the  "Transverse Enhancement" model (with error bands
shown as dotted red lines).
  The top (a) panel shows the  $\nu_\mu$  total QE cross section. 
The middle (b)  panel shows the $\bar{\nu}_\mu$ total QE cross section. 
The bottom (c) panel shows the ratio of  $\bar{\nu}_\mu$  and $\nu_{\mu}$  total QE cross sections.

\begin{figure}
\includegraphics[width=3.503 in,height=2.6in]{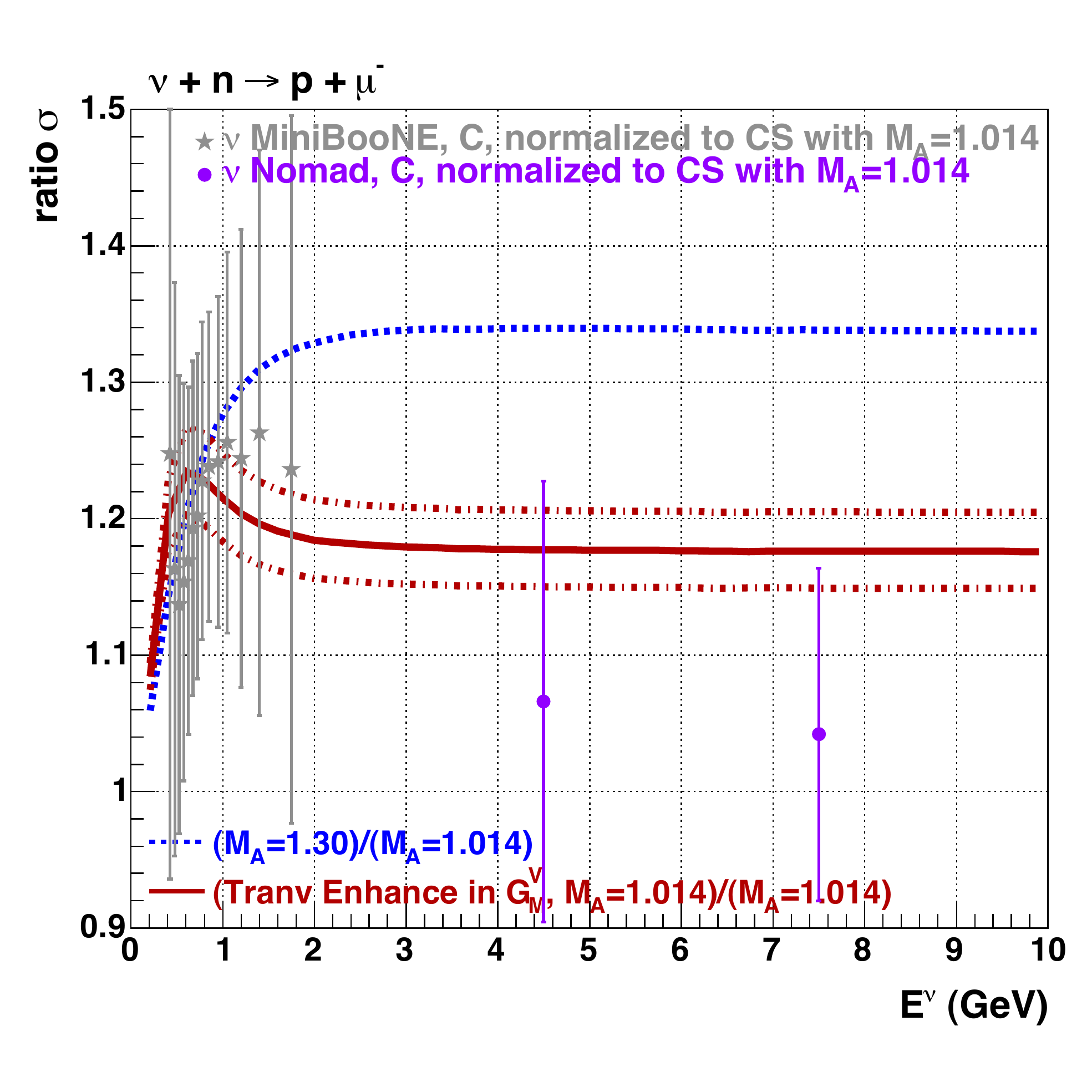}
\vspace{-0.15in}
\includegraphics[width=3.503 in,height=2.6in]{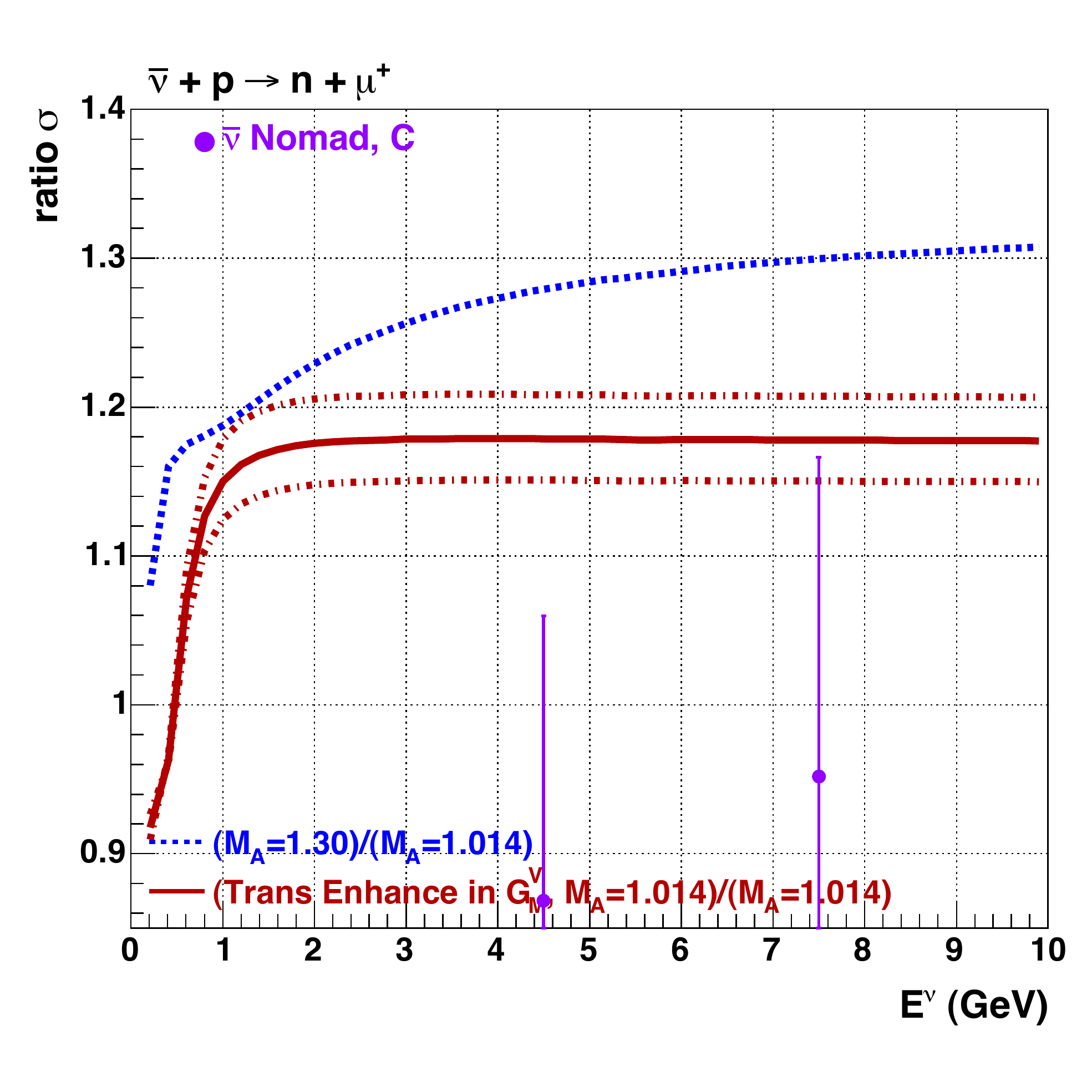}
\vspace{-0.15in}
\includegraphics[width=3.503 in,height=2.6in]{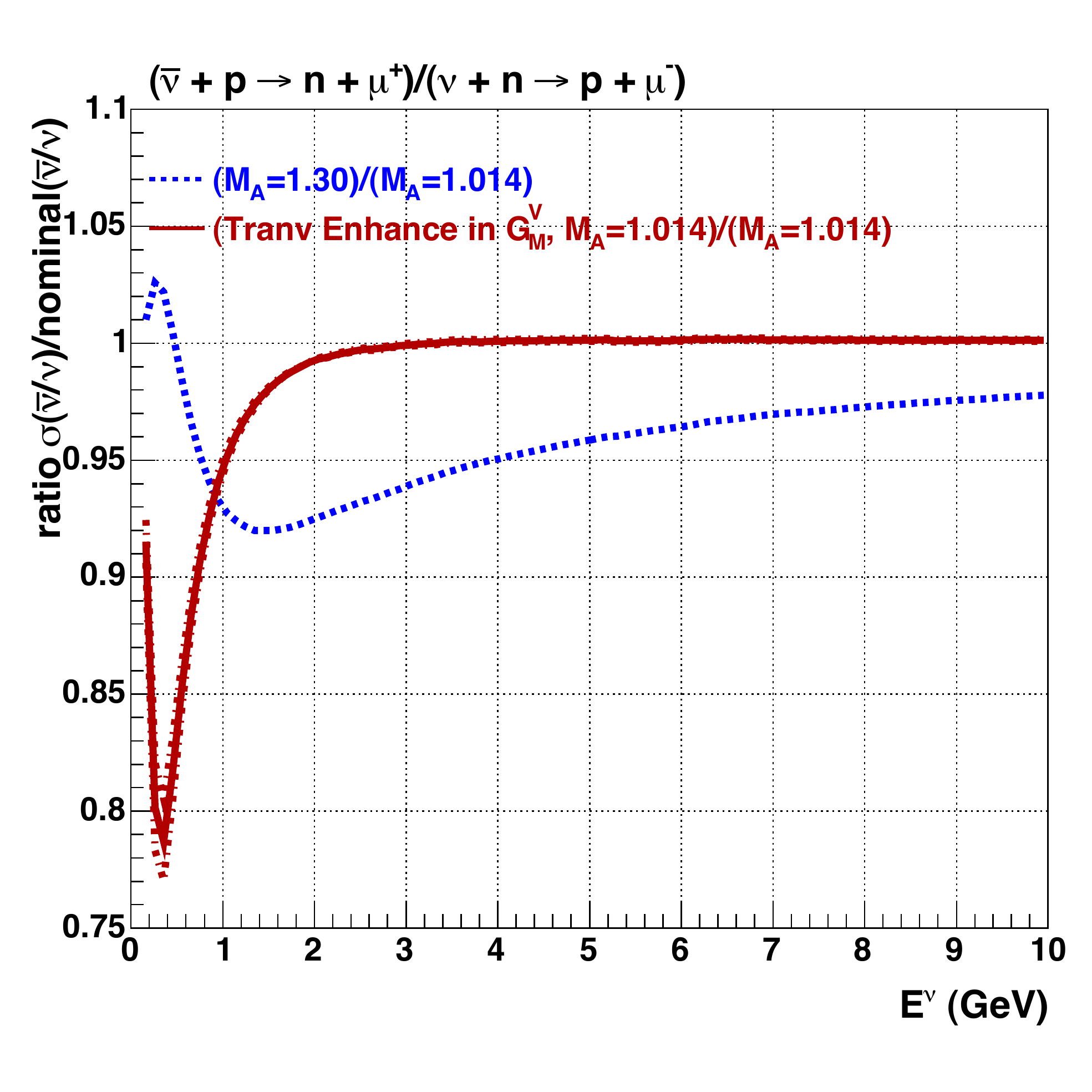}
\vspace{-0.1in}
\caption{The ratios of predicted and measured total  QE cross section  to the predictions of the 
 "Independent Nucleon  ($M_A$=1.014)"  model as a function
energy.
%  Here, the nominal
%model is the (over $\nu$) QE scattering  QE cross section on carbon  with Pauli suppression %(but no Fermi motion),   $M_A=1.014~GeV$  and $M_V=0.8426~GeV$. 
The ratios for the predictions of the "Larger $M_A$  ($M_A$=1.3) model" and "Transverse Enhancement model"  are shown. 
The data points are the ratios for the  measurements of MiniBooNE\cite{MiniBooNE} (gray stars)  and NOMAD\cite{NOMAD} (purple circles).
%
%The blue dashed  line is  the ratio for the prediction of the  "Larger $M_A$  ($M_A$=1.3)", and the red  line is the ratio of the predictions for the %"Transverse Enhancement" model (with error bands
%shown as dotted red lines).
Top (a): The ratio for $\nu_{\mu}$  total QE cross sections.  Middle (b): The ratio for  $\bar{\nu}_\mu$  QE cross sections. Bottom (c): The $\bar{\nu}_\mu/\nu_{\mu}$  total QE cross section ratio divided by the corresponding ratio for the "Independent Nucleon  ($M_A$=1.014)" model).  }
\label{ratiototal}
\end{figure}

Figure~\ref{ratiototal} shows the  ratio of the predictions for total  QE cross section  to the predictions of the  "Independent Nucleon  ($M_A$=1.014)"  model as a function
energy.  The blue dashed  line is  the ratio of the predictions  for the  "Larger $M_A$  ($M_A$=1.3)" model, and the red  line is ratio for the "Transverse Enhancement" model (with error bands
shown as dotted red lines).
 The top (a) panel shows the ratio of the predictions for  the $\nu_\mu$  total QE cross section. 
The middle (b) panel shows the ratio of the predictions for the  $\bar{\nu}_\mu$ total QE cross section. The bottom (c) panel shows the predicted  $\bar{\nu}_\mu/\nu_{\mu}$ cross section ratio divided by  the predicted  $\bar{\nu}_\mu/\nu_{\mu}$   ratio  for  the "Independent Nucleon  ($M_A$=1.014)" model.
The data points are measurements from MiniBooNE\cite{MiniBooNE} and NOMAD\cite{NOMAD}.

As shown in Fig.~\ref{total} (a) (top), and Fig.~\ref{ratiototal} (a)(top), at low $\nu_{\mu}$ energies the  "Transverse Enhancement" model  (red  line) predicts  QE cross sections at a level similar to the  "Larger $M_A$  ($M_A$=1.3)" model (blue dashed  line).
Both the "Larger $M_A$  ($M_A$=1.3)" model and the   "Transverse Enhancement" model predictions are  in  agreement with the MiniBooNE QE $ \nu_{\mu}$ cross sections.
However, at higher $\nu_{\mu}$ energies the  "Transverse Enhancement" model 
predicts  QE cross sections which are lower than the prediction of the "Larger $M_A$  ($M_A$=1.3)".  The lower QE $\nu_{\mu}$  cross sections at high energy  
are consistent with the NOMAD measurements (within experimental errors).

Similarly, for  $\bar{\nu}_\mu$ scattering  the "Transverse Enhancement"  model 
predicts total QE cross section which are lower than the predictions of the  "Larger $M_A$  ($M_A$=1.3)" model as shown in Fig.~\ref{total} (b) (middle) and Fig.~\ref{ratiototal} (b) (middle). 
The lower QE cross $\bar{\nu}_\mu$  sections 
are consistent with the NOMAD measurements (within experimental errors).    

  \section{Conclusion}
  
  We parametrize the  enhancement  in the transverse QE cross section observed
  in QE electron scattering on nuclear targets as a correction
  to the magnetic form factors of bound nucleons.  Within models of MEC, 
MEC  processes contribute only to the transverse QE response function  and do not enhance the longitudinal and axial response functions.  We find that the  QE cross sections for  $\nu_{\mu},\bar{\nu}_\mu$ QE scattering predicted by the  "Transverse Enhancement" model   
agree with the MiniBooNE low energy neutrino QE cross sections, and are also consistent with  QE cross sections measured by NOMAD at higher energies.

   The simple two parameter parametrization of $Q^2$  dependence of the transverse enhancement as a correction to the proton and neutron magnetic form factors can easily
be incorporated into existing Monte Carlo generators\cite{GENIE}.
            
At present,  $\nu_{\mu},\bar{\nu}_\mu$ experiments use the "Large $ M_A$"  model to predict 
${\cal W}_1^{Qelastic}$, ${\cal W}_2^{Qelastic}$, and ${\cal W}_3^{Qelastic}$ for  neutrino QE scattering on nuclear targets.
A large increase in  $M_A$ is contrary to theoretical expectations\cite{ma-nuclear,Tsushima_03}.

The differential and total QE cross sections predicted in the  "Larger $M_A$  ($M_A$=1.3)" model
  are  similar to the predictions of "Transverse Enhancement"  model only at low
    $\nu_{\mu}$ energies. 
    %For $ \bar{\nu}_\mu$ the "Transverse Enhancement"  model
%prediction for the  total   $ \bar{\nu}_\mu$ QE cross section is smaller than the  prediction of  %"Larger $M_A$  ($M_A$=1.3)" model at all energies, as shown in Fig.\ref{ratiototal}.

%Figure 9
% 
%
%Figure 10
%
 %2.6133  to become 3.503
%Figure 11
 \begin{figure}
\includegraphics[width=3.503 in,height=2.6in]{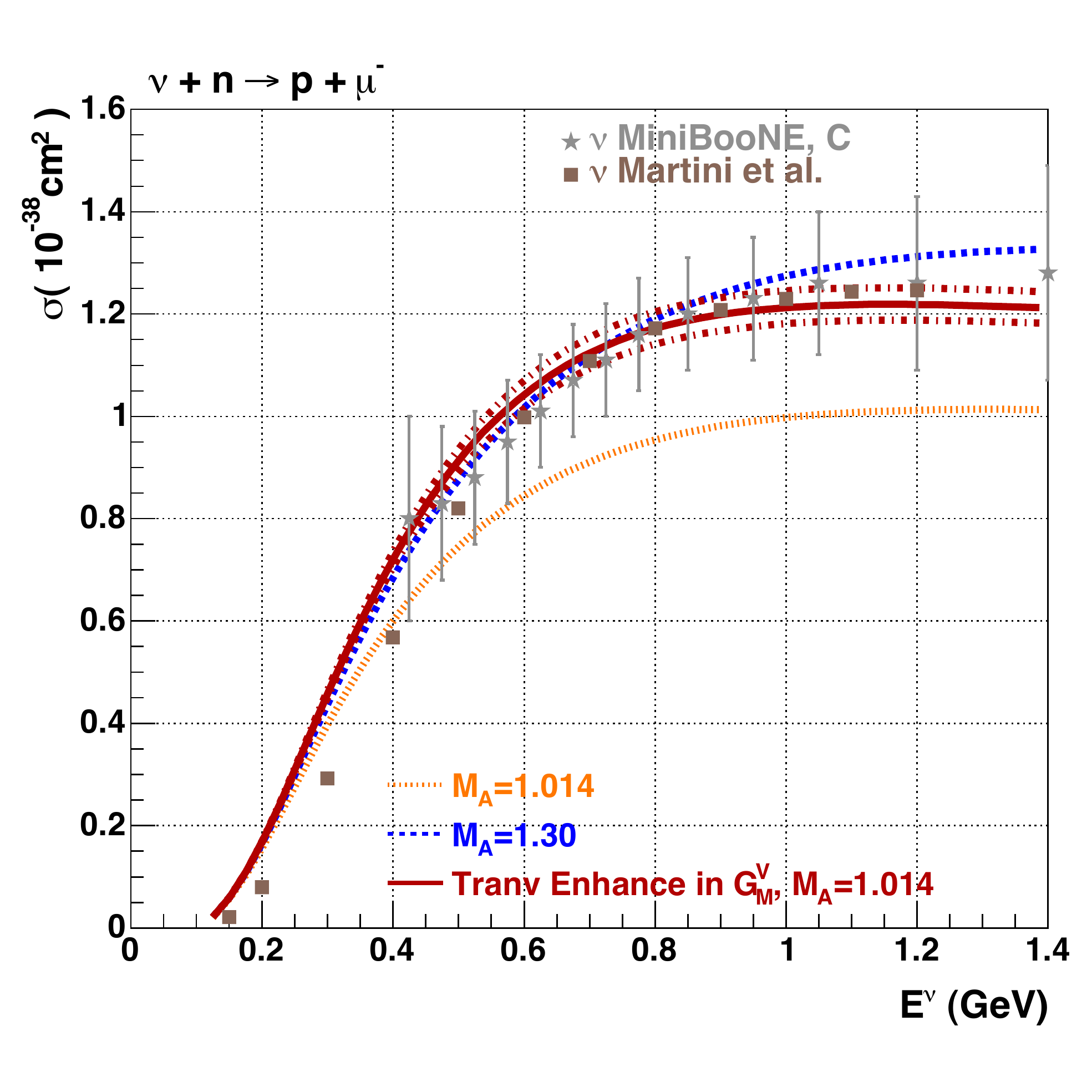}
\vspace{-0.15in}
\includegraphics[width=3.503 in,height=2.6in]{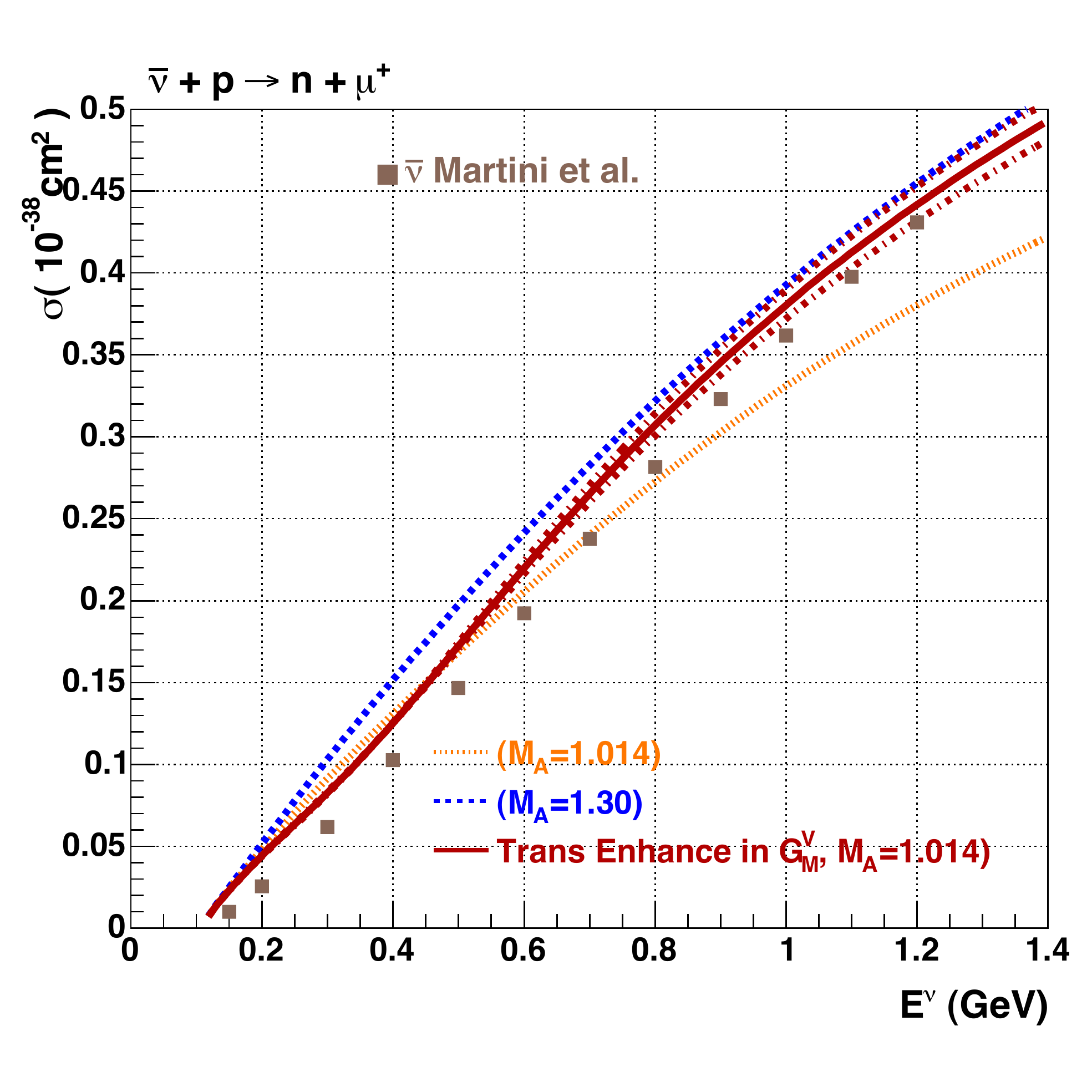}
%\vspace{-0.3in}
%\includegraphics[width=2.6133 in,height=2.6in]{f9nub_nu_enu_meson_lagr.pdf}
%\vspace{-0.05in}
\caption{Comparison to the QE cross section predicted by the  "QE+np-nh RPA"  MEC model of Martini {\em et al.}\cite{MEC5}  (Predictions for this model have only been published for neutrino energies less than 1.2 GeV). 
The predictions for the "Independent Nucleon (MA=1.024)"  model,
"Larger $M_A$  ($M_A$=1.3) model", and "Transverse Enhancement model"  are shown. 
 The  grey squares are the predictions of the MEC model of  Martini {\em et al.} \cite{MEC5} 
%The  orange dotted line is the prediction of the  "Independent Nucleon  ($M_A$=1.014)" model.  The  blue dashed  line is  the prediction of the  "Larger %$M_A$  ($M_A$=1.3)" model. The red  line is the prediction of the  "Transverse Enhancement" model (with error bands
%shown as dotted red lines).
  The data points are measurements from MiniBooNE\cite{MiniBooNE} (grey stars).
Top (a): $\nu_{\mu}$  total QE cross section.  Middle (b): $\bar{\nu}_\mu$ total  QE cross section.
% Bottom (c):  QE $\bar{\nu}_\mu/\nu_{\mu}$  total cross section ratio.
  }
\label{totalMEC}
\end{figure}
% 
%Figure 12
\begin{figure}
\includegraphics[width=3.503 in,height=2.6in]{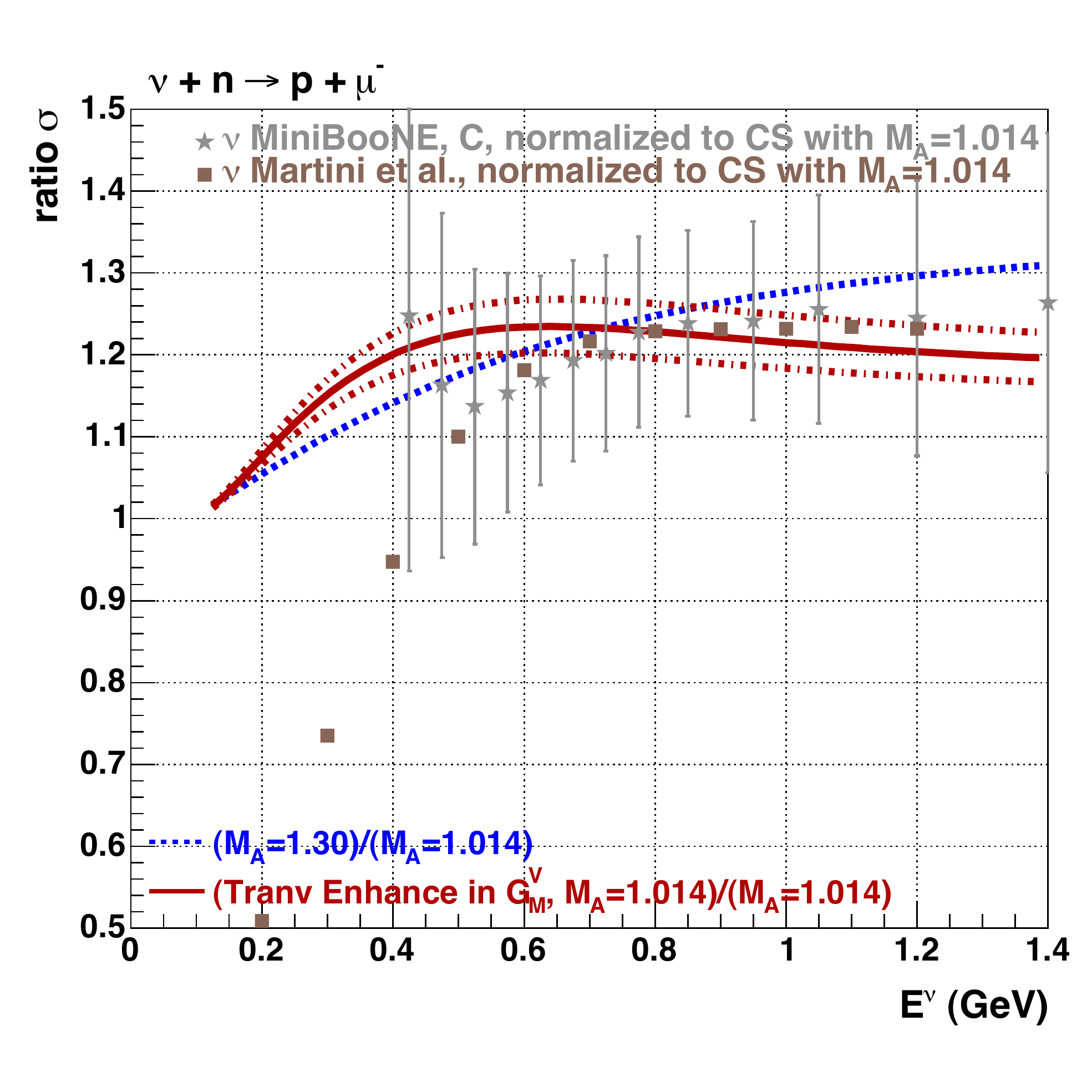}
\vspace{-0.15in}
\includegraphics[width=3.503 in,height=2.6in]{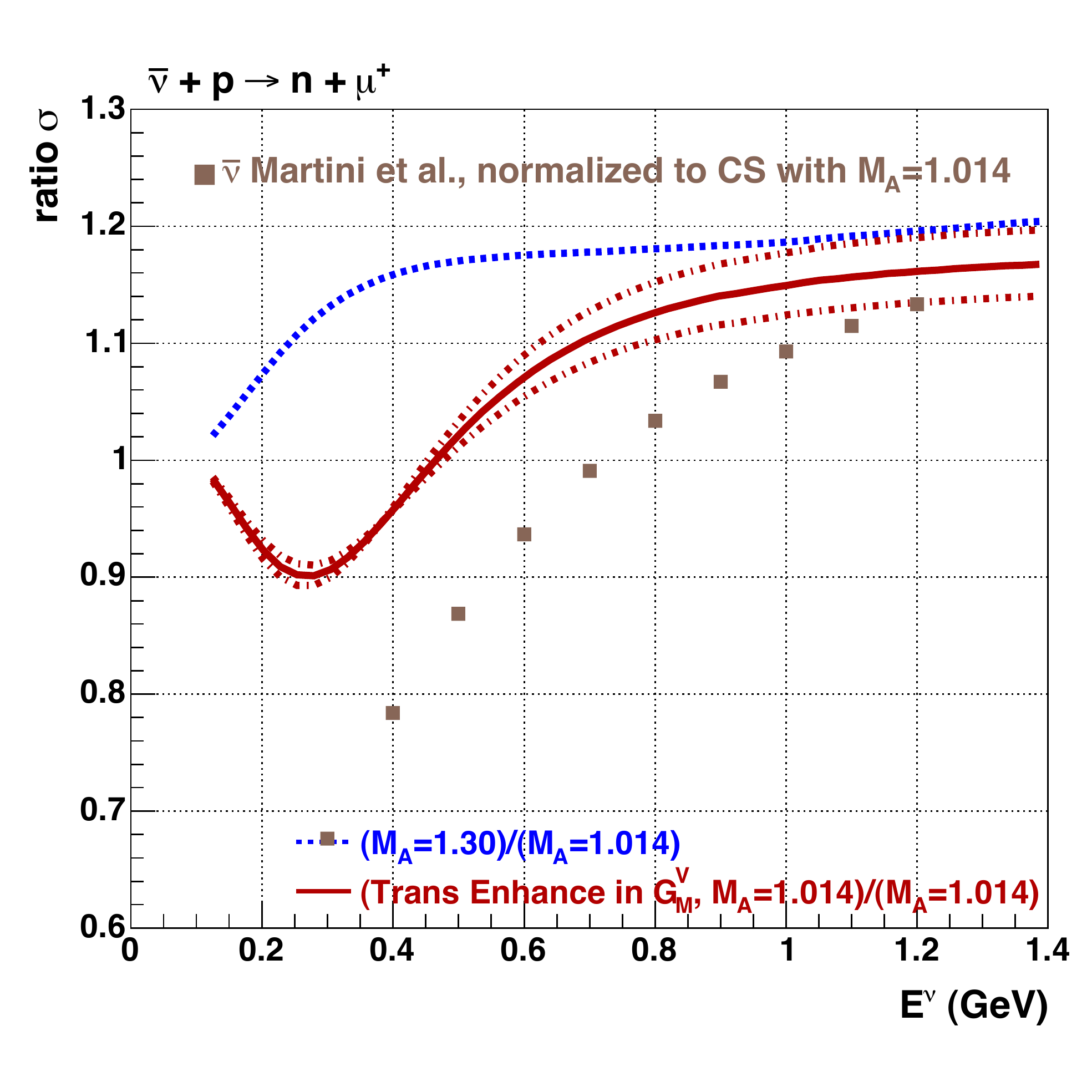}
%\vspace{-0.3in}
%\includegraphics[width=2.5 in,height=2.6in]{f10nub_nu_enu_meson_lagr.pdf}
%\vspace{-0.05in}
\caption{Comparison to the QE cross section predicted by the  "QE+np-nh RPA"  MEC model of Martini {\em et al.}\cite{MEC5}  (Predictions for this model have only been published for neutrino energies less than 1.2 GeV). Shown are the ratios for  the predictions of the "Larger $M_A$  ($M_A$=1.3) model" and "Transverse Enhancement model" to the  "Independent Nucleon  ($M_A$=1.014)"  QE cross section as a function
energy.
%  Here, the nominal
%model is the (over $\nu$) QE scattering  QE cross section on carbon  with Pauli suppression %(but no Fermi motion),   $M_A=1.014~GeV$  and $M_V=0.8426~GeV$. The grey squares are  the ratios for the prediction of the  Martini {\em et al.}
%The ratios for predictions for the 
%"Larger $M_A$  ($M_A$=1.3) model", and "Transverse Enhancement model"  are shown. 
The  grey squares are the ratios for  the predictions of the MEC model of  Martini {\em et al.}\cite{MEC5} 
%
 % The  blue dashed  line is  ratio for  the prediction of the  "Larger $M_A$  ($M_A$=1.3)" model.  The red  line is the ratio for the predictions for the %"Transverse Enhancement" model (with error bands
%shown as dotted red lines). 
The data points are the ratios for the  measurements of MiniBooNE\cite{MiniBooNE} (grey stars).
Top (a): The ratio for $\nu_{\mu}$  total QE cross sections.  Bottom (b): The ratio for  $\bar{\nu}_\mu$  QE cross sections. 
%(c): The $\bar{\nu}_\mu/\nu_{\mu}$  total QE cross section ratio divided by the corresponding ratio for the "Independent Nucleon  ($M_A$=1.014)" %model). 
 }
\label{ratioMEC}
\end{figure}

  The assumption made in the "Transverse Enhancement" model is that the enhancements in the transverse response functions  in $\bar{\nu}_\mu/\nu_{\mu}$ scattering
are  the same as measured in electron scattering,  and that there is no
additional  enhancement  in the longitudinal or axial response functions (as  
 expected in MEC models\cite{MEC4}).
  Since we only uses parameters from electron scattering
   data, our analysis is purely phenomenological, and does not rely on
   a specific MEC model.   Because in electron scattering  the transverse enhancement  is only significant at low values of $Q^2$, the contribution of "Transverse Enhancement" to the total neutrino QE cross section is  energy dependent, thus resolving  much of the apparent discrepancy between the low energy and high energy neutrino QE cross sections on nuclear targets.
   
    In an earlier publication, Martini, Ericson, Chanfray, and  Marteau\cite{MEC5}  calculated the contribution of meson exchange currents   to the differential and total QE cross sections for  $\bar{\nu}_\mu/\nu_{\mu}$ energies less than 1.2 GeV.
In the comparison with our model, we use  the Martini $et~al$ 
predictions with the  random phase approximation ( "QE+np-nh RPA").   
%  $\nu_{\mu}$ scattering at this energy range
 For  the range $0.5<E<1.2$, the  predictions of  Martini $et~al$  are similar to the predictions of the  "Transverse Enhancement" model as shown in Figures~\ref{totalMEC}, and \ref{ratioMEC}. 
 %For the case of  $\bar{\nu}_\mu$ scattering there are difference in the predictions of the two models.
 For  $E<0.5$, the  predictions of  Martini $et~al$ are lower than the predictions of the  "Transverse Enhancement" model. However, for such
 low energies, the predictions are  sensitive to differences in the modeling of Pauli blocking in the two models.  
 The predictions of the  
   Martini $et~al$  model  for $\bar{\nu}_\mu/\nu_{\mu}$ scattering for 
 energies greater than 1.2 GeV  have not yet been published.  
 
\begin{figure}
\includegraphics[width=3.503 in,height=2.6in]{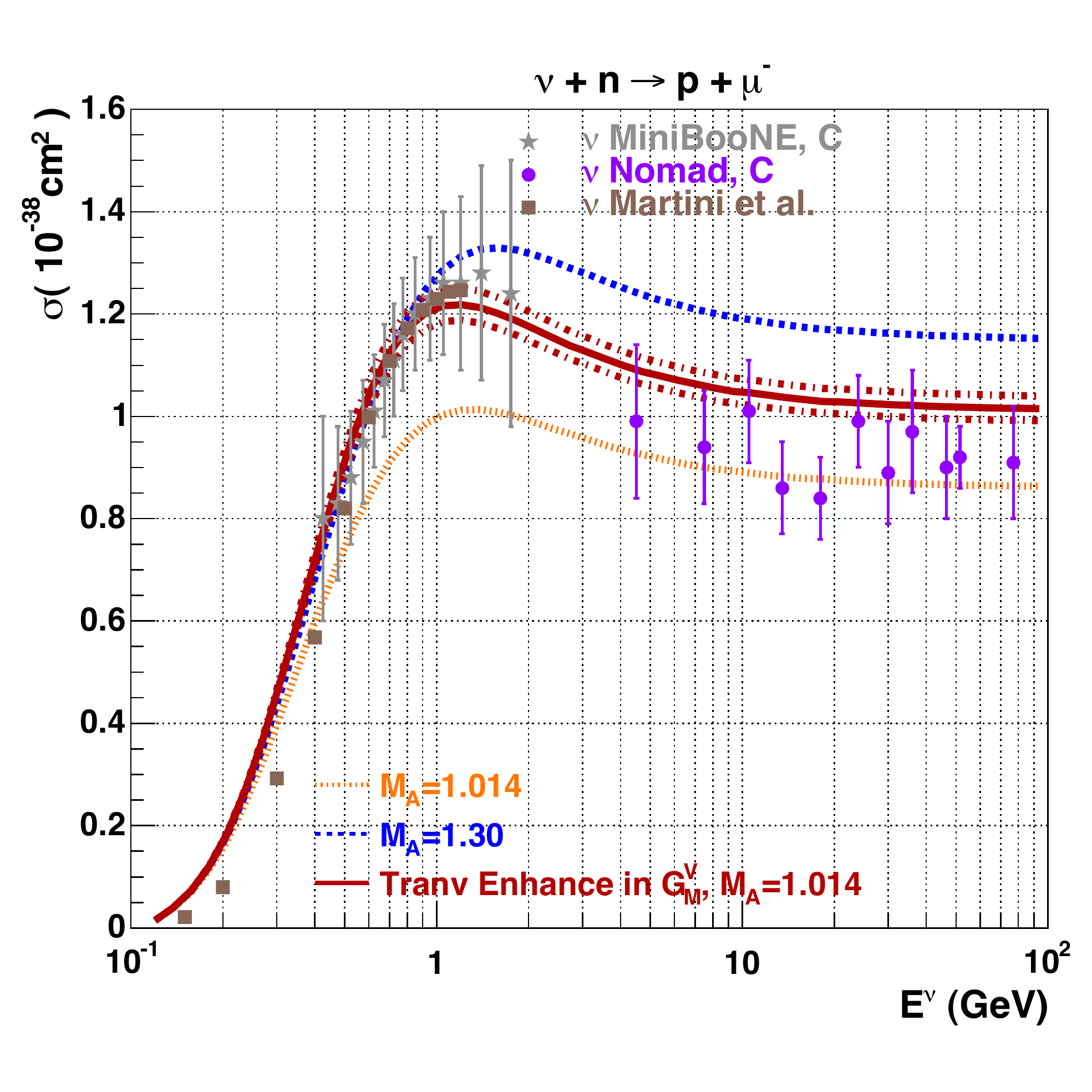}
\includegraphics[width=3.503 in,height=2.6in]{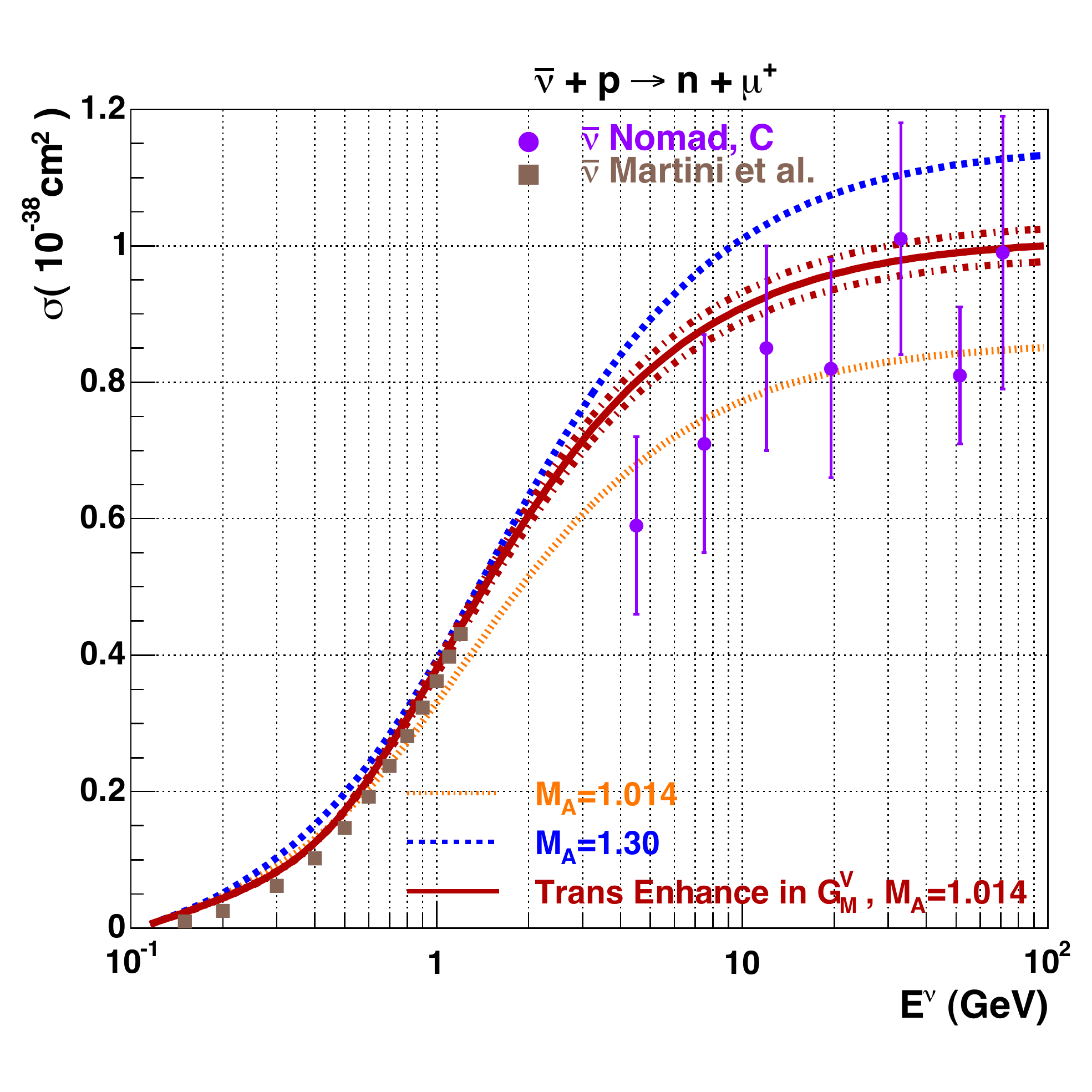}
\caption{Comparison of predictions for the   $\nu_{\mu}$, $\bar{\nu}_\mu$  total QE cross section section at high energies for
the "Independent Nucleon (MA=1.024)"  model, the 
"Larger $M_A$  ($M_A$=1.3) model",  the "Transverse Enhancement model",  and  the 
  "QE+np-nh RPA"  MEC model of Martini {\em et al.}\cite{MEC5} (Predictions for this model have only been published for neutrino energies less than 1.2 GeV). 
  The data points are the ratios for the  measurements of MiniBooNE\cite{MiniBooNE} (gray stars)  and NOMAD\cite{NOMAD} (purple circles)
  }
\label{totalMEClog}
\end{figure}

\begin{figure}
\includegraphics[width=3.5in,height=2.8in]{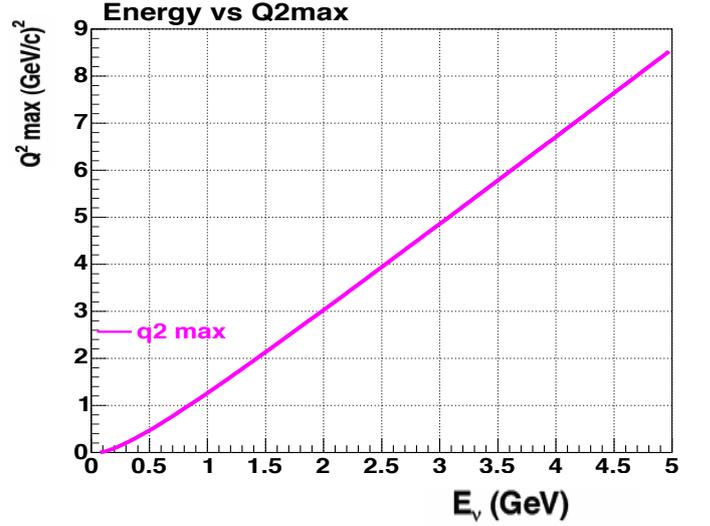}
\caption{The maximum accessible $Q^2$ for QE events as a function of neutrino energy. }
\label{q2max}
\end{figure}

\begin{figure}
\includegraphics[width=3.503in,height=2.8in]{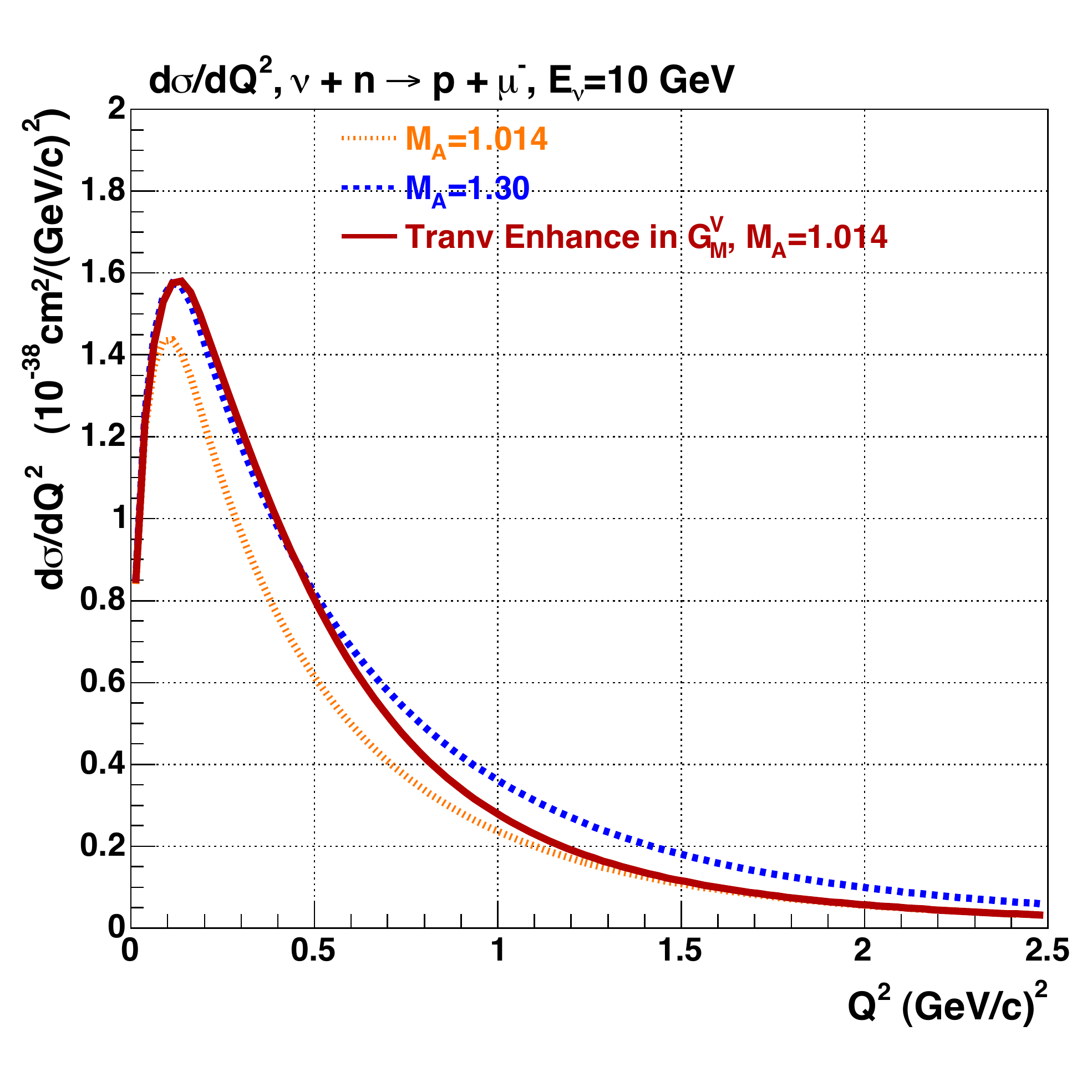}
\vspace{-0.15in}
\includegraphics[width=3.503 in,height=2.8in]{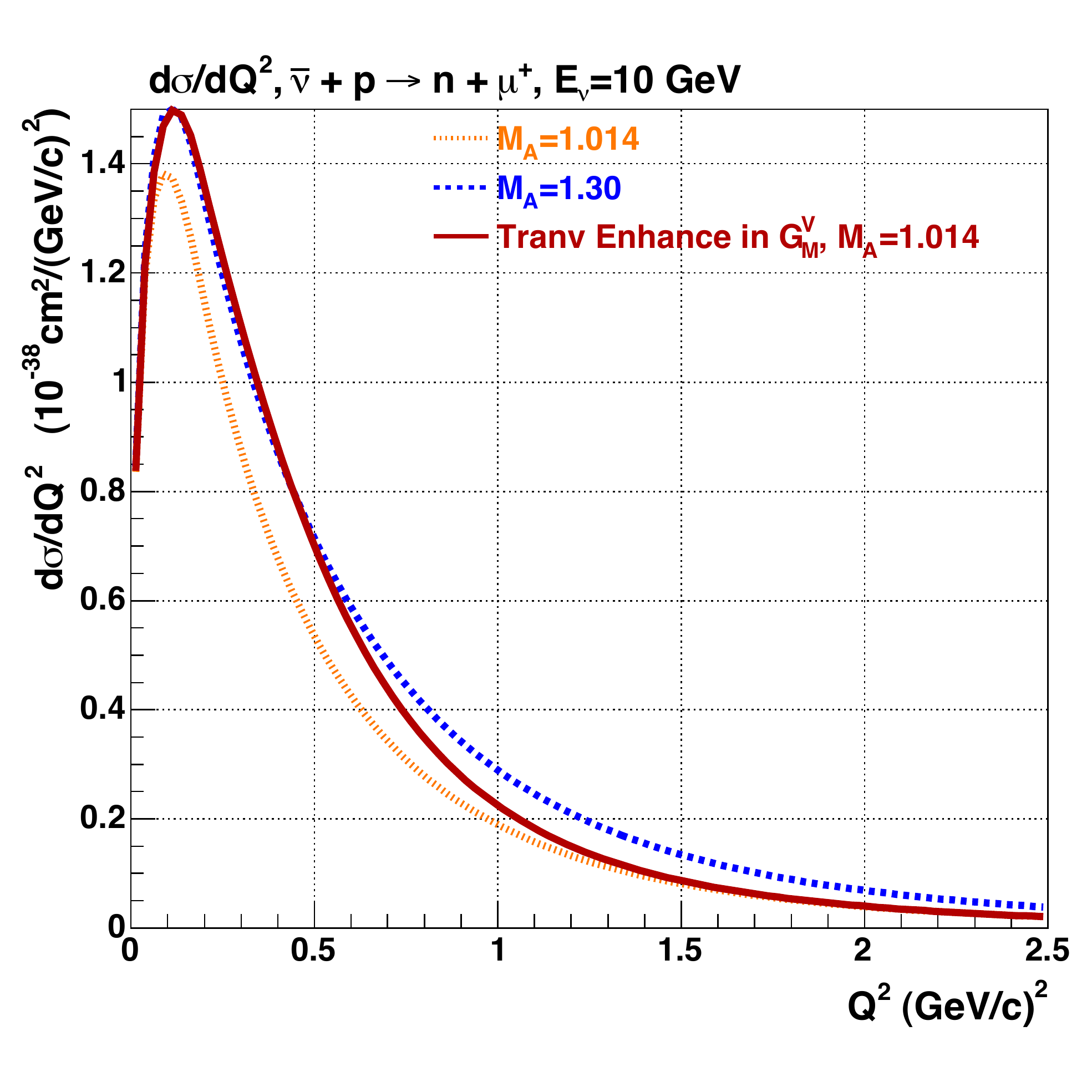}
\vspace{-0.1in}
\caption{Same as figure~\ref{diff1} for $\nu_{\mu},\bar{\nu}_\mu$  energies of 10.0 GeV.  }
\label{diff10}
\end{figure}

\begin{figure}
\includegraphics[width=3.503in,height=2.8in]{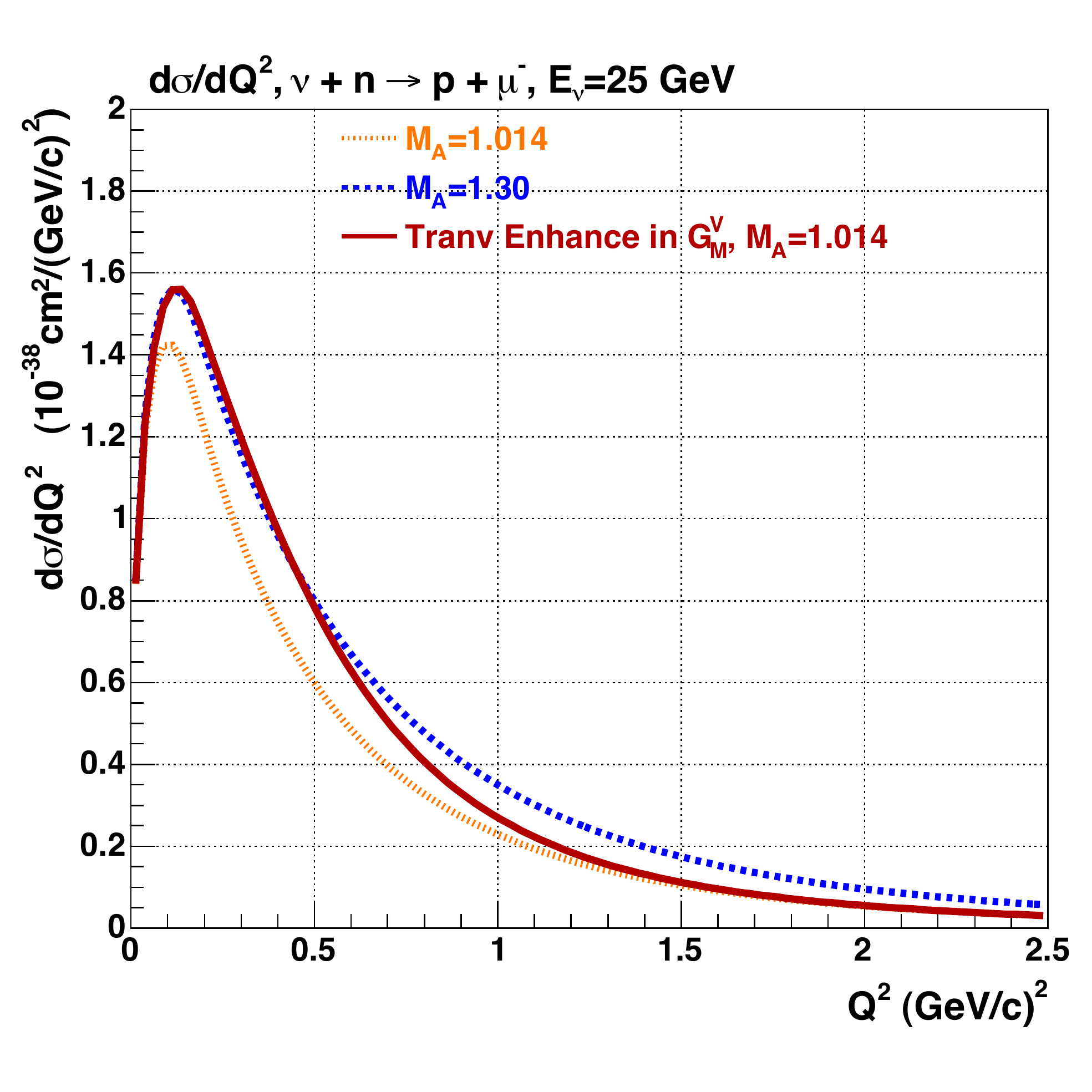}
\vspace{-0.15in}
\includegraphics[width=3.503in,height=2.8in]{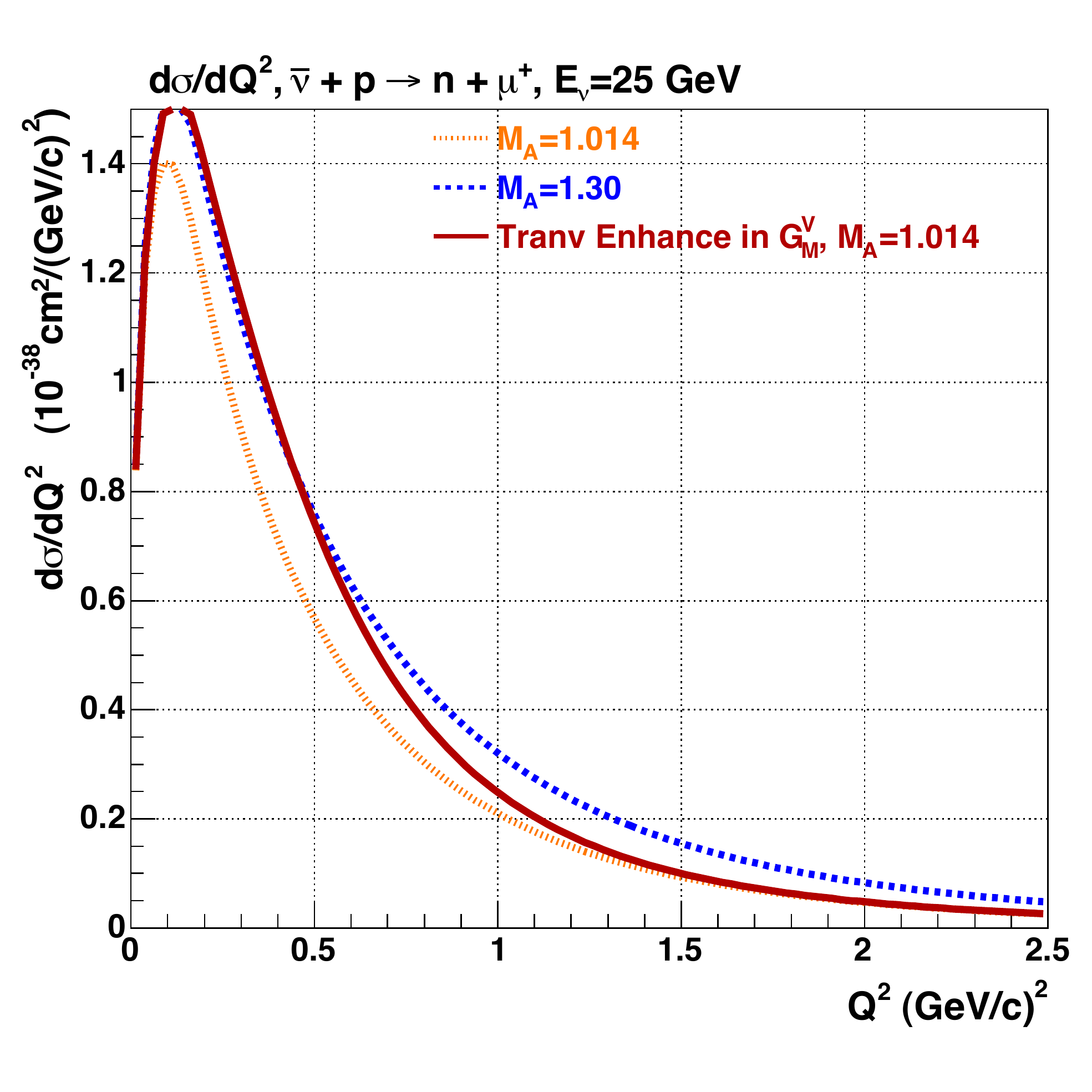}
\vspace{-0.1in}
\caption{ Same as figure~\ref{diff1} for $\nu_{\mu},\bar{\nu}_\mu$  energies of 25.0 GeV. }
\label{diff25}
\end{figure}

\begin{figure}
\includegraphics[width=3.5 in,height=2.6in]{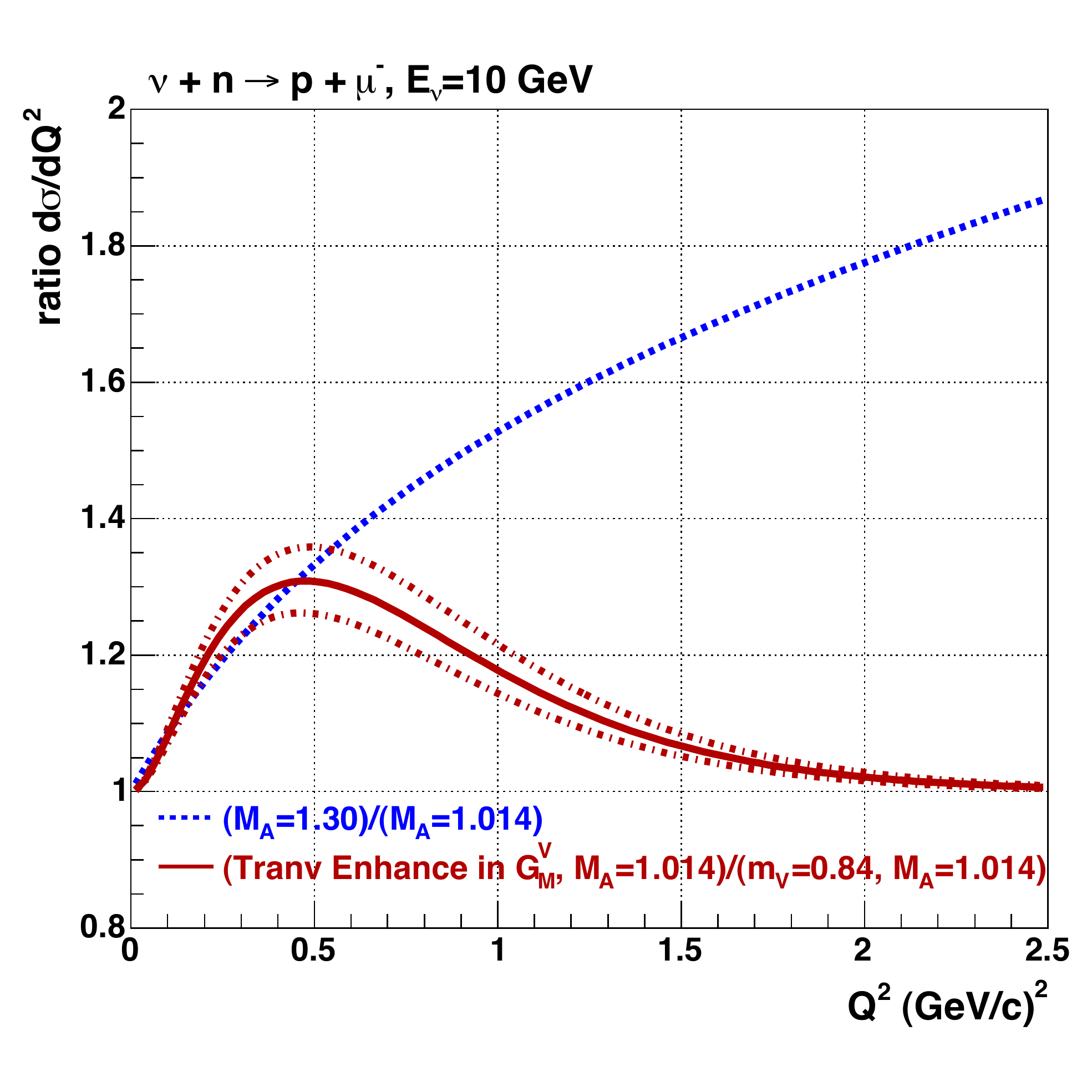}
\vspace{-0.15in}
\includegraphics[width=3.5in,height=2.6in]{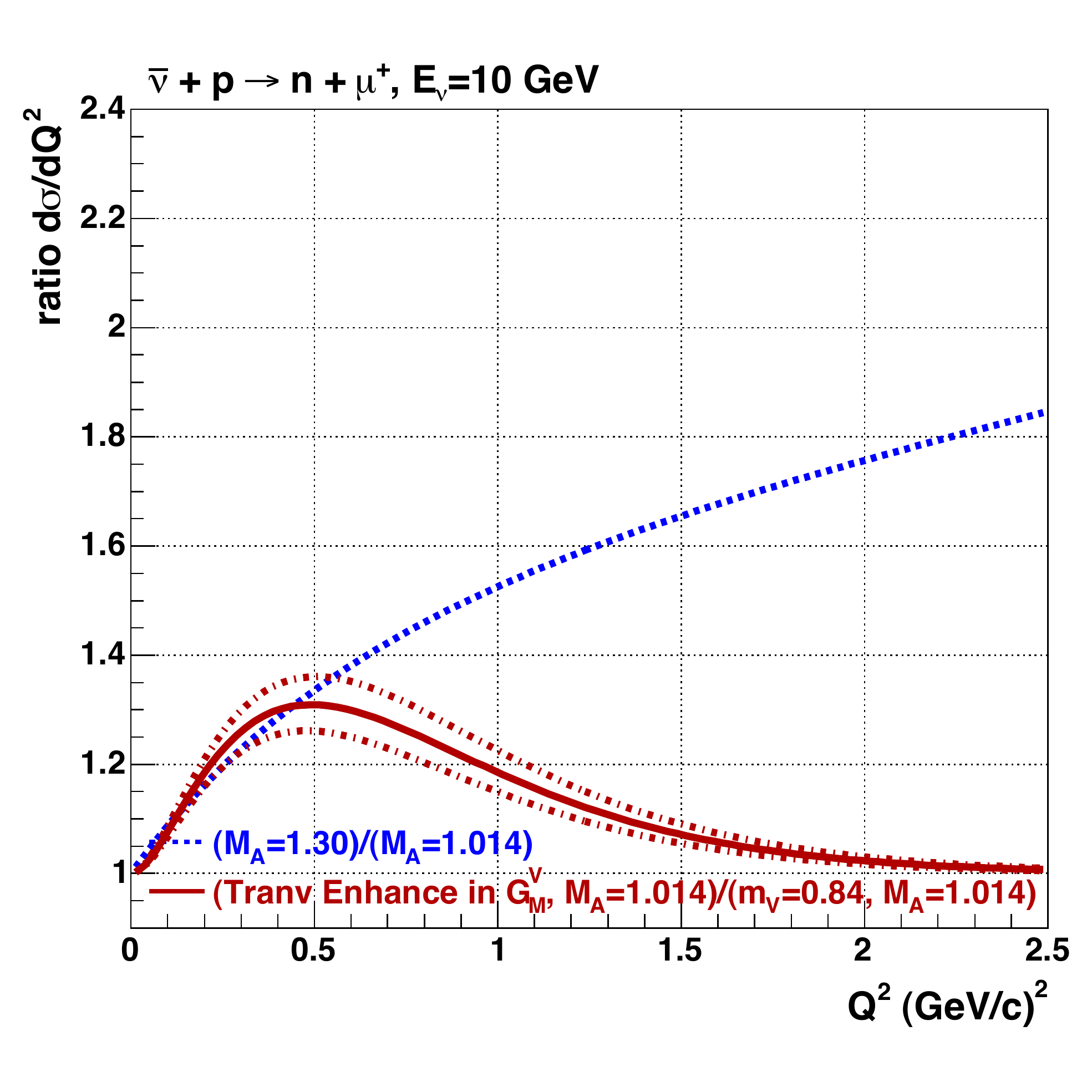}
\vspace{-0.15in}
\includegraphics[width=3.5in,height=2.6in]{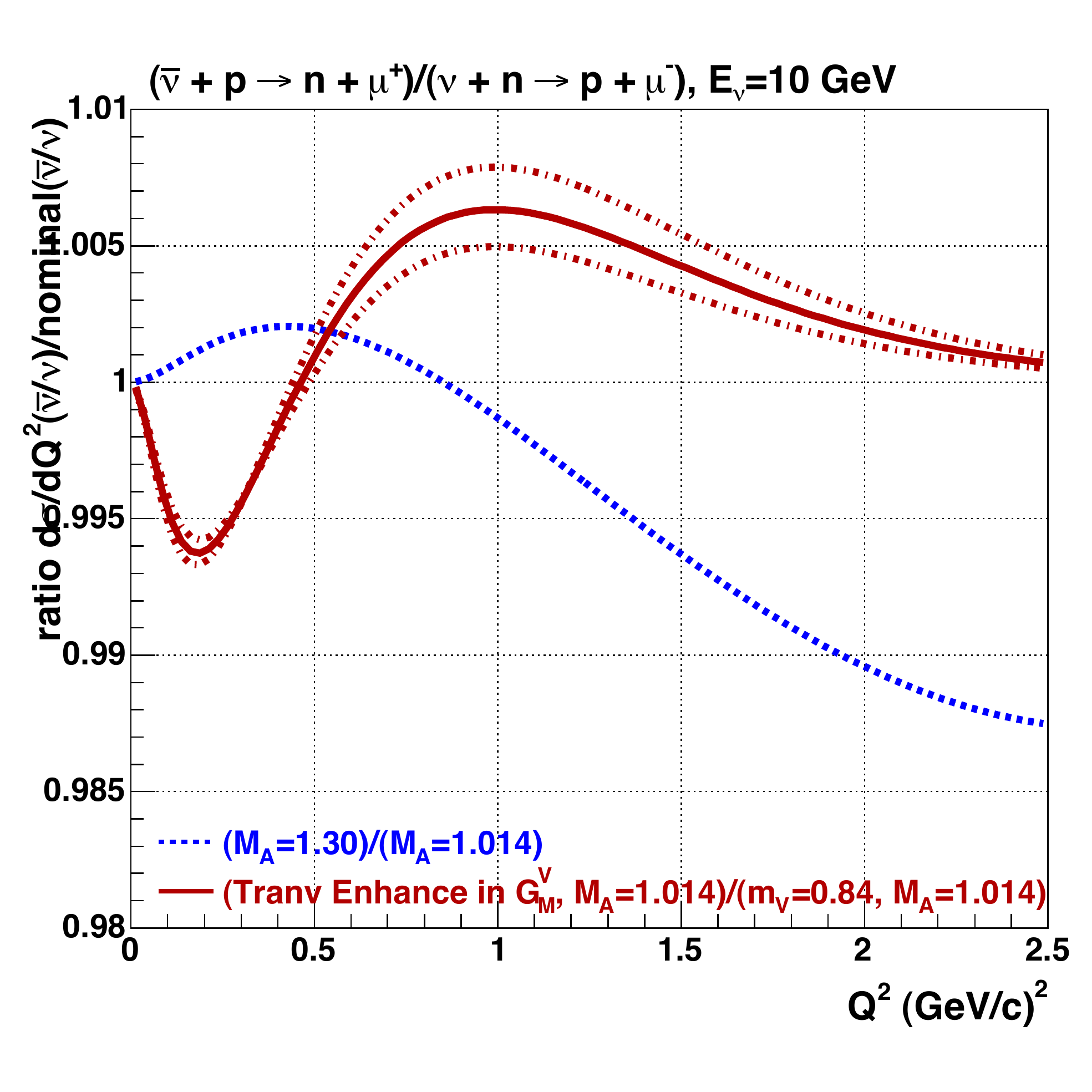}
\vspace{-0.1in}
\caption{Same as figure~\ref{ratio1} for $\nu_{\mu},\bar{\nu}_\mu$  energies of 10   GeV.}
\label{ratio10}
\end{figure}

\begin{figure}
\includegraphics[width=3.5in,height=2.6in]{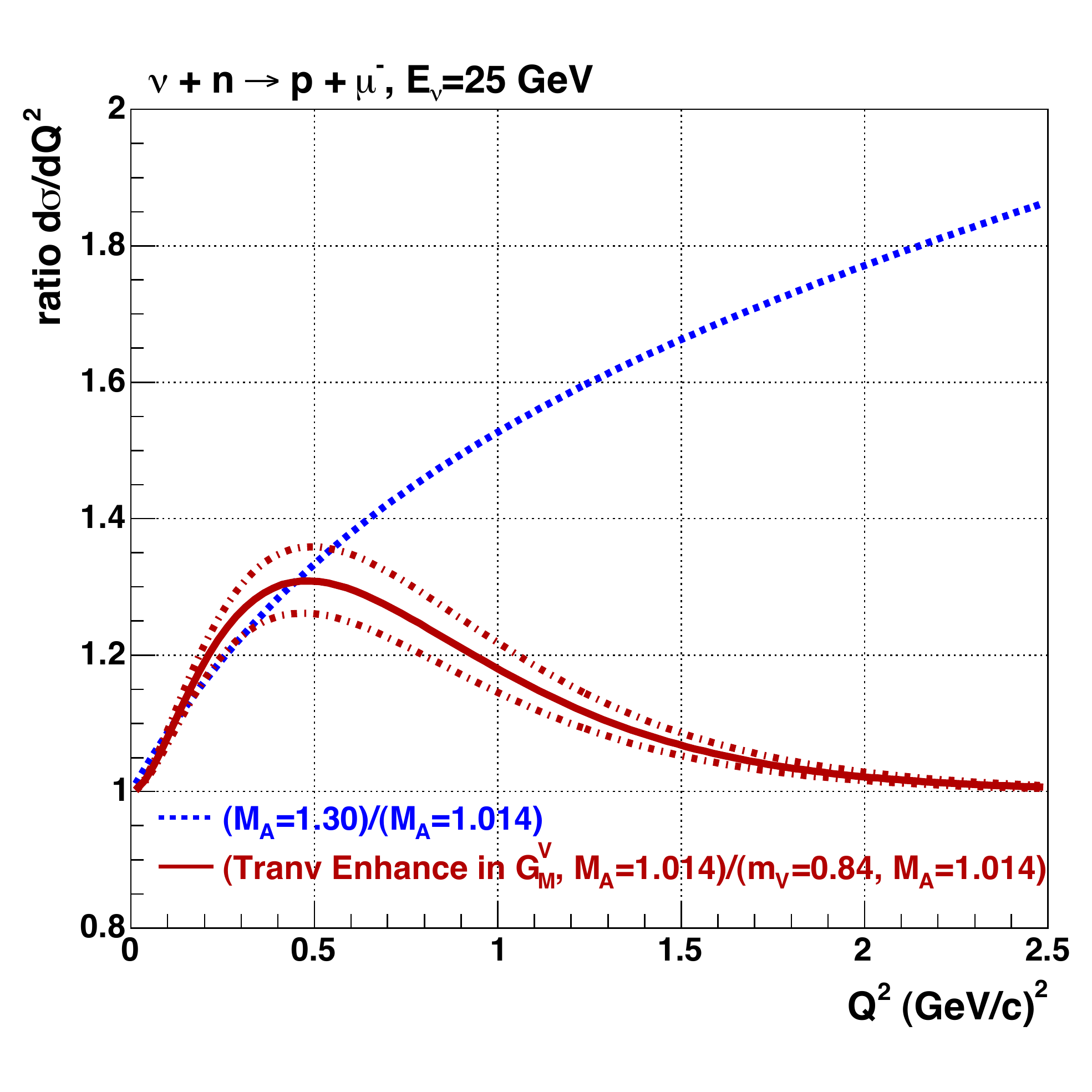}
\vspace{-0.15in}
\includegraphics[width=3.5in,height=2.6in]{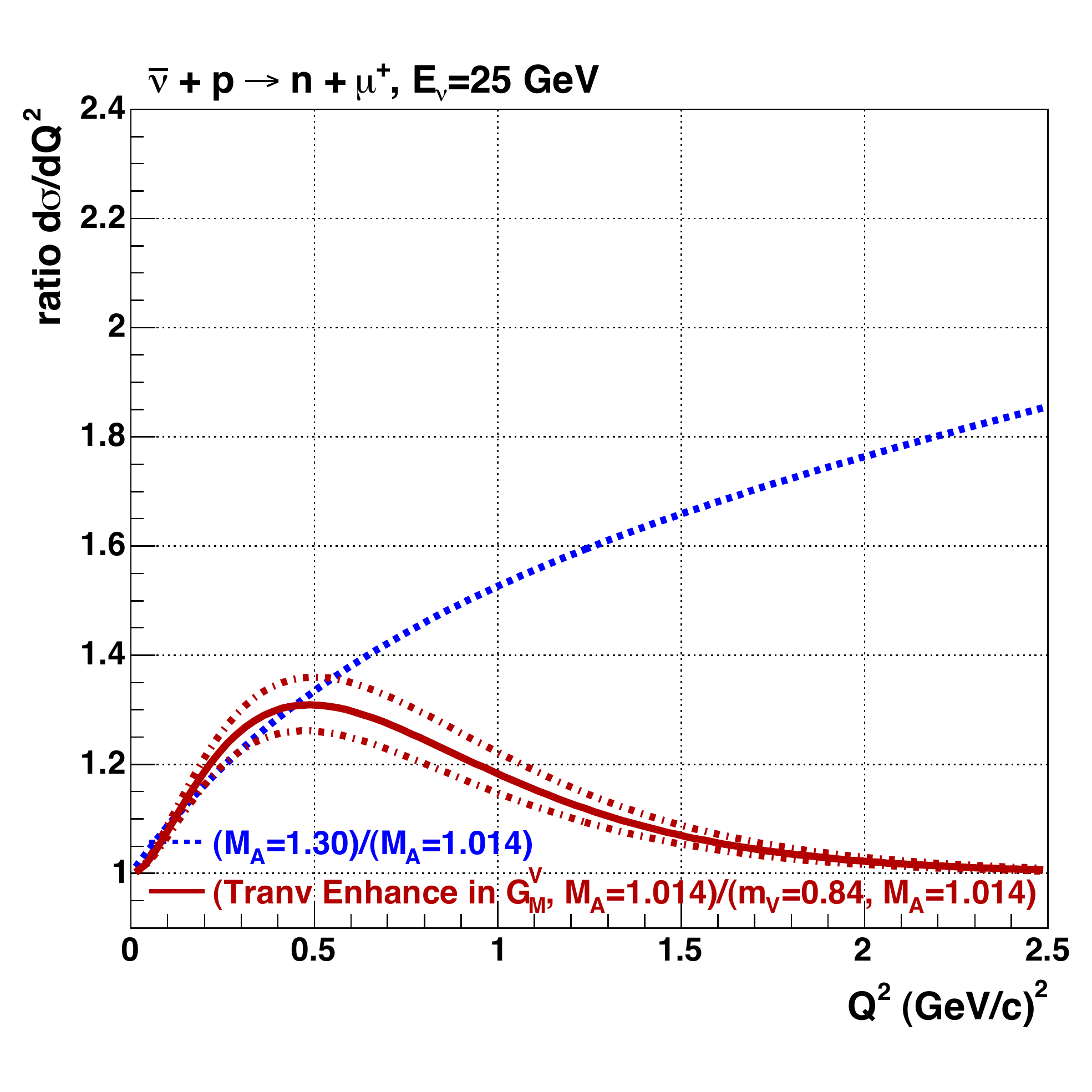}
\vspace{-0.15in}
\includegraphics[width=3.5in,height=2.6in]{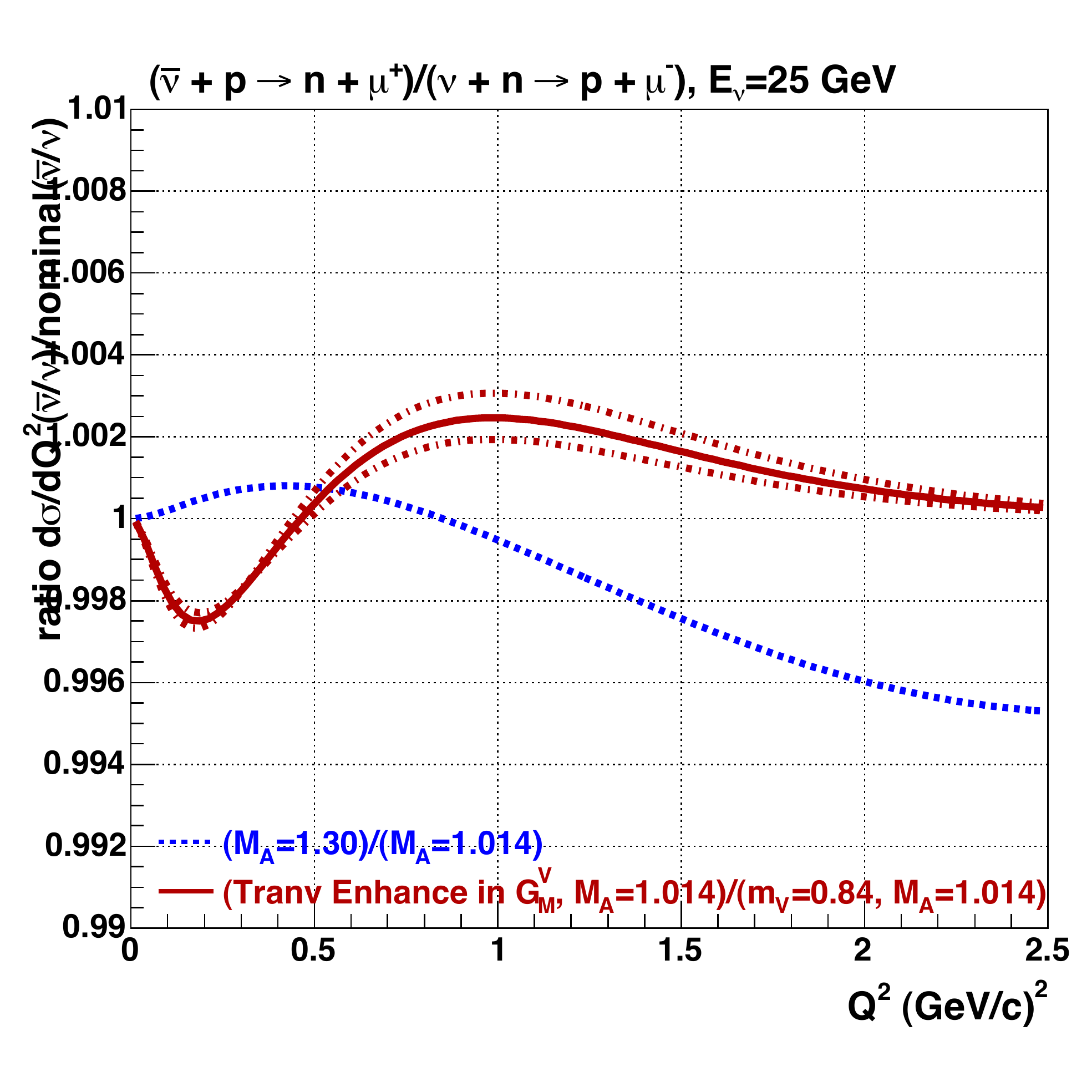}
\vspace{-0.1in}
\caption{ Same as figure~\ref{ratio1} for $\nu_{\mu},\bar{\nu}_\mu$  energies of 25  GeV.}
\label{ratio25}
\end{figure}

 Figure~\ref{totalMEClog} shows a comparison of the various model for a larger energy range (0.1 to 100 GeV). The energy dependence
 for the predictions of the transverse enhancement model originates from the energy dependence of the maximum
 accessible $Q^2$  ($Q^2_{max}$) for QE scattering, as shown in Fig.\ref{q2max}. The lower energies have lower $Q^2_{max}$ where the transverse enhancement is large, while higher energies have a higher $Q^2_{max}$, where the transverse enhancement is small. The differential cross section 
at high energy is almost independent of energy, as shown in Fig. \ref{diff10}, \ref{diff25}, \ref{ratio10} and \ref{ratio25}.   

    %
 %Figure 
%
%Figure 13
   \section{Testing the model with neutrino data}
  
The MINERvA high statistics neutrino experiment\cite{minerva}  at Fermilab is currently taking data with a fully active
scintillator target calorimeter in the NUMI beam (with both neutrinos and antineutrinos).
The QE differential cross sections would be measured as a function of $Q^2$
at a variety of neutrino energies within one single experiment, and compared to
the predictions of various models.

      \section{Appendix:  $\nu_{\mu},\bar{\nu}_\mu$  nucleon/nucleus scattering}
    At a fixed value of the  final state invariant mass $W$, the differential cross section for
$\nu_{\mu},\bar{\nu}_\mu$  scattering at incident energy E  is given\cite{adler2} by: 
%\begin{widetext}
\begin{eqnarray} 
\label{crosseqn}
\di
\frac{d\si}{dQ^2 dW}=  \frac{G^2}{2\pi}\cos^2\theta_C\frac{W}{M} \Biggl\{  
\frac{1}{2E^2}{\cal W}_1 \left[Q^2+m_\mu^2\right] \nonumber \\
%+{\cal W}_2  \left[ 2(1-\frac {\nu}{E} )  - \frac {1}{2E^2} (Q^2+m_\mu^2) \right]  \\ \nonumber 
+ {\cal W}_2  +{\cal W}_2  \left[ -\frac {\nu}{E}   - \frac {1}{4E^2} (Q^2+m_\mu^2) \right]  \\ \nonumber 
\pm {\cal W}_3 \left[ \frac{Q^2}{2ME} -  \frac{\nu}{4E}  \frac { Q^2+m_\mu^2}{ME}  \right] \\ \nonumber 
+\frac{{\cal W}_4}{M^2}m_\mu^2 \frac{(Q^2+m_\mu^2)}{4E^2} -\frac{{\cal W}_5}{ME} m_\mu^2 
 \Biggr\}
\end{eqnarray}
%.
Here, $\frac{G^2}{2\pi}\cos^2\theta_C = 80\times 10^{-40}~cm^2/GeV^2$.
The final state muon mass places  the following
kinematic limits\cite{reno}  on $x=Q^2/2M\nu$ and $y=\nu/E$:
\begin{eqnarray}
 && \frac{m_\mu^2}{2M(E_\nu -m_\mu)}\ \leq \  x\ \leq\ 1\ ,\\
&&a\ -\ b\ \leq \ y\ \leq \ a\ +\ b \ ,
\end{eqnarray}
where the quantities $a$ and $b$ are
\begin{eqnarray*}
a & = & \Biggl[1-m_\mu^2\Biggl(\frac{1}{2ME_\nu x}+\frac{1}
{2E_\nu ^2}\Biggr)\Biggr]  
%\\ \nonumber &&\times (2+Mx/E_\nu)^{-1}\\ 
/(2+Mx/E_\nu)\ ,\\
b & =&  \Biggl[\Biggl(1-\frac{m_\mu^2}{2ME_\nu
  x}\Biggr)^2-\frac{m_\mu^2}{E_\nu ^2}\Biggr]^{1/2}  
%  \\ && \times (2+Mx/E_\nu )^{-1}\ .
/(2+Mx/E_\nu )\ .
\end{eqnarray*}

Or alternatively,  for a fixed energy and  $Q^2$, there is a maximum value of
$W$ which is given by\cite{paschos}:
\[
\begin{array}{r} \di 
W_{+}^2(Q^2)=\Biggl[ \frac14 s^2 a_-^2\left(\frac{m_\mu^4}{s^2}-2\frac{m_\mu^2}{s}\right) 
- \left(Q^2+\frac12 m_\mu^2 a_+^2 \right)^2 
\\  \di
+ s \, a_{-} \left( Q^2 + \frac{m_\mu^2}{2} a_+ \right) \Biggr] \left/ \left[a_- (Q^2+m_\mu^2)\right], \right.
\end{array}
\]
where $s=2M E +M^2$, $a_{\pm}=1 \pm M^2/s$. 
For QE scattering, this corresponds to a minimum and maximum accessible $Q^2$ for
a given neutrino energy.  The maximum accessible $Q^2$ ($Q^2_{max}$)  for QE events as a function of neutrino energy is shown in Fig.~\ref{q2max}.

   %
      %Figure 14
%
%2.6133 to become 3.503
%Figure f15

%2.6133  to become 3.503
 % Figure 16
%
%
% Figure 817
\subsection {Quasielastic   $\nu_{\mu},\bar{\nu}_\mu$   scattering}
A theoretical framework for  quasi-elastic ($\nu_{\mu},\bar{\nu}_\mu$)-Nucleon Scattering
has been given by Llewellyn Smith~\cite{Lle_72}. Here, 
we use the notation of Llewellyn Smith 
(except that $F_V^2$ in our notation is equal to $\xi_{ls} F_V^2$
in Llewellyn Smith's notation, where $\xi_{ls}=(\mu_p-1-\mu_n$)).
In addition,  we use $Q^2$ while Llewellyn Smith uses $q^2$ where  
$$q^2 = q^2_0 - \vec q_3^2 = -4E_0E^\prime \sin^2{\theta\over 2} =
-Q^2 \; .$$

The hadronic current for QE $\nu_{\mu},\bar{\nu}_\mu$  scattering is given by~\cite{Lle_72}
\begin{eqnarray*}
 \lefteqn{<p(p_2)|J_{\lambda}^+|n(p_1)>  =   } \nonumber \\
& \overline{u}(p_2)\left[
  \gamma_{\lambda}{\cal F}_1^V(q^2)
  +\frac{\D i\sigma_{\lambda\nu}q^{\nu}{\cal F}_2^V(q^2)}{\D 2M} \right. \nonumber \\
 & \left. ~~~~~~~~~~~+\gamma_{\lambda}\gamma_5{\cal F}_A(q^2)
+\frac{\D q_{\lambda}\gamma_5{\cal F}_P(q^2)}{\D M} \right]u(p_1),
\end{eqnarray*}
where $q=k_{\nu}-k_{\mu}$, and 
$M=(m_p+m_n)/2$.  Here, $\mu_p$ and $\mu_n$ are the 
proton and neutron magnetic moments.
We assume that there are no second class currents, so the scalar
form factor  ${\cal F}_V^3$ and the tensor form factor ${\cal F}_A^3$
need not be included. 
Using the above current, the QE cross section is
\begin{eqnarray*}
\label{QEeqn}
 \lefteqn{ \frac{d\sigma^{\nu,~\overline{\nu}}}{dQ^2} = 
  \frac{M^2G_F^2cos^2\theta_c}{8{\pi}E^2_{\nu}}\times }  \nonumber \\
&\left[A(Q^2) \mp \frac{\D (s-u)B(Q^2)}{\D M^2} + \frac{\D C(Q^2)(s-u)^2}{\D M^4}\right],
\end{eqnarray*}
where $ s-u = 4ME_{\nu} -Q^2 - m_\mu^2$.
\begin{eqnarray}
A(Q^2)&= & \frac{m_\mu^2+Q^2} {M^2} 
 \Biggl\{ \left( 1+\tau \right) |{\cal F}_A|^2   -\left(1-\tau \right)|{\cal F}_1^V|^2 \nonumber \\
 &+&\tau \left(1-\tau \right) |{\cal F}_2^V|^2  + 4\tau {\cal F}_1^V{\cal F}_2^V  \Biggl\}  \nonumber \\
  &-&  \frac{m_\mu^2+Q^2} {M^2}  \frac{m_\mu^2}{4M^2}    \Biggl\{ \left(|{\cal F}_1^V+{\cal F}_2^V|^2  \right)\nonumber \\
 & +&({\cal F}_A+2{\cal F}_P)^2  - 4(1+\tau){\cal F}_P^2 \Biggl\} 
 % \nonumber 
 %\\
 \end{eqnarray} 
 \begin{eqnarray}
B(Q^2)& =& 4\tau {\cal F}_A({\cal F}_1^V+{\cal F}_2^V)= 4\tau {\cal F}_A {\cal G}_M^V,
%\nonumber \\ 
 \end{eqnarray} 
 \begin{eqnarray}
C(Q^2)& =& \frac{1}{4}\left(|{\cal F}_A|^2 + |{\cal F}_1^V|^2 +
 \tau \left|{\cal F}_2^V\right|^2\right)\nonumber \\
&= & \frac{1}{4}\left(|{\cal F}_A|^2 +|F_V(Q^2)|^2\right) 
\end{eqnarray} 
Where $\tau=Q^2/4M^2$.
% Although we have
%have not shown terms of order $(m_\mu/M)^2$, and 
%terms including ${\cal F}_P(Q^2)$ (which
%is multiplied by  $(m_\mu/M)^2$), these terms
%${\cal F}_P(q^2)$ and  terms of order  $(m_l/M)^2$ 
%are included in our calculations~\cite{Lle_72}.)
The form factors $ F_1^V(Q^2)$ and  $F_2^V(Q^2)$
are given by:
$${\cal F}_1^V(Q^2)=
\frac{{\cal G}_E^V(Q^2)+\frac{\D Q^2}{\D 4M^2}{\cal G}_M^V(Q^2)}{1+\frac{\D Q^2}{\D 4M^2}},
$$
$$
{\cal F}_2^V(Q^2) =\frac{{\cal G}_M^V(Q^2)-{\cal G}_E^V(Q^2)}{1+\frac{\D Q^2}{\D 4M^2}}.
$$
From conserved vector current (CVC)  $ {\cal G}_E^V(Q^2)$ and $ {\cal G}_M^V(Q^2)$ 
are related to the electron scattering form factors
$G_E^p(Q^2)$, $G_E^n(Q^2)$, $G_M^p(Q^2)$, and $G_M^n(Q^2)$:
$$ 
{\cal G}_E^V(Q^2)=G_E^p(Q^2)-G_E^n(Q^2), 
$$
$$
{\cal G}_M^V(Q^2)=G_M^p(Q^2)-G_M^n(Q^2). 
$$
We also define 
  $$ |  {\cal F}_V(Q^2)|^{2}=
\frac{[{\cal G}_E^V(Q^2)]^2+ \tau [{\cal G}_M^V(Q^2)]^2}{1+\tau}.$$
The axial form
factor ${\cal F}_A$  can be approximated by the dipole form
$$
{\cal F}_A(q^2)=\frac{g_A}{\left(1+\frac{\D Q^2}{\D M_A^2}\right)^2 },
$$
Where $g_A=-1.267$. 

The pseudoscalar form factor ${\cal F}_P$
is related to ${\cal F}_A$ by PCAC and is given by: 
$$
{\cal F}_P(q^2)=\frac{2M^2{\cal F}_A(q^2)}{M_{\pi}^2+Q^2}.
$$
In the expression for the QE cross section,
${\cal F}_P(q^2)$ is multiplied by  $(m_\mu/M)^2$. 
Therefore, 
in $\nu_{\mu},\bar{\nu}_\mu$  interactions, this effect 
is very small except at very low energy, below 0.2~GeV.

% ${\cal F}_A(q^2)$ needs to be extracted from 
%QE neutrino scattering. At low $Q^2$,
%${\cal F}_A(q^2)$ can also be extracted from pion
%electroproduction data.

In the dipole approximation, 
  $${\cal G}_M^V (Q^2)\approx 4.706~G_D^V(Q^2).$$
  %
% $$ |F_V(Q^2)|^{2}=
%\frac{({\cal G}_E^V(Q^2))^2+ \tau ({\cal G}_M^V(Q^2))^2}{1+\tau},
%$$
In our analysis we  apply  $BBBA2007_{25}$ corrections\cite{quasi}  to the
dipole parametrization of the electromagnetic form factors as described in reference\cite{quasi}.

By comparing equations \ref{crosseqn} and \ref{QEeqn}
and using the  following expressions:
 $$ {\cal F}_1^V(Q^2) +F_2^V(Q^2) = {\cal G}_M^V(Q^2),$$
 $$| {\cal F}_1^V(Q^2)|^2 + \tau | {\cal F}_2^V(Q^2)|^2 = |{\cal F}_V(Q^2)|^2,$$

we obtain the following  relationships between the structure functions and form factors for $\nu_\mu,\bar{\nu}_\mu$ 
QE scattering on free nucleons:

$${\cal W}^{\nu-vector}_{1-Qelastic} =\delta(\nu-\frac{Q^2}{2M})\tau |{\cal G}_M^V (Q^2)|^2$$
$${\cal W}^{\nu-axial}_{1-Qelastic} = \delta(\nu-\frac{Q^2}{2M})(1+\tau)|{\cal F}_A (Q^2)|^2$$
$${\cal W}^{\nu-vector}_{2-Qelastic} =
%2\frac {M}{\nu}
 \delta(\nu-\frac{Q^2}{2M})|{
 \cal F}_V (Q^2)|^2$$
$${\cal W}^{\nu-axial}_{2-Qelastic} =
%2\frac {M}{\nu} 
\delta(\nu-\frac{Q^2}{2M})|{\cal F}_A (Q^2)|^2$$
$${\cal W}^{\nu}_{3-Qelastic} = 
%2\frac {M}{\nu} 
\delta(\nu-\frac{Q^2}{2M})|2 {\cal G}_M^V(Q^2) {\cal F}_A (Q^2)|$$
 $${\cal W}^{\nu-vector}_{4-Qelastic} =
%2\frac {M}{\nu}
 \delta(\nu-\frac{Q^2}{2M})\frac{1}{4}(|{\cal F}_V (Q^2)|^2 - |{\cal G}_M^V (Q^2)|^2)$$
  $${\cal W}^{\nu-axial}_{4-Qelastic} =
%2\frac {M}{\nu}
 \delta(\nu-\frac{Q^2}{2M})\times \frac{1}{4} \times $$
$$ \left[ {\cal F}_A^2(Q^2)+(\frac{Q^2}{M^2}+4)|{\cal F}_p(Q^2)|^2 -({\cal F}_A(Q^2) +2{\cal F}_P(Q^2))^2 \right]$$
 $${\cal W}^{\nu-vector}_{5-Qelastic} =
%2\frac {M}{\nu}
 \delta(\nu-\frac{Q^2}{2M})\frac{1}{2}|{\cal F}_V (Q^2)|^2$$
  $${\cal W}^{\nu-axial}_{5-Qelastic} =
%2\frac {M}{\nu}
 \delta(\nu-\frac{Q^2}{2M})\frac{1}{2}|{\cal F}_A (Q^2)|^2$$
 
 The  vector part of ${\cal W}_4$  and ${\cal W}_5$ are related to the
vector part of  ${\cal W}_2$ and  ${\cal W}_1$ by the following expressions\cite{paschos}:
\begin{eqnarray} 
{\cal W}_4^{vector}&=&{\cal W}_2^{vector}\frac{ M^2\nu^2 }{Q^4}-{\cal W}_1^{vector}\frac{ M^2 }{Q^2}
\nonumber \\
{\cal W}^{vector}_5&=&{\cal W}_2^{vector}\frac{ M\nu }{Q^2}\nonumber  
\end{eqnarray}
Note that:
 $$\sigma_T^{vector}\propto  \tau |{\cal G}^V_M (Q^2)|^2;~~\sigma_T^{axial}\propto (1+\tau)|{\cal F}_A (Q^2)|^2$$
 $$\sigma_L^{vector}\propto {({\cal G}_E^V(Q^2))^2}; ~~~\sigma_L^{axial}= 0$$
 
 Therefore, for QE  $\nu_{\mu} $ and $\bar{\nu}_\mu$  scattering  only ${\cal G}_M^V$ contributes
 to the  vector part of the transverse  virtual
 boson absorption cross section. 
 % Note that ${\cal G}_M^V$
% also contributes to vector-axial interference term  $W^{\nu}_{3}$ 
 
%  \section{Appendix: Higher Energy QE Charged Current $\nu_{\mu},\bar{\nu}_\mu$  %Nucleon Scattering}

\end{document}